\documentclass{aastex}
\usepackage{epsfig}
\usepackage{amsmath}
\usepackage{mathrsfs}
\usepackage{subfigure}
\usepackage{textcomp}
\usepackage{ulem}
\usepackage{xfrac}
\usepackage{units}

\newcommand{\begeq}{\begin{equation}}
\newcommand{\fineq}{\end{equation}}
\newcommand{\begeqarray}{\begin{eqnarray}}
\newcommand{\fineqarray}{\end{eqnarray}}
\newcommand{\sigmaT}{\sigma_{\rm{T}}}
\newcommand{\lapprox}{\lower.4ex\hbox{$\;\buildrel <\over{\scriptstyle\sim}\;$}}
\newcommand{\gapprox}{\lower.4ex\hbox{$\;\buildrel <\over{\scriptstyle\sim}\;$}}
\newcommand{\Mach}{\mathscr{M}}

\slugcomment{accepted for publication in ApJ}
\shorttitle{Gas and Radiation Pressure Effects on Accretion Column Properties}
\shortauthors{West, Wolfram, \& Becker}

\begin{document}

\title{A NEW TWO-FLUID RADIATION-HYDRODYNAMICAL MODEL FOR X-RAY PULSAR ACCRETION COLUMNS}

\author{Brent F. West}
\affil{Unites States Naval Academy, Annapolis, MD; bwest@usna.edu}
\author{Kenneth D. Wolfram}
\affil{Naval Research Laboratory (retired), Washington, DC; kswolfram@gmail.com}
\author{Peter A. Becker}
\affil{Department of Physics and Astronomy, George Mason University, Fairfax, VA USA; pbecker@gmu.edu}

\begin{abstract}

Previous research centered on the hydrodynamics in X-ray pulsar accretion columns has largely focused on the single-fluid model, in which the super-Eddington luminosity inside the column decelerates the flow to rest at the stellar surface. This type of model has been relatively successful in describing the overall properties of the accretion flows, but it does not account for the possible dynamical effect of the gas pressure. On the other hand, the most successful radiative transport models for pulsars generally do not include a rigorous treatment of the dynamical structure of the column, instead assuming an ad hoc velocity profile. In this paper, we explore the structure of X-ray pulsar accretion columns using a new, self-consistent, ``two-fluid'' model, which incorporates the dynamical effect of the gas and radiation pressure, the dipole variation of the magnetic field, the thermodynamic effect of all of the relevant coupling and cooling processes, and a rigorous set of physical boundary conditions. The model has six free parameters, which we  vary in order to approximately fit the phase-averaged spectra in Her X-1, Cen X-3, and LMC X-4. In this paper, we focus on the dynamical results, which shed new light on the surface magnetic field strength, the inclination of the magnetic field axis relative to the rotation axis, the relative importance of gas and radiation pressure, and the radial variation of the ion, electron, and inverse-Compton temperatures. The results obtained for the X-ray spectra are presented in a separate paper.



\end{abstract}


\section{INTRODUCTION}

Accretion-powered X-ray pulsars are among the most luminous X-ray sources in the sky, and now number in the hundreds (e.g., Caballero \& Wilms 2012). The availability of the unprecedented resolution provided by modern X-ray observatories is opening up new areas for study involving the coupled formation of the continuum emission and the cyclotron absorption features observed in accretion-powered X-ray pulsar spectra. These sources are of special interest because of the unique combination of extreme physics, including strong gravity, relativistic velocities, high temperatures, strong magnetic fields, and locally super-Eddington radiation luminosities. Although these sources have been studied observationally and theoretically for over five decades, several fundamental issues remain unresolved by the current generation of models. One question that has received considerable attention in the past few years is the possible relation between the luminosity of the source and the energy of the fitted cyclotron absorption feature, driven by observations of correlated (or anticorrelated) variability between these two quantities observed on both pulse-to-pulse timescales, and on much longer timescales (e.g., Becker et al. 2012; Staubert et al. 2007; Staubert et al. 2014).

In the standard model for accretion-powered X-ray pulsars, originally developed by Lamb et al. (1973), the kinetic energy of the infalling gas is converted into observable radiation as the flow is channeled onto one or both magnetic poles by the strong magnetic field ($B \sim 10^{12}\,$G), forming ``hot spots'' on the stellar surface. The X-rays were initially assumed to emerge as fan-shaped beams, generated as the photons escaped through the vertical walls of the accretion column, but it soon became clear that a pencil beam component (representing escape through the column top) was sometimes necessary in order to obtain adequate agreement with the observed pulse profiles (Tsuruta \& Rees 1974; Bisnovatyi-Kogan \& Komberg 1976; Tsuruta 1975).

The typical X-ray pulsar spectrum is a combination of a power-law continuum, combined with an iron emission line and an apparent cyclotron absorption feature, terminating in a high-energy exponential cutoff. The earliest spectral models, based on the emission of blackbody radiation from the hot spots, were unable to reproduce the observed nonthermal power-law continuum. The observation of the putative cyclotron absorption features led to the development of more sophisticated models, based on a static slab geometry, in which the emitted spectrum is strongly influenced by cyclotron scattering (e.g., M\'esz\'aros \& Nagel 1985a,b; Nagel 1981; Yahel 1980a,b). While the magnetized slab models are able to roughly fit the shape of the observed cyclotron absorption features, a remaining problem was the inability to reproduce the observed nonthermal power-law X-ray continuum.

The pioneering literature from the 1970's established the basic theoretical framework for the accretion of matter as the fundamental mechanism powering the emission from hot spots at the magnetic poles in X-ray pulsars (e.g., Pringle \& Reese 1972; Davidson 1973; Lamb et al. 1973; Basko \& Sunyaev 1976). Later work by Wang \& Frank (1981) and Langer \& Rappaport (1982) improved our understanding of the details of the fluid flow and its relation to the radiation production. The Wang \& Frank (1981) model is based upon a dipole field geometry, and comprises two adjacent flow zones, separated in radius. The upper region is a single-fluid, 2D regime in which the field-aligned, inflowing free-fall plasma is decelerated by radiation pressure. The lower 1D ``collisional regime'' is located just above the stellar surface, and is a two-fluid zone in which the deceleration is created by a strong gradient in the gas pressure. The main weakness of the model is the lack of a detailed treatment of the radiation spectrum, which results in the inability of the model to either predict observed X-ray spectra, or to properly account for the exchange of energy between the radiation and the gas. Hence their dynamical results cannot be viewed as  self-consistent.

The model of Langer \& Rappaport (1982) focuses solely on low-luminosity sources ($\dot M \lapprox 10^{16} \, {\rm g \, s}^{-1}$), in which the radiation field exerts negligible pressure on the infalling material. Their two-fluid dipole model investigates the field-aligned hydrodynamics between the stellar surface and the upper boundary, which is assumed to be a classical, gas-mediated shock. Although X-ray spectra are computed, the lack of coupling between the hydrodynamics and the radiative transfer means that the results are not necessarily self-consistent. In particular, their model is unable to describe how the characteristic power-law shape of the observed X-ray spectra is developed, nor can it conclusively establish the conditions under which a discontinuous shock is expected to form. The results obtained by Langer \& Rappaport (1982) suggest that most of the escaping radiation consists of cyclotron line photons, in the low-luminosity sources that they treated. However, we find in Paper~II that in the high-luminosity sources, the observed spectrum is dominated by Comptonized bremsstrahlung.

It became clear in later work that the power-law continuum was the result of a combination of bulk and thermal Comptonization occurring inside the accretion column. The first physically-motivated model based on these principles that successfully described the shape of the X-ray continuum in accretion-powered pulsars was developed by Becker \& Wolff (2007, hereafter BW07). This new model allowed for the first time the computation of the X-ray spectrum emitted through the walls of the accretion column based on the solution of a fundamental radiation transport equation. While the BW07 model has demonstrated success in reproducing the observed X-ray spectra for several higher luminosity sources, the model is nonetheless quite simplified from a physical perspective, and it does not include, for example, a thermodynamic calculation of the electron temperature variation, or a hydrodynamical calculation of the variation of the bulk inflow (accretion) velocity.

Kawashima et al. (2016) developed a 2D accretion model in spherical coordinates for a neutron star with canonical mass $M_* = 1.4\,M_\odot$, although they did not assume that the flow follows the magnetic field exactly. Their model includes the existence of a radiation-dominated shock located approximately 3\,km above the stellar surface, and the emission of fan-beam radiation at and below the sonic surface. The model exhibits an exponential increase in the gas density as the material enters the extended sinking regime, in agreement with Basko \& Sunyaev (1976). However, the Kawashima et al. (2016) model does not include radiative transfer, or the Compton exchange of energy between the photons and gas. Hence, although the general features of the model provide some interesting clues regarding the hydrodynamical behavior of the flow, it does not provide a self-consistent picture of the relationship between the hydrodynamics and the formation of the observed phase-averaged X-ray spectra.

The availability of copious high-quality spectral data for accretion-powered X-ray pulsars, combined with the lack of a fully self-consistent radiation-hydrodynamical model, has motivated us to investigate the importance of additional radiative and hydrodynamical processes beyond the scope of those considered by BW07. The complexity of the resulting mathematical model precludes the analytical treatment carried out by BW07, and we must therefore solve the problem within the context of a detailed numerical simulation. The new simulation described here includes the implementation of a realistic dipole geometry, rigorous physical boundary conditions, and a self-consistent treatment of the energy transfer between electrons, ions, and radiation. We refer to the formalism as a ``two-fluid'' model, due to the explicit treatment of the separate dynamical effects of the gas and radiation pressure, which is analogous to the two-fluid treatments of cosmic-ray acceleration in supernova-driven shock waves (e.g., Becker \& Kazanas 2001).

This is the first in a series of two papers in which we describe in detail the new coupled radiative-hydrodynamical model. The integrated approach involves an iteration between an ODE-based hydrodynamical code that determines the dynamical structure, and a PDE-based radiation transport code that computes the X-ray spectrum. The iterative process converges rapidly to yield a self-consistent description of the dynamical structure over the full length of the accretion column, as well as the energy distribution in the emergent radiation field. In this paper (Paper I), we focus on solving the coupled hydrodynamical conservation equations to determine the column structure, and in Paper~II we present the results for the X-ray spectra.

The flow velocity and electron temperature profiles computed here are used as input for the spectral analysis conducted in Paper II, which focuses on solving the fundamental photon transport equation in a dipole geometry using the {\it COMSOL} multiphysics environment. The linkage between the two simulation components is carried by the inverse-Compton temperature profile, which depends on the shape of the radiation energy distribution. The inverse-Compton temperature profile, which is an output from the {\it COMSOL} environment, is used as an input to a {\it Mathematica} code that computes the accretion column structure by solving the ODEs. The output velocity and electron temperature profiles computed using {\it Mathematica} are then used as input to the {\it COMSOL} simulation, and the process is repeated until the inverse-Compton and electron temperature profiles converge, as discussed in detail below. In Paper II we present and discuss the phase-averaged X-ray spectra computed using our model for Her X-1, Cen X-3, and LMC X-4, and compare the results with the observational data in order to determine the model parameters for sources covering a wide range of luminosities.

The paper is organized as follows. In Section~2, we discuss the relation between the accretion disk and the pulsar magnetosphere, and the approximations we will use to treat the effect of the cyclotron resonance on the electron scattering occurring in the strong magnetic field. We also discuss the equation of state used to describe the thermodynamics of the coupled gas and radiation. In Section~3 we introduce the conservation relations for mass, momentum, and energy, and we discuss the fundamental energy exchange processes that couple the electrons with the ions and the radiation field. In Section~4 we derive the fundamental boundary conditions operative at the top of the accretion column, at the stellar surface, and at the thermal mound surface. In Section~5 we describe the procedure used to solve the coupled set of conservation relations to obtain a self-consistent description of the radiative and hydrodynamical structure of the accretion column. The new model is applied to three sources in Section~6, and in Section~7 we discuss our results and describe our plans for future research.

\section{PHYSICAL BACKGROUND}

The analytical model developed by BW07 has proven to be quite useful in the physical interpretation of the X-ray spectra observed from a number of accretion-powered X-ray pulsars, including Her X-1, Cen X-3, and LMC X-4 (BW07; Wolff et al. 2016), by providing an alternative to the commonly used ad hoc mathematical forms, such as power-laws, exponential cutoffs, and Gaussian emission and absorption features. In addition to providing good spectra fits, the BW07 model also yields meaningful estimates for key source parameters, such as the electron temperature $T_e$, the hot-spot radius $r_0$, and the scattering cross-sections for photons propagating either perpendicular or parallel to the magnetic field axis, denoted by $\sigma_\perp$ and $\sigma_\parallel$, respectively. However, the success of the BW07 model leads to further questions about how the underlying assumptions built into the model may be affecting the estimates for the fitting parameters. This is a multi-faceted question since a number of different idealizations and assumptions had to be incorporated into the BW07 model in order to made an analytical solution tractable. We shall discuss these assumptions below, and relate them to the work presented in this paper.

In the BW07 model, the column radius $r_0$ is treated as a constant, so that the accretion column is cylindrical. This is perhaps a reasonable assumption near the base of the column, but if the height of the column becomes a significant fraction of the stellar radius, which we shall see is true in the case of our new models, then the effects of the dipole curvature of the magnetic field cannot be ignored. Beyond the cylindrical geometry, the mathematical formalism employed by BW07 also incorporates two additional idealizations in order to make the problem amenable to analytical solution. The first is that the actual physical profile of the accretion velocity, $v$, was replaced with the ad hoc form $v \propto \tau$, where $\tau$ is the scattering optical depth measured upward from the stellar surface. This profile correctly results in the stagnation of the flow at the stellar surface, but it does not merge smoothly with the free-fall velocity profile that characterizes the infalling material above the top of the accretion column.

The second key assumption made by BW07 is that the electrons in the accretion column comprise an isothermal distribution, with no vertical variation of the temperature. This constant temperature assumption is required in order to separate the transport equation for the radiation field, which is almost certainly wrong at some level, but it's not clear a priori how much variation in the temperature is expected, since Compton scattering is likely to regulate the temperature and cool the electrons, whereas bulk compression and the Coulomb transfer of kinetic energy from the protons will tend to heat the electrons. There are also additional effects due to the heating and cooling that occur via bremsstrahlung and cyclotron emission and absorption. The entire accretion scenario over the full length of the accretion column, including the dynamics, the energy transfer, and the solution for the radiation field, is in reality far more complicated than could be represented by the idealized mathematical model developed by BW07.

\begin{figure}[htbp]
\centering
\includegraphics[width=4in]{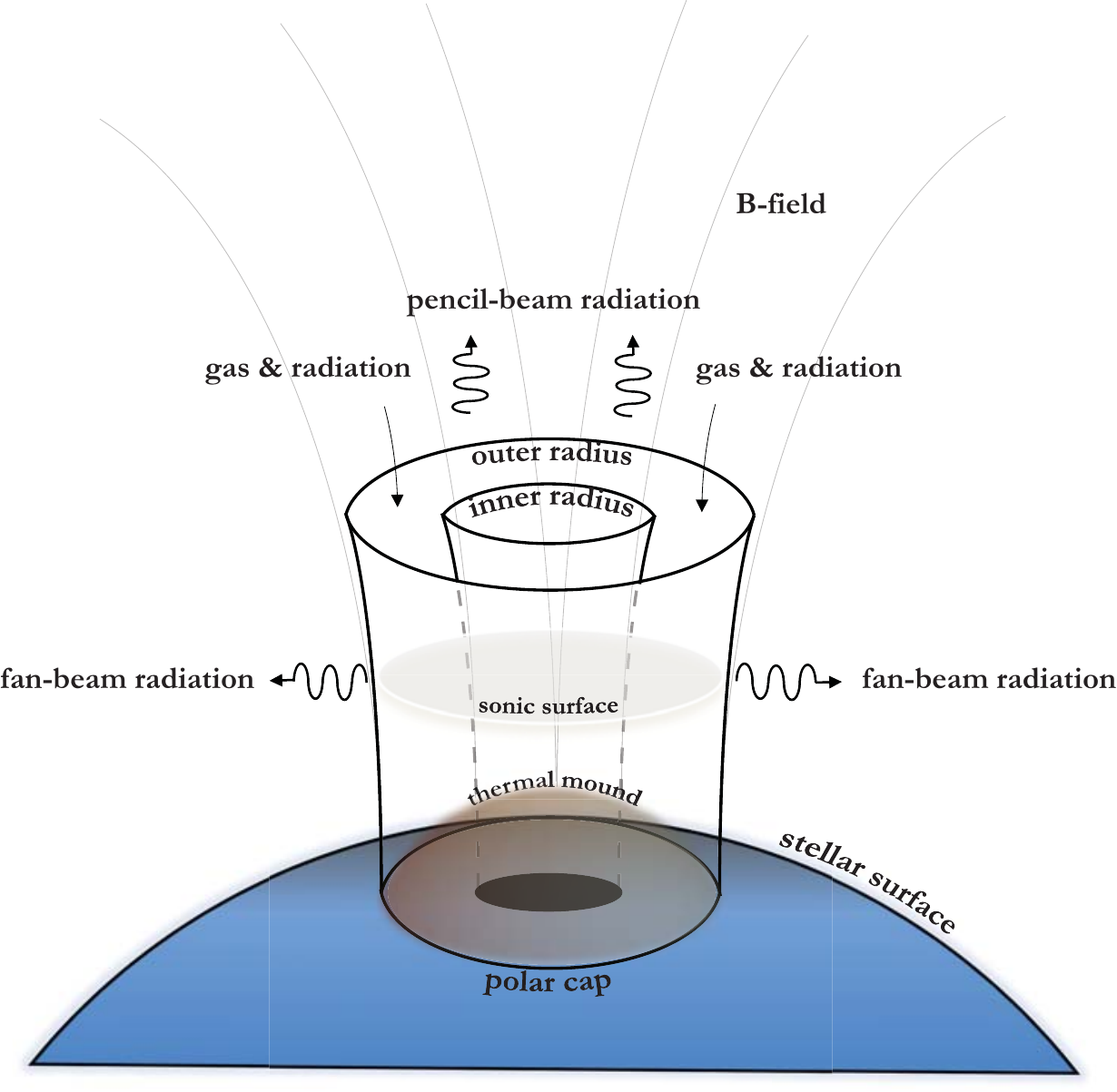}
\caption[Accretion Column]{Accretion column formation in the two-fluid model. Ions and electronsenter at the top of the column as coupled and interacting fluids. X-ray photons are produced in the column and escape through the top and the sides as pencil and fan beam components, respectively. Also indicated are the thermal mound surface (where the absorption optical depth in the parallel direction equals unity, $\tau_{\parallel}^{\rm{abs}}=1$), and the radiation sonic surface, where the radiation Mach number $\Mach_r = 1$.}
\label{fig:column}
\end{figure}

Our goal here is to relax some of the key assumptions incorporated into the BW07 model, and reexamine the resulting structure of the accretion column using a more realistic physical description. The problem is quite complex because of the dominant role the radiation pressure plays in mitigating the accretion velocity as the infalling material decelerates towards the stellar surface. Hence one must employ a self-consistent methodology in which the nonlinear coupling between the radiation spectrum and the flow dynamics is treated explicitly. In the present paper, we will model X-ray pulsar accretion flows in a dipole geometry, including the vertical variation of the electron temperature, and the thermodynamic effects of all of the relevant coupling mechanisms (see Figure~\ref{fig:column}). We also incorporate the dynamical effect of the individual pressure components due to the ions, the electrons, and the radiation, and we allow for the possible presence of a hollow cavity within the accretion column.

\subsection{Accretion Power and X-ray Luminosity}
\label{subsection:accretion power}

The ultimate power source for the observed X-ray emission from accretion-powered pulsars is gravity, and therefore the total power available is equal to the accretion luminosity, defined by
\begin{equation}
L_{\rm acc} \equiv \frac{G M_* \dot M}{R_*} \ ,
\label{eqn:accretion luminosity definition}
\end{equation}
where $G$ is the gravitational constant, $\dot M$ denotes the accretion rate, and $M_*$ and $R_*$ are the stellar mass and radius, respectively. If no kinetic or thermal energy enters the star (Lenzen \& Tr\"{u}mper 1978), then the X-ray luminosity $L_{\rm X}$ is given by the relation
\begin{equation}
L_{\rm X} = L_{\rm acc} \ ,
\label{eqn:observed luminosity definition}
\end{equation}
although we note that Basko \& Sunyaev (1976) have argued that some energy may diffuse down into the star.

Becker et al. (2012) have shown there is a critical luminosity, $L_{\rm crit}$, below which the pressure of the radiation alone is insufficient to bring the matter to rest, and therefore Coulomb interactions must cause the final deceleration to stagnation at the stellar surface. Additionally, very low-luminosity sources ($L_{\rm X} \lesssim 10^{34}\, \textrm{erg s}^{-1}$) can potentially exhibit the presence of a gas-mediated (discontinuous) shock downstream from the (smooth) radiation shock (Langer \& Rappaport 1982). The precise locations of the radiation and gas shocks largely depend upon the source luminosity and the upstream and downstream boundary conditions. Hence, in order to fully understand the hydrodynamic and thermodynamic processes that determine the structure of the accretion column over the full range of observed luminosities ($L_{\rm X} \sim 10^{34-38}\, \textrm{erg s}^{-1}$), it is essential to include the effect of gas pressure in the model. In this paper, we focus on treating three well-known luminous X-ray pulsars, and we defer discussion of low-luminosity sources, such as X Persei, to a later paper.

\subsection{Pulsar Magnetosphere}
\label{subsection:conical geometry}

The magnetic field surrounding a neutron star is well approximated by a dipole configuration, with spherical vector components given by (e.g., Jackson 1962)
\begin{equation}
B_r = -\frac{B_* R_*^3 \cos \theta}{r^3} \ , \quad
B_\theta = -\frac{B_* R_*^3 \sin \theta}{2 r^3} \ , \quad
B_\phi = 0 \ ,
\end{equation}
where the polar angle $\theta$ is measured from the magnetic field axis, $B_*$ denotes the field strength measured at the magnetic pole on the surface of the star, and $R_*$ is the stellar radius. The magnitude of the field, $|B|$, varies with the spherical radius $r$ according to
\begin{equation}
|B|=\frac{B_* R_*^3}{2 r^3} \sqrt{1 + 3 \cos^2 \theta} \ ,
\label{eqn:Bmag}
\end{equation}
and therefore the field strength decreases by a factor of two between the magnetic pole ($\theta=0$) and the magnetic equator ($\theta=\pi/2$), so that
\begin{equation}
B_{\rm eq} = \frac{1}{2} B_* \ ,
\label{eqn:dipolevar}
\end{equation}
where $B_{\rm eq}$ denotes the magnitude of the field at the stellar surface along the magnetic equator.

In the scenario considered here, the accreting gas is entrained onto magnetic field lines from the surrounding disk, and fed onto the magnetic poles of the star. The detailed density distribution inside the accretion column is influenced by a variety of unknown geometrical factors, such as the angle between the star's rotation and magnetic axes (Lamb et al. 1973; Ghosh et al. 1977, Elsner \& Lamb 1977). In some sources, the entrainment of matter from the disk results in a partially filled column, but in other sources, such as Her X-1, an alternate accretion mode seems to be at play, in which the gas is introduced into the polar cap region from a dense atmosphere concentrated above the cap, and the accretion column is completely filled (Boroson et al. 2001).

\begin{figure}[htbp]
\centering
\includegraphics[width=2.0in]{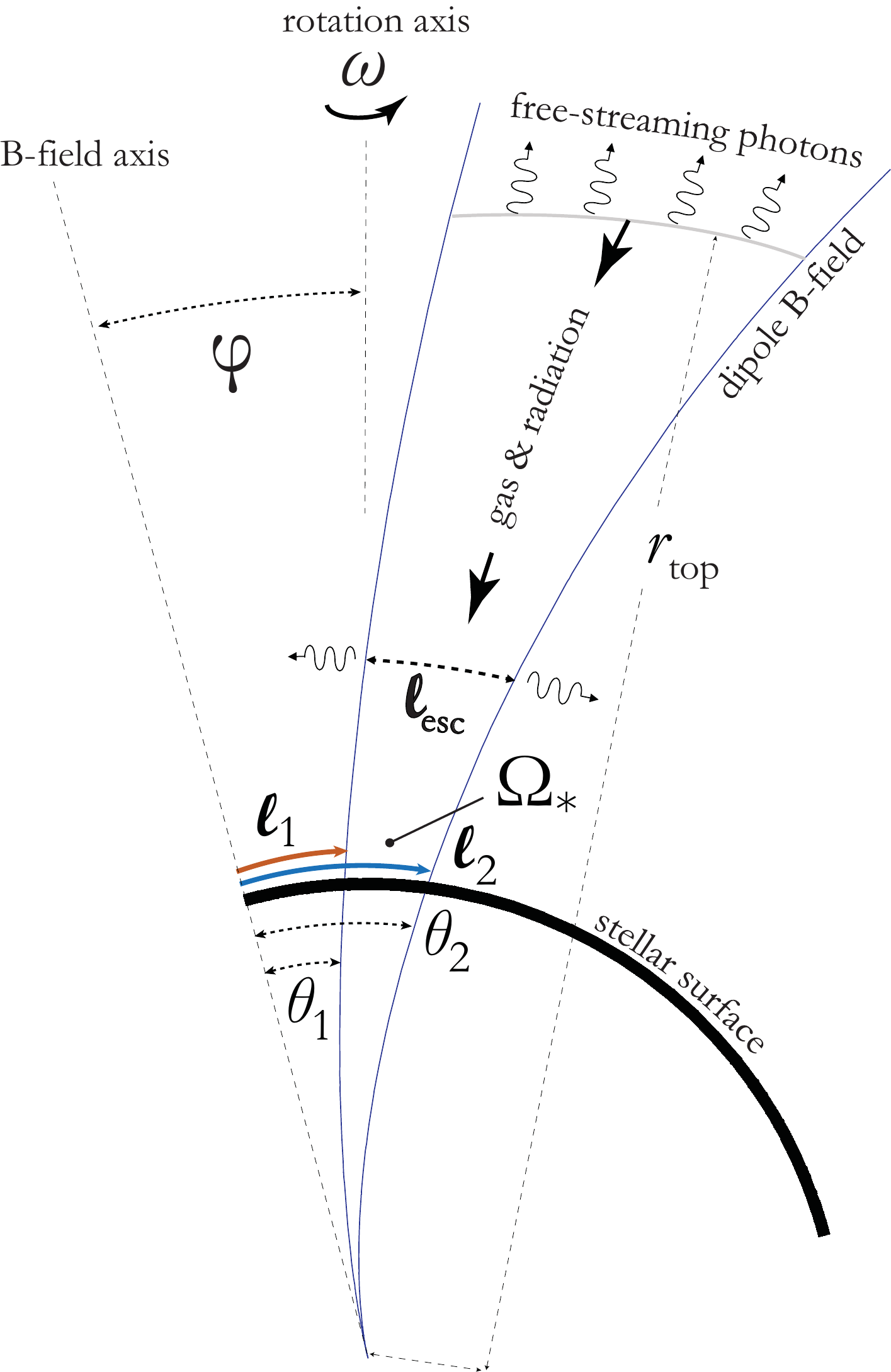}
\caption[Dipole Model Geometry]{Geometry of the dipole accretion column. The inner and outer arc-radii of the accretion column at the stellar surface are denoted by $\ell_1$ and $\ell_2$, respectively, with associated surface angles $\theta_1$ and $\theta_2$. The magnetic field axis is tilted by angle $\varphi$ with respect to the rotation axis. The fan component is formed by photons diffusing through the side walls, and the pencil component is formed by photons that free-stream through the upper surface of the column at radius $r_{\rm top}$.}
\label{fig:dipole geometry}
\end{figure}

We define the physical extent of the accretion column at the stellar surface using the polar angles $\theta_1$ and $\theta_2$, which are measured from the magnetic field axis and delineate the inner and outer boundaries of the dipole accretion column at the stellar surface, respectively. The corresponding inner and outer arc-length surface radii, denoted by $\ell_1$ and $\ell_2$, respectively, are given by (see Figure~\ref{fig:dipole geometry})
\begin{equation}
\ell_1 = \theta_1 \, R_* \ , \ \ \ \
\ell_2 = \theta_2 \, R_* \ .
\label{eqn:thetas}
\end{equation}
Note that the column is partially hollow if $0 < \ell_1 < \ell_2$, and it is completely filled if $\ell_1 = 0$. The solid angle subtended by the accretion column at the stellar surface, $\Omega_*$, is related to $\theta_1$ and $\theta_2$ via
\begin{equation}
\Omega_* = 2 \pi (\cos{\theta_1}-\cos{\theta_2}) \ .
\label{eqn:solid angle at stellar surface}
\end{equation}
The variable solid angle, $\Omega(r)$, subtended by the accretion column at radius $r$ increases in proportion to $r$ in the dipole field geometry, so that
\begin{equation}
\Omega(r) = \frac{r}{R_*} \ \Omega_* \ .
\label{eqn:solid angle}
\end{equation}

Lamb et al. (1973) provide some insight into the upper limit of the outer polar cap arc-radius $\ell_2$. The stellar surface \textquotedblleft hot spot" has an area which must be less than or equal to $\pi (R_*/R_{\rm A})R_*^2$, and therefore
\begin{equation}
\Omega_2 R_*^2 \lesssim \pi \left(\frac{R_*}{R_{\rm A}} \right) R_*^2 \ ,
\label{eqn:Alfven radius constraints}
\end{equation}
where $R_{\rm A}$ is the Alfv\'{e}n radius and $\Omega_2$ is the solid angle subtended by a filled polar cap of radius $\ell_2$ on the surface of the star, given by
\begin{equation}
\Omega_2 \equiv 2 \pi (1-\cos{\theta_2}) \ .
\label{eqn:theta2}
\end{equation}
It follows that the solid angle of the polar cap is restricted by the condition
\begin{equation}
\Omega_2 \lesssim \pi \, \frac{R_*}{R_{\rm A}}  \ .
\label{eqn:Omega Star Constraint}
\end{equation}
Since the stellar radius is much larger than the polar cap radius $(R_* \gg \ell_2)$, and the Alfv\'en radius is much larger that the stellar radius $(R_{\rm{A}} \gg R_*)$, we can use the small angle approximation, $\sin \theta \approx \theta$, along with Equations~(\ref{eqn:thetas}), (\ref{eqn:theta2}), and (\ref{eqn:Omega Star Constraint}), to conclude that the outer polar cap arc-radius, $\ell_2$, is constrained by the condition
\begin{equation}
\ell_2 \leq R_* \left(\frac{R_*}{R_{\rm{A}}} \right)^{1/2} \ .
\label{eqn:maximum outer radius constraint}
\end{equation}

In the dipole field geometry, the radius $r$ along a field line (which is also a flow streamline in the pulsar application) is a function of the angle $\theta$ measured from the magnetic pole via
\begin{equation}
r(\theta) = R_{\rm eq} \sin^2\theta \ ,
\label{eqn:rtheta}
\end{equation}
where $R_{\rm eq}$ is the radius of the field line in the magnetic equatorial plane ($\theta=\pi/2$). The field lines connected with the inner and outer surfaces of the accretion column have magnetic equatorial radii $R_{\rm eq}$ equal to $R_1$ and $R_2$, respectively, where $R_1 > R_2$ (see Figure~\ref{fig:turnover}). We can relate $R_1$ and $R_2$ to the corresponding polar angles, $\theta_1$ and $\theta_2$, respectively, by setting $r=R_*$ in Equation~(\ref{eqn:rtheta}), which yields
\begin{equation}
R_1 = \frac{R_*}{\sin^2\theta_1} \ , \ \ \ \
R_2 = \frac{R_*}{\sin^2\theta_2} \ .
\label{eqn:R1andR2}
\end{equation}

We define the coordinate $z$ as the altitude above the magnetic equatorial plane, measured along the field line that connects with the {\it outer} wall of the accretion funnel, and with the {\it inner} accretion radius in the Keplerian disk. We therefore have
\begin{equation}
z(\theta) = r(\theta) \cos\theta = R_2 \sin^2\theta \cos\theta \ ,
\label{eqn:zheight}
\end{equation}
where the final result follows from Equation~(\ref{eqn:rtheta}). The dipole field reaches its maximum altitude, $z_c$, at the critical angle $\theta = \theta_c$, and then turns over to extend downward towards the disk. By setting the derivative of Equation~(\ref{eqn:zheight}) with respect to $\theta$ equal to zero, we find that the critical angle is given by
\begin{equation}
\theta_c = \cos^{-1}\left(\frac{1}{\sqrt{3}}\right) = 54.74^\circ\ .
\label{eqn:thetac}
\end{equation}
The corresponding maximum altitude is therefore
\begin{equation}
z_c = r_c \cos \theta_c = \frac{2}{3 \sqrt{3}} \, R_2 \ ,
\label{eqn:zc}
\end{equation}
where the corresponding spherical radius, $r_c$, is related to the magnetic equatorial plane radius, $R_2$, via (see Equation~(\ref{eqn:rtheta}))
\begin{equation}
r_c = \frac{2}{3} \ R_2 \ .
\label{eqn:rcritical}
\end{equation}
A fundamental geometrical restriction of our model is that the spherical radius at the top of the accretion funnel, denoted by $r_{\rm top}$, must be below the dipole turnover radius, $r_c$, associated with the outer-wall field line. Hence we must satisfy the condition
\begin{equation}
r_{\rm top} \le r_c \ .
\label{eqn:rtopmax}
\end{equation}

\subsection{Entrainment from the Disk}
\label{subsection:entrainment}

\begin{figure}[htbp]
\centering
\includegraphics[width=\textwidth]{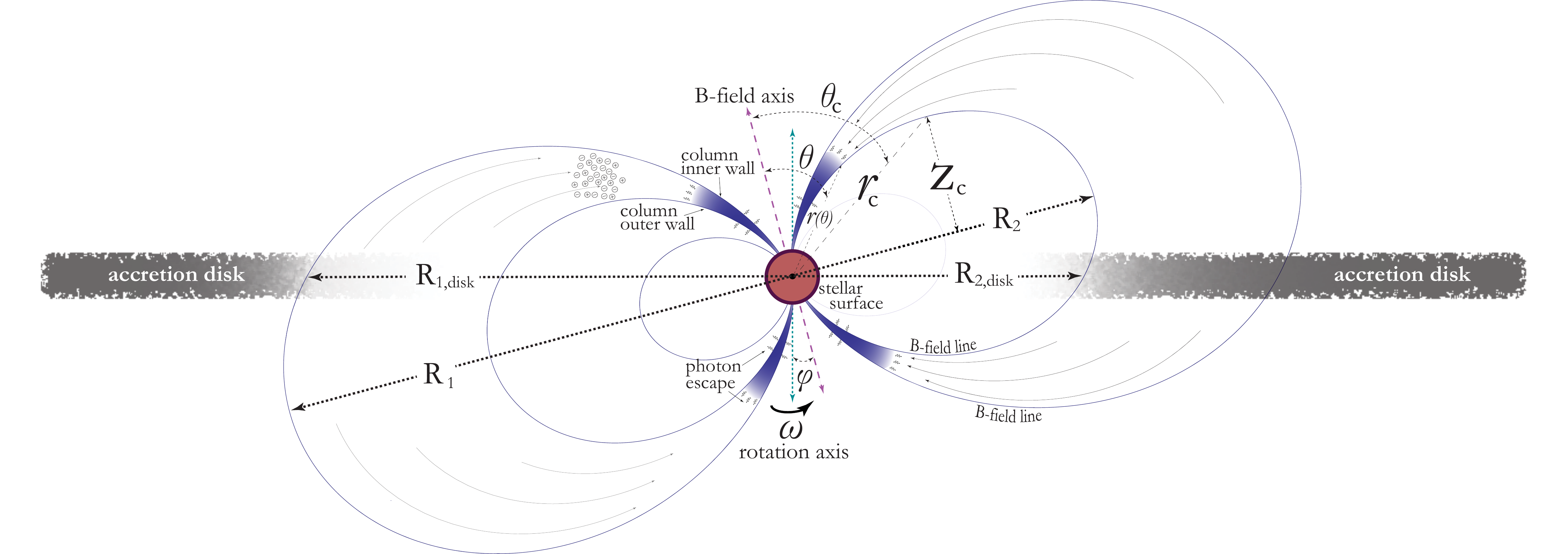}
\caption[Turnover height]{Dipole magnetic field of a neutron star is shown at an inclination $\varphi$ with respect to the rotation axis. The maximum height of a dipole field line above the magnetic equatorial plane occurs at the critical polar angle $\theta=\theta_c=54.74^\circ$. The outer wall of the accretion funnel corresponds to the field line that crosses through the plane of the dipole field at radius $R_2$, and through the plane of the accretion disk at radius $R_{2,\,{\rm disk}}$.}
\label{fig:turnover}
\end{figure}

The pulsations observed from an X-ray pulsar result from a misalignment between the magnetic and rotation axes of the star. The angle between these two axes is denoted by $\varphi$ in our model. The misalignment causes the magnetic field at the surface of the star in the plane of the accretion disk, $B_{\rm disk}$, to sweep between minimum and maximum values during the star's rotation, as observed from a standard reference direction, which we take to be the direction to the companion star. Based on Equation~(\ref{eqn:Bmag}), we find that
\begin{equation}
B_{\rm disk} = \frac{B_*}{2} \sqrt{1 + 3 \sin^2 \alpha} \ ,
\label{eqn:Beqmag}
\end{equation}
where
\begin{equation}
\alpha \equiv \frac{\pi}{2} - \theta
\label{eqn:maglat}
\end{equation}
represents the magnetic latitude in the accretion disk (in the direction towards the companion star), which varies between $\pm \varphi$ as the star rotates, such that $-\varphi \leq \alpha \leq \varphi$ (see Figure~\ref{fig:turnover}).

Matter is picked up from the disk and entrained onto the magnetic field lines at the Alfv\'en radius, $R_{\rm A}$, located where the pressure of the magnetic field balances the ram pressure of the accreting gas (Lamb et al. 1973). Outside this radius, the magnetic field of the neutron star is effectively shielded, and therefore it does not significantly influence the flow structure. Inside the Alfv\'en radius, the strong magnetic field channels the plasma onto the magnetic poles of the star. Due to the complex structure of the pulsar magnetosphere and the uncertainties regarding its interaction with the matter in the disk, it is difficult to precisely compute the value of $R_{\rm A}$ (e.g., Romanova et al. 2003). However, a useful estimate is provided by Lamb et al. (1973), who find that
\begin{equation}
R_{\rm A} \sim 2.6 \times 10^8 \ {\rm cm} \left(\frac{B_{\rm disk}}{10^{12} \,\rm G}\right)^{4/7} \left(
\frac{R_*}{10\,\textrm{km}} \right)^{10/7} \left(\frac{M_*}{M_\odot}
\right)^{1/7} \left(\frac{L_{\rm X}}{10^{37} \, \textrm{erg s}^{-1}}
\right)^{-2/7}  \xi \ ,
\label{eqn:Alfven radius formula}
\end{equation}
where $\xi$ is a constant of order unity. Based on Equations~(\ref{eqn:Beqmag}) and (\ref{eqn:Alfven radius formula}), we observe that the oscillation of the disk-plane surface magnetic field, $B_{\rm disk}$, in the direction towards the companion star, will generate a corresponding oscillation in the Alfv\'en radius, $R_{\rm A}$, in the same direction. Since the matter is picked up from the accretion disk and entrained onto the magnetic field lines at radius $R_{\rm A}$, it follows that the pick-up radius in the disk oscillates between minimum and maximum values as the star rotates.

We denote the radii where the magnetic field lines connected with the inner and outer walls of the accretion column cross the accretion disk as $R_{1,{\rm disk}}$, and $R_{2,{\rm disk}}$, respectively, where $R_{1,{\rm disk}} > R_{2,{\rm disk}}$. The corresponding radii at which these field lines cross the equatorial plane of the magnetic dipole are $R_1$ and $R_2$, respectively. By setting the magnetic-equatorial crossing radius, $R_{\rm eq}$, equal to either $R_1$ or $R_2$ in Equation~(\ref{eqn:rtheta}), we find that the corresponding disk-crossing radii for the two field lines in question are given by
\begin{equation}
\begin{split}
R_{1,\,{\rm disk}} &= R_1 \sin^2 \theta = R_1 \cos^2 \alpha \ , \\
R_{2,\,{\rm disk}} &= R_2 \sin^2 \theta = R_2 \cos^2 \alpha \ ,
\label{eqn:R_equator}
\end{split}
\end{equation}
where $\alpha$ is the magnetic latitude in the accretion disk, and the final results follow from Equation~(\ref{eqn:maglat}). Equations~(\ref{eqn:R_equator}) indicate that the disk-crossing radii $R_{1,\,{\rm disk}}$ and $R_{2,\,{\rm disk}}$ oscillate as the star rotates and $\alpha$ varies between $\pm \varphi$. By combining Equations~(\ref{eqn:R1andR2}) and (\ref{eqn:R_equator}), we can eliminate $R_1$ and $R_2$ to express the disk-crossing radii in terms of the angles $\theta_1$ and $\theta_2$, which yields
\begin{equation}
R_{1,\,{\rm disk}} = R_* \left(\frac{\cos\alpha}{\sin\theta_1}\right)^2  \ , \qquad
R_{2,\,{\rm disk}} = R_* \left(\frac{\cos\alpha}{\sin\theta_2}\right)^2 \ .
\label{eqn:R_equator2}
\end{equation}

Equations~(\ref{eqn:R_equator2}) allow us to study the variation of the two disk-crossing radii as the star spins and the disk-plane latitude $\alpha$ oscillates between $\pm \varphi$. This is important because the matter is picked up from the disk at the Alfv\'en radius, $R_{\rm A}$, and therefore material is fed onto the inner and outer walls of the accretion column when $R_{1,\,{\rm disk}}=R_{\rm A}$ and $R_{2,\,{\rm disk}}=R_{\rm A}$, respectively. For intermediate values, the matter is fed into the central part of the column, between the inner and outer walls. Hence, as the star rotates, matter is cyclically fed into the entire volume of the accretion column.

In order to close the system and ensure that we are generating self-consistent models for the pulsar accretion column and its connection with the surrounding accretion disk, we must therefore set $R_{1,\,{\rm disk}}$ and $R_{2,\,{\rm disk}}$ equal to the maximum and minimum values for the oscillating Alfv\'en radius, which is obtained by combining Equations~(\ref{eqn:Alfven radius formula}) and (\ref{eqn:Beqmag}). Essentially, we must find that during the spin of the star, $R_{\rm A}$ varies in the range
\begin{equation}
R_{2,\,{\rm disk}} \lesssim R_{\rm A} \lesssim R_{1,\,{\rm disk}} \ .
\label{eqn:R_Avar}
\end{equation}
We should emphasize that our model does not include a complete description of the entire pulsar magnetosphere and the associated accretion disk, and therefore we must interpret expressions such as Equations~(\ref{eqn:R_Avar}) as approximations, rather than strict quantitative relations. However, these expressions are nonetheless valuable in assessing the overall validity of our model and the related parameters, which we will discuss in more detail in Section~\ref{subsection:Pick-Up Radius}.


\subsection{Quantization and Electron Scattering Cross-Section}
\label{subsection:scattering cross section}

Quantum mechanical effects play an important role in the strong magnetic fields ($B \sim 10^{12}\,$G) inherent to X-ray pulsars because the cyclotron energy, $\epsilon_c$, separating the ground state from the first excited Landau level,
\begin{equation}
\epsilon_c = \frac{e B h}{2 \pi m_e c} \approx 11.57 \left(
\frac{B}{10^{12}\,\rm G} \right)\,{\rm keV} \ ,
\label {eqn:epsilonc}
\end{equation}
is in the range $\epsilon_c \sim 10-50\,$keV, where $m_e$, $e$, $h$, and $c$ denote the electron mass and charge, Planck's constant, and the speed of light, respectively. The resulting cyclotron absorption feature can be clearly identified in many X-ray pulsar spectra (e.g., White et al. 1983).

The strong magnetic field inside the accretion column differentiates the photons into ordinary and extraordinary polarization modes. In the case of the ordinary mode, the electric field vector is oriented in the plane formed by the magnetic field and the photon propagation direction. In the case of the extraordinary mode, the electric field vector is aligned perpendicular to this plane. The details of the photon-electron scattering process depend on the relationship between the photon energy $\epsilon$ and the cyclotron energy $\epsilon_{\rm{cyc}}$, and also on the propagation direction and polarization state of the photon (Ventura 1979; Chanan et al. 1979; Nagel 1980).

In the ordinary polarization mode (m=1), the scattering cross-section is given by
\begin{equation}
\sigma_s^{\rm m=1} = \sigma_{\rm T}\left[\sin^2 \theta_s + f_s(\epsilon)\,
\cos^2 \theta_s \right] \ ,
\label{eqn:normal mode}
\end{equation}
and the extraordinary mode (m=2) scattering cross-section can be written as
\begin{equation}
\sigma_s^{\rm m=2} = \sigma_{\rm T} f_s(\epsilon) + \Psi \ ,
\label{eqn:extraordinary mode}
\end{equation}
where $\sigma_{\rm T}$ is the Thomson cross-section, $\theta_s$ is the angle between the photon propagation direction and the magnetic field, $\Psi$ is the resonant contribution, and the function $f_s(\epsilon)$ is defined in terms of the cyclotron energy, $\epsilon_{\rm{cyc}}$, by
\begin{equation}
f_s(\epsilon) \equiv
\begin{cases}
1 \ , & \epsilon \geq \epsilon_{\rm{cyc}} \ , \\
\left(\epsilon / \epsilon_{\rm{cyc}} \right)^2 \ , & \epsilon < \epsilon_{\rm{cyc}} \ .
\end{cases}
\label{eqn:f function for scattering cross section}
\end{equation}
A complete treatment of the energy and angular dependence of the scattering of the ordinary and extraordinary mode photons is beyond the scope of this paper, and therefore we follow Wang \& Frank (1981) and BW07 by splitting the photons into two populations: those propagating either parallel or perpendicular to the magnetic field direction.

Photons propagating perpendicular to the magnetic field ($\theta_s = \pi / 2$) are dominated by the ordinary polarization mode (m=1) if $\epsilon$ is below the cyclotron energy, $\epsilon_{\rm{cyc}}$, because in this case the resonant portion of Equation~(\ref{eqn:extraordinary mode}) makes no contribution, and we find that $\sigma_s^{\rm m=2} < \sigma_s^{\rm m=1} = \sigma_{\rm T}$. In this situation, we can therefore set the perpendicular scattering cross-section equal to the Thomson value (Ventura 1979; Becker 1998),
\begin{equation}
\sigma_\perp = \sigma_{\rm T} \ .
\label{eqn:perpendicular cross section}
\end{equation}

For photons propagating parallel to the magnetic field ($\theta_s = 0$), with energy $\epsilon < \epsilon_{\rm{cyc}}$, both modes see the Thomson cross-section reduced by the ratio $(\epsilon/\epsilon_{\rm{cyc}})^2$. In this case, we follow Arons et al. (1987) and remove the energy dependence of the parallel scattering cross-section by replacing $\epsilon$ with the radius-dependent mean photon energy, $\bar\epsilon(r)$, so that $\sigma_\parallel(r)$ varies as
\begin{equation}
\sigma_\parallel \approx
\begin{cases}
\sigmaT \ , & \bar{\epsilon} \geq \epsilon_{\rm{cyc}} \ , \\
\sigma_{\rm T}\left(\bar{\epsilon} / \epsilon_{\rm{cyc}} \right)^2 \ , & \bar{\epsilon} < \epsilon_{\rm{cyc}} \ .
\end{cases}
\label{eqn:parallel}
\end{equation}

In our computational approach, the value of $\sigma_\parallel$ is obtained as part of an iterative parameter variation procedure in which we self-consistently compute the radiation spectrum and the hydrodynamic structure of the accretion column, and attempt to fit the observational spectral data with adherence to the appropriate boundary conditions (see Section~\ref{sec:BOUNDARY CONDITIONS}). However, as a check on the validity of the model parameters, we will refer to Equation~(\ref{eqn:parallel}) in our discussion in Section~7 in order to verify that the resulting values for $\sigma_\parallel$ are physically reasonable. We also require that the angle-averaged cross-section, $\overline{\sigma}$, used in the solution of the photon transport equation, must satisfy the constraint $\sigma_\parallel < \overline\sigma < \sigma_\perp$ (Canuto et al. 1971; BW07).

\subsection{Equation of State}
\label{subsection:equation of state}

The magnetic field near the surface of an accreting neutron star is so large that the cyclotron energy given by Equation~(\ref{eqn:epsilonc}) becomes comparable to the thermal energy of the electrons. Consequently, the electron energy distribution is quantized in the direction perpendicular to the magnetic field, and therefore the electrons possess a one-dimensional Maxwellian distribution along the magnetic field direction, with a mean thermal energy equal to $(1/2) \, k T_e$, where $k$ is Boltzmann's constant. On the other hand, the proton energy is not quantized, and therefore the protons are described by a three-dimensional Maxwellian distribution, with a mean thermal equal to $(3/2) \, k T_i$. The ion and electron internal energy densities are therefore given by
\begin{equation}
U_i = \frac{3}{2} \, n_i k T_i \ , \ \ \ \
U_e = \frac{1}{2} \, n_e k T_e \ ,
\label{eqn:eos3}
\end{equation}
where $n_i$ and $n_e$ denote the ion and electron number densities, respectively. In principle, the ion and electron temperatures $T_i$ and $T_e$ are not necessarily equal, and therefore in our two-temperature model we implement separate energy equations for each species, including a term describing their Coulomb coupling.

The magnetic field pressure is orders of magnitude stronger than either the gas pressure or the radiation pressure in an X-ray pulsar accretion column, and therefore the charged particles are constrained to follow the curved dipole magnetic field as the plasma flows downwards towards the stellar surface. Charge neutrality ensures that $n_i=n_e$ at all locations. From the point of view of the accretion hydrodynamics, the relevant pressure is the total pressure parallel to the local magnetic field direction, given by the sum of the electron, ion, and radiation components,
\begin{equation}
P_{\rm tot} = P_e + P_i + P_r \ ,
\label{eqn:eos}
\end{equation}
where
\begin{equation}
P_i = n_i k T_i \ , \qquad
P_e = n_e k T_e \ ,
\label{eqn:pressures}
\end{equation}
denote the ion and electron pressures, respectively. The radiation pressure, $P_r$, is not given by a thermal formula since the X-ray pulsar radiation field is nonthermal. Hence the radiation pressure must be computed using a conservation relation. The pressure components are related to their corresponding energy densities via
\begin{equation}
P_i = (\gamma_i-1) \, U_i \ , \quad P_e = (\gamma_e-1) \, U_e \ , \quad
P_r = (\gamma_r-1) \, U_r \ ,
\label{eqn:eos2}
\end{equation}
where it follows from Equations~(\ref{eqn:eos3}), (\ref{eqn:pressures}), and (\ref{eqn:eos2}) that $\gamma_e=3$ and $\gamma_i=5/3$. The ratio of specific heats for the radiation is $\gamma_r=4/3$.

\section{CONSERVATION EQUATIONS}

Our self-consistent model for the hydrodynamics and the radiative transfer occurring in X-ray pulsar accretion flows is based on a fundamental set of conservation equations governing the flow velocity, $v(r)$, the bulk fluid mass density, $\rho(r)$, the radiation energy density, $U_r(r)$, the ion energy density, $U_i(r)$, the electron energy density, $U_e(r)$, and the total energy transport rate, $\dot E(r)$. The mathematical model can be reduced to a set of five first-order, coupled, nonlinear ordinary differential equations satisfied by $v$, $\dot E$, and the ion, electron, and radiation sound speeds, $a_i$, $a_e$, and $a_r$, respectively, defined by
\begin{equation}
a_i^2 = \frac{\gamma_i P_i}{\rho} \ , \quad
a_e^2 = \frac{\gamma_e P_e}{\rho} \ , \quad
a_r^2 = \frac{\gamma_r P_r}{\rho} \ ,
\label{eqn:soundspeeds}
\end{equation}
where the ion and electron temperatures, $T_i$ and $T_e$, are related to the respective sound speeds via (see Equations~(\ref{eqn:pressures}))
\begin{equation}
a_i^2 = \frac{\gamma_i k T_i}{m_{\rm tot}} \ , \ \ \ \
a_e^2 = \frac{\gamma_e k T_e}{m_{\rm tot}} \ .
\label{eqn:temperatureconversion}
\end{equation}
Here, $m_{\rm tot} = m_e + m_i$ denotes the total particle mass, assuming the accreting gas is composed of pure, fully-ionized hydrogen, with $n_e=n_i$ for charge neutrality. There is no corresponding relation for the radiation sound speed since the radiation distribution inside the accretion column is not expected to approach a blackbody.

Solving the five coupled conservation equations to determine the radial profiles of the quantities $v$, $\dot E$, $a_i$, $a_e$, and $a_r$ requires an iterative approach, because the rate of Compton energy exchange between the photons and the electrons depends on the relationship between the electron temperature, $T_e$, and the inverse-Compton temperature, $T_{\rm IC}$, which in turn is determined by the shape of the radiation distribution. In order to achieve a self-consistent solution for all of the flow variables, while taking into account the feedback loop between the dynamical calculation and the radiative transfer calculation, the simulation must iterate through a specific sequence of steps. The steps required in a single iteration are (1) the computation of the dynamical structure of the accretion column by solving the five conservation equations, (2) calculation of the associated radiation distribution function by solving the radiative transfer equation, (3) computation of the inverse-Compton temperature profile from the radiation distribution, and then (4) re-computation of the dynamical structure, etc. The iterative process is discussed in detail in Section~\ref{subsection:convergence}. Here we describe the physics contained in each of the coupled conservation equations that form the core of the dynamical model.

\subsubsection{Mass Flux}

In the one-dimensional case considered here, the cross-sectional structure of the accretion column is not considered in detail, and all of the densities and temperatures represent averages across the column at a given radius $r$. Hence the mass continuity equation can be written in dipole geometry as (e.g., Langer \& Rappaport 1982)
\begin{equation}
\frac{\partial \rho(r)}{\partial t} = - \frac{1}{A(r)}\frac{\partial}{\partial r}
\Big[A(r) \rho(r) v(r) \Big] \ ,
\label{eqn:mass flux}
\end{equation}
where $v < 0$ denotes the radial inflow velocity, and the cross-sectional area of the column, $A(r)$, is related to the solid angle, $\Omega(r) = (r / R_*) \Omega_*$, by
\begin{equation}
A(r) = r^2 \Omega(r) = \frac{r^3 \Omega_*}{R_*} \ .
\label{eqn:area}
\end{equation}
In a steady state, we see from Equation (\ref{eqn:mass flux}) that $A(r) \rho(r) v(r) $ is a conserved quantity, i.e. the mass accretion rate $\dot M$ is conserved and is related to the density $\rho$ and velocity $v$ via
\begin{equation}
\dot M = \Omega \, r^2 \rho |v| \ ,
\label{eqn:dipolarmassflux}
\end{equation}
which can be combined with Equation~(\ref{eqn:solid angle}) to obtain for the mass density
\begin{equation}
\rho =  \frac{\dot M R_*}{\Omega_* r^3 |v|} \ .
\label{eqn:massdensity}
\end{equation}
This algebraic relation for the density is used to supplement the set of differential conservation equations in our hydrodynamical model for the column structure.

We assume that the accreting gas is composed of pure, fully-ionized hydrogen, and therefore the electron and ion number densities are given by
\begin{equation}
n_e = n_i = \frac{\rho}{m_{\rm tot}} = \frac{\dot M R_*}{m_{\rm tot}
\Omega_* r^ 3 |v|} \ ,
\label{eqn:numberdensity}
\end{equation}
where $m_{\rm tot} = m_e + m_i$.

\subsubsection{Total Energy Flux}

The total energy flux in the radial direction, averaged over the column cross-section at radius $r$, is given by
\begin{equation}
F(r) = \frac{1}{2}\rho v^3 + v(P_i + P_e + P_r + U_i + U_e
+ U_r) - \frac{c}{n_e \sigma_\parallel}\frac{\partial P_r}{\partial r}
- \frac{GM_*\rho v}{r} \ ,
\label{eqn:energyflux}
\end{equation}
where the energy flux is defined to be negative for energy flow in the downward direction, and the accretion velocity $v$ is negative ($v<0$). The terms on the right-hand side of Equation~(\ref{eqn:energyflux}) represent the kinetic energy flux, the enthalpy flux, the radiation diffusion flux, and the gravitational energy flux, respectively. The total energy flux $F$ is related to the total energy transport rate in the radial direction, denoted by $\dot E$, via
\begin{equation}
\dot E(r) = A(r) \, F(r) \ \ \propto \ \ {\rm erg \
s}^{-1} \ ,
\label{eqn:energytransport}
\end{equation}
where the column cross-sectional area $A(r)$ is given by Equation~(\ref{eqn:area}).

We can derive a first-order differential equation for the radiation sound speed, $a_r$, by substituting for the energy densities and pressures in Equation~(\ref{eqn:energyflux}) using Equations~(\ref{eqn:eos2}) and (\ref{eqn:soundspeeds}), substituting for the electron number density $n_e$ using Equation~(\ref{eqn:numberdensity}), and substituting for $F$ using Equation~(\ref{eqn:energytransport}). After some algebra, we obtain in the steady state case
\begin{align}
\frac{d a_r}{d r} = \frac{3 a_r}{2 r} &+ \frac{a_r}{2
v} \frac{d v}{d r} - \frac{1}{2}
\frac{\sigma_\parallel \gamma_r \dot M}{m_{\rm tot} c a_r \Omega r^2}
\bigg(\frac{\dot E}{\dot M} + \frac{v^2}{2}  \nonumber \\
&+  \frac{a^2_i}{\gamma_i-1}+\frac{a^2_e}{\gamma_e-1}
+ \frac{a^2_r}{\gamma_r-1} - \frac{G M_*}{r} \bigg) \ .
\label{eqn:dardr}
\end{align}

\subsubsection{Ion and Electron Energy Equations}
\label{section:gasenergyequation}

The variation of the internal energy density of the ionized gas is influenced by adiabatic heating, energy exchange between the ions and electrons, and the emission and absorption of radiation. Averaging over the cross-section of the column at radius $r$, the energy equations for the ions and electrons can be written as
\begin{equation}
\frac{DU_i}{Dt} = \gamma_i \frac{U_i}{\rho}
\frac{D\rho}{Dt} + \dot U_i \ , \ \ \ \ \ \ \
\frac{DU_e}{Dt} = \gamma_e \frac{U_e}{\rho}
\frac{D\rho}{Dt} + \dot U_e \ ,
\label{eqn:gasenergydensity}
\end{equation}
respectively, where the first terms on the right-hand side represent adiabatic compression, the final terms represent thermal coupling with other species, and the comoving (Lagrangian) time derivative $D/Dt$ is defined by
\begin{equation}
\frac{D}{Dt} \equiv \frac{\partial}{\partial t} + v \frac{\partial}
{\partial r} \ .
\label{eqn:lagrangian}
\end{equation}

The thermal coupling terms appearing in Equations~(\ref{eqn:gasenergydensity}) represent the net heating due to a variety of combined processes, which are broken down as follows,
\begin{eqnarray}
\dot U_e & = & \dot U_{\rm brem}^{\rm emit}
+ \dot U_{\rm brem}^{\rm abs}
+ \dot U_{\rm cyc}^{\rm emit}
+ \dot U_{\rm cyc}^{\rm abs}
+ \dot U_{\rm Comp} + \dot U_{\rm ei} \ ,  \nonumber \\
\dot U_i & = & - \dot U_{\rm ei} \ .
\label{eqn:Udottotal}
\end{eqnarray}
The terms in the expression for $\dot U_e$ denote, respectively, bremsstrahlung (free-free) emission and absorption, cyclotron emission and absorption, photon-electron Comptonization, and electron-ion Coulomb energy exchange. The ions do not radiate appreciably, and therefore they only experience adiabatic compression and Coulomb energy exchange (see Langer \& Rappaport 1982).  In our sign convention, a heating term is positive and a cooling term is negative. These energy transfer rates are discussed in more detail in Section~\ref{section:energy exchange}.

In a steady state, Equations~(\ref{eqn:gasenergydensity}) can be written as
\begin{equation}
\frac{dU_i}{dr} = \gamma_i \frac{U_i}{\rho}
\frac{d\rho}{dr} + \frac{\dot U_i}{v} \ , \ \ \ \ \ \ \
\frac{dU_e}{dr} = \gamma_e \frac{U_e}{\rho}
\frac{d\rho}{dr} + \frac{\dot U_e}{v} \ .
\label{eqn:gasenergydensity1}
\end{equation}
We can derive equivalent differential equations satisfied by the electron and ion sound speeds by using Equations~(\ref{eqn:eos2}), (\ref{eqn:soundspeeds}), and (\ref{eqn:massdensity}) to substitute for the energy and mass densities in Equations~(\ref{eqn:gasenergydensity1}), obtaining
\begin{equation}
\frac{d a_i}{d r}=-(\gamma_i-1)\left(
\frac{a_i}{2 v}\frac{d v}{d r} + \frac{3
a_i}{2 r}\right)-\frac{1}{2}\gamma_i(\gamma_i-1)
\frac{\Omega r^2}{\dot M} \frac{\dot U_i}{a_i} \ ,
\label{eqn:daidr}
\end{equation}
\begin{equation}
\frac{d a_e}{d r}=-(\gamma_e-1)\left(
\frac{a_e}{2 v}\frac{d v}{d r} + \frac{3
a_e}{2 r}\right)-\frac{1}{2}\gamma_e(\gamma_e-1)
\frac{\Omega r^2}{\dot M} \frac{\dot U_e}{a_e} \ .
\label{eqn:daedr}
\end{equation}
\subsubsection{Momentum Equation}

The ionized, accreting gas is constrained to spiral around the magnetic field lines by the Lorentz force. Since there is no component of the Lorentz force parallel to the local $B$-field, the remaining acceleration in the parallel direction is due to the pressure gradient and the gravitational field of the neutron star. If we average over the cross-section of the accretion column at radius $r$, then the comoving acceleration in the radial direction can be written as (e.g., Langer \& Rappaport 1982),
\begin{equation}
\frac{Dv}{Dt} = - \frac{1}{\rho}\frac{\partial P_{\rm tot}}{\partial r}
- \frac{GM_*}{r^2} \ ,
\label{eqn:conservation of momentum equation}
\end{equation}
where $P_{\rm tot}=P_r+P_i+P_e$ is the total pressure, and the Lagrangian time derivative $D/Dt$ is defined by
Equation~(\ref{eqn:lagrangian}). Substituting for the mass density $\rho$ and the pressure components $P_i$, $P_e$, and $P_r$ using Equations~(\ref{eqn:massdensity}) and (\ref{eqn:soundspeeds}), respectively, we can derive a first-order differential equation satisfied by the fluid velocity $v$ involving the sound speeds $a_i$, $a_e$, and $a_r$, and the energy transport rate $\dot E$. After some algebra, the result obtained in a steady state is
\begin{align}
\frac{d v}{d r} = &\frac{v}{v^2-(a_i^2+a_e^2)}
\bigg \{ \frac{3(a_i^2+a_e^2)}{r} -\frac{G M_*}{r^2}
+ \frac{\sigma_\parallel \dot M}{m_{\rm tot} c \, \Omega r^2}
\bigg(\frac{\dot E}{\dot M}+\frac{v^2}{2} \nonumber \\
&+ \frac{a_i^2}{\gamma_i-1}
+ \frac{a_e^2}{\gamma_e-1} + \frac{a_r^2}{\gamma_r-1}
- \frac{G M_*}{r} \bigg)
+ \frac{\Omega r^2}{\dot M} \left [ (\gamma_i-1) \dot U_i
+ (\gamma_e-1) \dot U_e \right ] \bigg \} \ ,
\label{eqn:dudr}
\end{align}
where we have also made use of Equations~(\ref{eqn:dardr}), (\ref{eqn:daidr}), and (\ref{eqn:daedr}).

\subsubsection{Radiative Losses}

The value of the energy transport rate $\dot E$ (Equation~(\ref{eqn:energytransport})) varies as a function of the radius $r$ in response to the escape of radiation energy through the walls of the accretion column, perpendicular to the magnetic field direction. In our one-dimensional model, all quantities are averaged over the cross-section of the column, and therefore we use an escape-probability formalism to account for the diffusion of radiation through the walls of the column. We therefore utilize a total energy conservation equation of the form
\begin{equation}
\frac{\partial}{\partial t}\left(\frac{1}{2}\,\rho v^2 + U_i + U_e + U_r
- \frac{G M_* \rho}{r} \right)
= - \frac{1}{A(r)}\frac{\partial}
{\partial r} \Big[A(r) F(r) \Big] + \dot U_{\rm esc} \ ,
\label{eqn:energy1}
\end{equation}
where the total energy flux $F(r)=\dot E(r)/A(r)$ (see Equation~(\ref{eqn:energytransport})), and the energy escape rate per unit volume is given by
\begin{equation}
\dot U_{\rm esc} = - \frac{U_r}{t_{\rm esc}} \ , \ \ \ \ \
t_{\rm esc} = \frac{\ell_{\rm esc}}{w_\perp} \ .
\label{eqn:energyescape}
\end{equation}
Here, $t_{\rm esc}(r)$ represents the mean escape time for photons to diffuse across the column and escape through the walls, $w_\perp(r)$ is the perpendicular diffusion velocity, and $\ell_{\rm esc}(r)$ denotes the perpendicular escape distance across the column at radius $r$, computed using (see Figure~\ref{fig:dipole geometry})
\begin{equation}
\ell_{\rm esc}(r) = (\ell_2 - \ell_1) \left(\frac{r}{R_*}\right)^{3/2} \ ,
\label{eqn:escape distance}
\end{equation}
so that at the stellar surface, we obtain $\ell_{\rm esc}=\ell_2-\ell_1$, as required. The perpendicular diffusion velocity $w_\perp$ cannot exceed the speed of light, and therefore we compute it using the constrained formula
\begin{equation}
w_\perp = \min\Big(c,\frac{c}{\tau_\perp}\Big) \ ,
\ \ \ \ \
\tau_\perp = n_e \sigma_\perp \ell_{\rm esc} \ ,
\label{eqn:diffusion velocity}
\end{equation}
where $\tau_\perp$ denotes the perpendicular optical thickness of the column at radius $r$.

In a steady state, Equation~(\ref{eqn:energy1}) reduces to
\begin{equation}
\frac{1}{r^2 \Omega(r)} \frac{d \dot E}{d r}
= - \frac{U_r}{\ell_{\rm esc}} \, \min\Big(c,\frac{c}
{\tau_\perp}\Big) \ .
\label{eqn:dEdr}
\end{equation}
By combining Equations~(\ref{eqn:eos2}), (\ref{eqn:soundspeeds}), (\ref{eqn:dipolarmassflux}), and (\ref{eqn:dEdr}), we can obtain the final form for the energy transport differential equation,
\begin{equation}
\frac{d \dot E}{d r}
= - \frac{a_r^2 \dot M}{\gamma_r
(\gamma_r-1) \, \ell_{\rm esc} \, |v|} \, \min\Big(c,\frac{c}
{\tau_\perp}\Big) \ .
\label{eqn:dEdr2}
\end{equation}

\subsection{Energy Exchange Processes}
\label{section:energy exchange}

The energy exchange rates per unit volume introduced in Equations~(\ref{eqn:Udottotal}), denoted by $\dot U_{\rm brem}^{\rm emit}$, $\dot U_{\rm brem}^{\rm abs}$, $\dot U_{\rm cyc}^{\rm emit}$, $\dot U_{\rm cyc}^{\rm abs}$, $\dot U_{\rm Comp}$, and $\dot U_{\rm ei}$, describe a comprehensive set of heating and cooling processes experienced by the gas and radiation, including Coulomb coupling between the ions and electrons, the Compton exchange of energy between the electrons and photons, and the emission and absorption of radiation energy via thermal bremsstrahlung and cyclotron. In this section we provide additional details regarding the computation of these various rates.

\subsubsection{Bremsstrahlung Emission and Absorption}

Thermal bremsstrahlung emission plays a significant role in cooling the ionized gas, and in the case of luminous X-ray pulsars, it also provides the majority of the seed photons that are subsequently Compton scattered to form the emergent X-ray spectrum (BW07). Assuming a fully-ionized hydrogen composition for the accreting gas, with $n_e=n_i$, the total power per unit volume emitted by the electrons is given by (see Rybicki \& Lightman 1979, Equation (5.14)),
\begin{equation}
\dot U_{\rm brem}^{\rm emit} = - \left(\frac{2 \pi k T_e}{3 m_e}\right)^{1/2}
\frac{2^5 \pi e^6}{3 h m_e c^3} \, n_e^2 \ ,
\label{eqn:5.14a}
\end{equation}
where we have set the Gaunt factor equal to unity. The negative sign appears in Equation~(\ref{eqn:5.14a}) because this term represents a cooling process in which heat is removed from the electrons. We can write an equivalent expression for the bremsstrahlung cooling rate in terms of the electron sound speed, $a_e$, by using Equation~(\ref{eqn:temperatureconversion}) to eliminate the electron temperature $T_e$ in Equation~(\ref{eqn:5.14a}), thereby obtaining, in cgs units,
\begin{equation}
\dot U_{\rm brem}^{\rm emit} = - 3.2 \times 10^{16} \, a_e \, \rho^2 \ ,
\label{eqn:Udotbrem}
\end{equation}
where we have also used Equation~(\ref{eqn:numberdensity}).

The electrons in the accretion column also experience heating due to free-free absorption of low-frequency radiation, which can play an important role in regulating the temperature of the gas. The heating rate per unit volume due to thermal bremsstrahlung absorption, integrated over photon frequency, is given by
\begin{equation}
\dot U_{\rm brem}^{\rm abs} = U_r \, \alpha_{\rm R} \, c \ ,
\label{eqn:bremsstrahlung absorption}
\end{equation}
where $\alpha_{\rm R}$ is the Rosseland mean absorption coefficient for fully ionized hydrogen, expressed in cgs units by (Rybicki \& Lightman 1979)
\begin{equation}
\alpha_{\rm R} = 1.7 \times 10^{-25} \, T_e^{-7/2} n_e^2 \ \ \propto
\ \ {\rm cm}^{-1} \ .
\label{eqn:absorptioncoefficient}
\end{equation}
Note that we have set the Gaunt factor equal to unity and assumed that the gas is composed of fully-ionized hydrogen. By combining Equations~(\ref{eqn:bremsstrahlung absorption}) and (\ref{eqn:absorptioncoefficient}) and substituting for $U_r=P_r/(\gamma_r-1)$, $n_e$, and $T_e$, using Equations~(\ref{eqn:soundspeeds}), (\ref{eqn:temperatureconversion}), and (\ref{eqn:numberdensity}), respectively, we obtain, in cgs units
\begin{equation}
\dot U_{\rm brem}^{\rm abs} = 9.8 \times 10^{62} \, a_e^{-7}
\, a_r^2 \, \rho^3
\ .
\label{eqn:Udot bremsstrahlung absorption term}
\end{equation}
The sign of this quantity is positive since it represents a heating process for the electrons.

\subsubsection{Cyclotron Emission and Absorption}
\label{subsubsection:cyclotron emission}

The electrons in the accretion column also experience heating and cooling due to the emission and absorption of thermal cyclotron radiation. At any given time, most of the electrons are found in the ground state, but they can be excited to the first Landau level via collisions, or via the absorption of radiation at the cyclotron energy, $\epsilon_{\rm{cyc}}$. At the densities and temperatures prevalent in pulsar accretion columns, radiative excitation is followed immediately by radiative de-excitation back to the ground state, so that in net terms, cyclotron absorption can be interpreted as a resonant scattering process, which results in no net change in the angle-averaged photon distribution (Nagel 1980; Arons et al. 1987). Hence, on average, cyclotron absorption does not result in the net heating of the gas, due to the rapid radiative de-excitation, and we therefore set $\dot U_{\rm cyc}^{\rm abs}=0$ in our dynamical calculations. However, near the surface of the accretion column, photons scattered out of the outwardly directed beam are not replaced, and this leads to the formation of the observed cyclotron absorption features, in a process that is very analogous to the formation of absorption lines in the solar spectrum (Ventura et al. 1979). The formation of the cyclotron absorption features is further considered in Paper~II.

While cyclotron absorption does not result in the net heating of the gas, due to the rapid radiative de-excitation, cyclotron emission will cool the gas. In this process, kinetic energy is converted into excitation energy via collisions, and the subsequent emission of cyclotron radiation removes heat from the electrons. To compute the cyclotron cooling rate, $\dot U_{\rm cyc}^{\rm emit}$, we begin by writing down the cyclotron emissivity, $\dot n_\epsilon^{\rm cyc}$, which gives the production rate of cyclotron photons per unit volume per unit energy. Using Equations~(7) and (11) from Arons et al. (1987), we have
\begin{equation}
\dot n_{\epsilon}^{\rm cyc} = 2.1\times10^{36} \rho^2
B_{12}^{-3/2} H \left(\frac{\epsilon_{c}}{k T_e}\right)
e^{-\epsilon_{\rm{cyc}}/k T_e} \delta \left(\epsilon-\epsilon_{\rm{cyc}}
\right) \ ,
\label{eqn:cyclotron1}
\end{equation}
where $B_{12}=B/(10^{12}\,\rm G)$ and $H(x)$ is a piecewise function defined by
\begin{equation}
H(x) \equiv
\begin{cases}
0.15 \, \sqrt{7.5} \ , & x \ge 7.5 \ , \\
0.15 \, \sqrt{x} \ , & x < 7.5 \ .
\end{cases}
\label{eqn:Hfunction}
\end{equation}
The total cyclotron cooling rate is obtained by multiplying Equation~(\ref{eqn:cyclotron1}) by the photon energy $\epsilon$ and integrating over all energies, which yields, in cgs units,
\begin{equation}
\dot U_{\rm cyc}^{\rm emit} = - 2.1 \times 10^{36} \rho^2
B_{12}^{-3/2} \epsilon_{\rm{cyc}} \ H\left(\frac{\epsilon_{\rm{cyc}}}{kT_e} \right)
e^{-\epsilon_{\rm{cyc}}/kT_e} \ ,
\label{eqn:Udotcyc}
\end{equation}
where the negative sign indicates that this is a cooling process for the electrons.

\subsubsection{Compton Heating and Cooling}

Compton scattering plays a fundamental role in the formation of the emergent X-ray spectrum. It is also critically important in establishing the radial variation of the electron temperature profile through the exchange of energy between the photons and electrons. Equation~(7.36) from Rybicki \& Lightman (1979) gives the mean change in the photon energy $\epsilon$ during a single scattering as
\begin{equation}
\left<\Delta\epsilon\right> = \frac{\epsilon}{m_e c^2}
\left(4 k T_e - \epsilon \right) \ ,
\label{eqn:Compton3}
\end{equation}
and the associated mean rate of change of the photon energy is therefore
\begin{equation}
\left<\frac{d\epsilon}{dt}\right>\bigg|_{\rm Comp} = n_e \bar\sigma c \,
\left<\Delta\epsilon\right> \ ,
\label{eqn:Compton3b}
\end{equation}
where $(n_e \bar\sigma c)^{-1}$ denotes the mean-free time between scatterings for the photons, and $\bar\sigma$ is the angle-averaged electron scattering cross-section (BW07). The corresponding rate of change of the electron energy density due to Compton scattering can therefore be written as
\begin{equation}
\dot U_{\rm Comp} = - n_e \bar\sigma c \int_0^\infty \epsilon^2 f(r,\epsilon)
\, \left<\Delta\epsilon\right> \, d\epsilon \ ,
\label{eqn:Compton4}
\end{equation}
where the distribution function, $f(r,\epsilon)$, is the solution to the photon transport equation introduced in Paper II, which is related to the total radiation number density, $n_r$, and energy density, $U_r$, via
\begin{equation}
n_r(r) = \int_0^\infty \epsilon^2 f(r,\epsilon) \, d\epsilon \ , \ \ \ \ \
U_r(r) = \int_0^\infty \epsilon^3 f(r,\epsilon) \, d\epsilon \ .
\label{eqn:Compton5}
\end{equation}
Combining Equations~(\ref{eqn:Compton3}) and (\ref{eqn:Compton4}), we find that the net Compton cooling rate for the electrons is given by
\begin{equation}
\dot U_{\rm Comp} = \frac{n_e \bar\sigma c}{m_e c^2} \left[
\int_0^\infty \epsilon^4 \, f(r,\epsilon)
\, d\epsilon - 4 k T_e \int_0^\infty
\epsilon^3 f(r,\epsilon) \, d\epsilon \right] \ ,
\label{eqn:Compton6}
\end{equation}
which vanishes if the electron temperature, $T_e$, is equal to the inverse-Compton temperature, $T_{\rm IC}$, defined by
\begin{equation}
T_{\rm IC}(r) \equiv \frac{1}{4k}\frac{\int_0^\infty
\epsilon^4 f(r,\epsilon) \, d\epsilon}{\int_0^\infty \epsilon^3 f(r,\epsilon)
\, d\epsilon} \ .
\label{eqn:Compton7}
\end{equation}

In the present paper, we are primarily interested in the implications of Compton scattering for the heating and cooling of the gas, and its effect on the dynamical structure of the accretion column. The electron cooling rate can be rewritten as
\begin{equation}
\dot U_{\rm Comp} = n_e \bar\sigma c \ \frac{4 k T_e}{m_e c^2}
\left[g(r) - 1\right] U_r \ ,
\label{eqn:Udotcompton}
\end{equation}
where we introduce $g(r)$ as the temperature ratio function,
\begin{equation}
g(r) \equiv \frac{T_{\rm IC}}{T_e} \ .
\label{eqn:gfunction}
\end{equation}
The sign of $\dot U_{\rm Comp}$ depends on the value of $g$. If $g < 1$ (i.e. $T_{\rm IC} < T_e$), then the electrons experience Compton cooling; otherwise, the electrons are heated via inverse-Compton scattering. We can obtain the final form for the Compton cooling rate in terms of the mass density, $\rho$, the electron sound speed, $a_e$, and the radiation sound speed, $a_r$ by combining Equations~(\ref{eqn:pressures}), (\ref{eqn:eos2}), (\ref{eqn:soundspeeds}), and (\ref{eqn:Udotcompton}), which yields
\begin{equation}
\dot U_{\rm Comp} = \frac{4 \, \bar{\sigma} \, (g-1)}
{m_e c \gamma_e \gamma_r (\gamma_r-1)}
\ \rho^2 \, a_r^2 \, a_e^2 \ .
\label{eqn:Udotcompton2}
\end{equation}

\subsubsection{Electron-Ion Energy Exchange}

The electrons can also be heated or cooled via Coulomb collisions with the protons, depending on whether the electron temperature $T_e$ exceeds the ion temperature $T_i$. The net heating rate per unit volume for the electrons is given by (Langer \& Rappaport 1982)
\begin{equation}
\dot U_{\rm ei} = \frac{3}{2} \left(\frac{2}{\pi}\right)^{1/2}
\sigma_{\rm T} c^3 m_e n_e^2
\left(\frac{m_e}{m_i}\right)\left(\frac{T_i - T_e}
{T_{\rm eff}} \right) \left( \frac{m_e c^2}{k T_{\rm eff}} \right)^{1/2}
\mathrm{ln} \ \Lambda_{\rm Coul} \ ,
\label{eqn:electron-ion energy exchange rate}
\end{equation}
where
\begin{equation}
T_{\rm eff} \equiv T_e + \left ( \frac{m_e}{m_i} \right ) T_i \ ,
\label{eqn:Teff}
\end{equation}
and the Coulomb logarithm is given by
\begin{equation}
\mathrm{ln} \ \Lambda_{\rm Coul} = 5.41+\frac{1}{4}\mathrm{ln} \
\left(\frac{k T_{\rm eff}}{20 \; \mathrm{keV}} \frac{B}{10^{12} \; \mathrm{G}} \frac{10^{20} \; \mathrm{cm^{-3}} }{n_e}
\right) \ .
\end{equation}
We can further simplify Equation~(\ref{eqn:electron-ion energy exchange rate}) by substituting for $n_e$ using Equation (\ref{eqn:numberdensity}) and substituting for $T_e$ and $T_i$ using Equations~(\ref{eqn:temperatureconversion}), obtaining
\begin{eqnarray}
\dot U_{\rm ei} = \frac{3}{2} \left(\frac{2}{\pi}\right)^{1/2}
\sigma_{\rm T} c^4
\left ( \frac{m_e}{m_i} \right )
\left ( 1 + \frac{m_i}{m_e} \right )^{-5/2}
\left ( \frac{a_i^2}{\gamma_i} - \frac{a_e^2}{\gamma_e} \right )
\left ( \frac{a_e^2}{\gamma_e m_e} + \frac{a_i^2}{\gamma_i m_i} \right )^{-3/2}
\rho^2 \ \mathrm{ln} \ \Lambda_{\rm Coul}
\ ,
\label{eqn:electron-ion energy exchange rate2}
\end{eqnarray}
which in cgs units becomes
\begin{eqnarray}
\dot U_{\rm ei} = 2.42 \times 10^6
\left ( \frac{a_i^2}{\gamma_i} - \frac{a_e^2}{\gamma_e} \right )
\left ( \frac{a_e^2}{\gamma_e m_e} + \frac{a_i^2}{\gamma_i m_i} \right )^{-3/2}
\rho^2 \ \mathrm{ln} \ \Lambda_{\rm Coul}
\ .
\label{eqn:electron-ion energy exchange rate2 cgs}
\end{eqnarray}
Note that when $T_e=T_i$, the second factor in Equation (\ref{eqn:electron-ion energy exchange rate2 cgs}) is zero, and thus $\dot U_{ei}=0$, as expected. Based on the symmetry of the energy exchange between the particle species, we immediately conclude that the energy transfer rate per unit volume for the protons is given by $\dot U_i = - \dot U_{\rm ei}$ (see Equation (\ref{eqn:Udottotal})).

\section{BOUNDARY CONDITIONS}
\label{sec:BOUNDARY CONDITIONS}

In order to solve the coupled set of conservation equations, we must specify a variety of physical boundary conditions that fall into two major categories. The first category is the set of boundary conditions required to solve the system of dynamical equations using {\it Mathematica}, and the second category is the set of boundary conditions required to solve the partial differential equation for the photon distribution function $f$ using {\it COMSOL}. We will focus primarily on the first set of conditions here, and defer detailed discussion of the {\it COMSOL} boundary conditions to Paper~II.

As part of the dynamical model implemented in {\it Mathematica}, we need to impose boundary conditions based upon the physics occurring at the top of the accretion column ($r=r_{\rm top}$) and at the stellar surface ($r=R_*$). At the top of the column (Boundary~1), we impose conditions related to the flow velocity and its acceleration; the free-streaming radiation field; and the conservation of bulk fluid momentum. At the stellar surface (Boundary~2), we impose conditions related to the stagnation of the accretion velocity, and the attenuation of the total energy transport rate into the star.

\subsection{Boundary Conditions at the Upper Surface}

The upper surface of the dipole-shaped accretion funnel is located at radius $r=r_{\rm top}$, which must be below the radius corresponding to the turnover height of the dipole field, $r_c$, as discussed in Section~\ref{subsection:conical geometry} (see Equation~(\ref{eqn:rtopmax})). In analogy with the theory of stellar atmospheres, the top of the accretion column represents the last scattering surface for photon-electron interaction as photons travel out the top of the column, implying that the scattering optical depth from $r_{\rm top}$ to $r_c$ should equal unity. Defining the parallel scattering optical depth, $\tau_\parallel$, so that it increases in the downward direction for bulk fluid entering at the top of the column and flowing downward, from $\tau_\parallel=0$ at $r=r_{\rm top}$, we have
\begin{equation}
\tau_\parallel(r)= \int_{r}^{r_{\rm top}} n_e(r') \sigma_\parallel dr' \ .
\label{eqn:absorption optical depth at thermal mound is equal to unity}
\end{equation}
Since the top of the accretion column is the last scattering surface, we can also write
\begin{equation}
\int_{r_{\rm top}}^{r_c} n_e(r') \sigma_\parallel dr' = 1 \ ,
\label{eqn:lastscatter}
\end{equation}
where ${r_{\rm top}} < r_c$.

We can use Equation~(\ref{eqn:lastscatter}) to constrain the radius at the top of the accretion column, $r_{\rm top}$, as follows. We assume that the gas is in free-fall above $r_{\rm top}$, with velocity
\begin{equation}
v(r) = v_{\rm ff}(r) \equiv -\left({\frac{2 G M_*}{r}} \right)^{1/2} \ , \qquad r>r_{\rm top} \ .
\label{eqn:free fall velocity}
\end{equation}
Using Equation~(\ref{eqn:free fall velocity}) to substitute for $v$ in Equation~(\ref{eqn:numberdensity}) yields for the variation of the electron number density $n_e$ the result
\begin{equation}
n_e(r) = \frac{\dot M R_*}{m_{\rm tot}
\Omega_* r^ 3} \left({\frac{2 G M_*}{r}} \right)^{-1/2} \ ,
\label{eqn:numberdensityNEW}
\end{equation}
where $m_{\rm tot} = m_e + m_i$.

By utilizing Equation~(\ref{eqn:numberdensityNEW}) to substitute for the electron number density $n_e$ in Equation (\ref{eqn:lastscatter}) and carrying out the radial integration, we obtain the condition
\begin{equation}
\frac{2 \sigma_\parallel
\dot M}{3 m_{\rm tot} (2 G M_*)^{1/2}} \frac{R_*}{\Omega_*} \left(
r_{\rm top}^{-3/2} - r_c^{-3/2} \right) = 1 \ ,
\label{eqn:tau parallel at starting height}
\end{equation}
where the left-hand side is positive definite, since ${r_{\rm top}} < r_c$, and the dipole turnover radius $r_c$ is given by Equation~(\ref{eqn:rcritical}). By rearranging Equation~(\ref{eqn:tau parallel at starting height}), we can obtain an explicit expression for $r_{\rm top}$, given by
\begin{equation}
r_{\rm top} = \left [ r_c^{-3/2} + \frac{3 m_{\rm tot} (2 G M_*)^{1/2}} {2 \sigma_\parallel
\dot M} \frac{\Omega_*}{R_*} \right ]^{-2/3} \ .
\label{eqn:rtop}
\end{equation}
This relation allows us to self-consistently compute the value of $r_{\rm top}$ in terms of the parameters $\Omega_*$, $r_c$, and $\sigma_\parallel$ in our model.

At the top of the accretion column, the inflow velocity $v$ equals the local free-fall velocity, so that
\begin{equation}
v_{\rm top} \equiv v_{\rm ff}(r_{\rm top}) = -\left(\frac{2 G M_*}{r_{\rm top}}\right)^{1/2} \ .
\label{eqn:free-fall top}
\end{equation}
We also assume that at the top of the accretion column, the local acceleration of the gas is equal to the gravitational value, so that
\begin{equation}
v\,\frac{dv}{dr}\bigg|_{r=r_{\rm top}} = - \frac{GM}{r^2_{\rm top}} \ ,
\label{eqn:free-fall1}
\end{equation}
which implies that
\begin{equation}
\frac{dv}{dr}\bigg|_{r=r_{\rm top}} = \left(\frac{G M_*}{2 \, r_{\rm top}^3}\right)^{1/2} \ .
\label{eqn:free-fall acceleration top}
\end{equation}
By assuming pure gravitational acceleration at the top of the accretion column, we are implicitly neglecting the effects of the radiation pressure gradient, which will partially counteract the downward gravitational force. We revisit this issue in Section~7, where we conclude that this assumption is warranted, since most of the radiation escapes out the sides of the accretion column as a fan beam in the high-luminosity sources of interest here. However, in lower-luminosity sources, a larger fraction of the radiation may escape out the top of the column via a pencil-beam component, but even in this case, the effect of radiation deceleration at the top of the column is still likely to be negligible.

Although our calculation allows for the possibility of two-temperature flow, with unequal values of $T_i$ and $T_e$, in luminous X-ray pulsar accretion columns, not much deviation between the two temperatures is expected, because the thermal equilibration timescale is much smaller than the dynamical timescale (BW07). We will therefore assume that $T_i=T_e$ for the inflowing gas at the top of the column (Elsner \& Lamb 1977), so that
\begin{equation}
T_{i,{\rm top}} = T_{e,{\rm top}} \ .
\label{eqn:appendix ion temperature equal to electron temperature}
\end{equation}
The electron and ion sound speeds at the top of the column are given by (see Equation (\ref{eqn:temperatureconversion}))
\begin{equation}
a_{e,{\rm top}}^2 = \frac{\gamma_e k T_{e,{\rm top}}}{m_{\rm tot}} \ , \qquad
a_{i,{\rm top}}^2 = \frac{\gamma_i k T_{i,{\rm top}}}{m_{\rm tot}} \ ,
\label{eqn:appendix relate sound speed to temperature}
\end{equation}
and therefore our assumption that $T_{i,{\rm top}} = T_{e,{\rm top}}$ leads to the relation

\begin{equation}
a_{e,{\rm top}} = \left( \frac{\gamma_e}{\gamma_i}
\right)^{1/2} a_{i,{\rm top}} \ .
\label{eqn:electron sound speed to ion sound speed top}
\end{equation}

The radial component of the radiation energy flux, averaged over the cross-section of the column at radius $r$, is given by
\begin{equation}
F_r(r) = - \frac{c}{3 n_e \sigma_\parallel} \, \frac{dU_r}{dr} + \frac{4}{3}
\, v U_r \ ,
\label{eqn:radiation energy flux}
\end{equation}
where the first term on the right-hand side represents the upward diffusion of radiation energy parallel to the magnetic field, and the second term represents the downward advection of radiation energy towards the stellar surface (with $v<0$). The fact that the top of the accretion column is the last scattering surface implies that the photon transport makes a transition from diffusion to free streaming at $r=r_{\rm top}$, so that we make the following replacement in Equation~(\ref{eqn:radiation energy flux}),
\begin{equation}
- \frac{c}{3 n_e \sigma_\parallel} \, \frac{dU_r}{dr} \to
c \, U_r \ , \ \ \ \ \ r \to r_{\rm top} \ .
\label{eqn:free-streaming}
\end{equation}
By incorporating this transition into Equation~(\ref{eqn:radiation energy flux}), we see that the radiation energy flux at the upper surface is given by
\begin{equation}
F_r(r_{\rm top}) = \left ( c + \frac{4}{3} v_{\rm top} \right ) U_r(r_{\rm top}) \ .
\label{eqn:radiation flux and free-streaming}
\end{equation}

The form of the total energy transport rate is derived from Equation (\ref{eqn:energyflux}), using Equations~(\ref{eqn:eos2}), (\ref{eqn:soundspeeds}), (\ref{eqn:area}), and (\ref{eqn:dipolarmassflux}), which yields
\begin{equation}
\dot E(r) = A(r) F(r) = \dot M \left( \frac{F_r}{\rho |v|} -
\frac{v^{2}}{2}  -
\frac{a_i^2}{\gamma_i-1} -
\frac{a_e^2}{\gamma_e-1} +
\frac{G M_*}{r} \ \right ).
\end{equation}
The expression for the total energy transport rate at $r=r_{\rm top}$ is simplified once we implement the free-streaming boundary condition in Equation~(\ref{eqn:radiation flux and free-streaming}), and use Equations~(\ref{eqn:eos2}) and (\ref{eqn:soundspeeds}) to substitute for the radiation energy density $U_r$ in terms of the radiation sound speed $a_r$. The result obtained is
\begin{equation}
\dot E_{\rm top} \equiv \dot E \bigg|_{r=r_{\rm top}} =
-\dot M
\left [ \left( \frac{\gamma_i}{\gamma_i-1} +
\frac{\gamma_e}{\gamma_e-1} \right) \frac{a_{i,{\rm top}}^{2}}{\gamma_i} +\left(\frac{c}{v_{\rm top}}+ \frac{4}{3} \right)
\frac{a_{r,{\rm top}}^{2}}{\gamma_r (\gamma_r-1)} \right] \ ,
\label{eqn:startingenergyflux}
\end{equation}
where we have also utilized Equations~(\ref{eqn:free-fall top}) and (\ref{eqn:electron sound speed to ion sound speed top}).

\subsection{Boundary Conditions at the Stellar Surface}

The ionized gas flows downward after entering the top of the accretion funnel at radius $r=r_{\rm top}$, and eventually passes through a standing, radiation-dominated shock, where most of the kinetic energy is radiated away through the walls of the accretion column (Becker 1998). Below the shock, the gas passes through a sinking regime, where the remaining kinetic energy is radiated away (Basko \& Sunyaev 1976). Ultimately, the flow stagnates at the stellar surface, and the accreting matter merges with the stellar crust.

The surface of the neutron star is too dense for radiation to penetrate significantly (Lenzen \& Tr\"{u}mper 1978), and therefore the diffusion component of the radiation energy flux must vanish there. Furthermore, due to the stagnation of the flow at the stellar surface, the advection component should also vanish, and therefore we conclude that the radiation energy flux $F_r \to 0$ as $r \to R_*$. We refer to this as the ``mirror" surface boundary condition, which can be written as
\begin{equation}
F_r(r) \bigg|_{r=R_*} = 0 \ .
\label{eqn:radiation mirror condition}
\end{equation}

The stagnation of the flow at the stellar surface also implies there is no flux of kinetic energy into the star. Hence, at the stellar surface, the total energy transport rate, $\dot E$, reduces to the addition of (negative) gravitational potential energy to the star. The surface boundary condition for the total energy transport rate is therefore given by (see Equations~(\ref{eqn:energyflux}) and (\ref{eqn:energytransport}))
\begin{equation}
\dot E(r) \bigg|_{r=R_*} = \frac{G M_* \dot M}{R_*} \ .
\label{eqn:total surface energy flux}
\end{equation}
The stagnation boundary condition formally requires that $v=0$ at the stellar surface, where $r=R_*$. However, in practice, it is not possible to perfectly satisfy this condition due to the divergence of the mass density $\rho$ implied by stagnation. Therefore, we approximate stagnation at the stellar surface in our simulations using the condition
\begin{equation}
\lim_{r \to R_*} |v(r)| \lapprox \ 0.01\,c \ .
\label{eqn:surface stagnation}
\end{equation}

\subsection{Boundary Conditions at the Thermal Mound Surface}

As the flow decelerates near the base of the accretion column, the density increases and the opacity becomes dominated by free-free absorption, leading to the formation of a dense ``thermal mound'' (e.g., Davidson 1973). The thermal mound, with a temperature between  $10^7\,$K and $10^8\,$K, is the source of the blackbody seed photons that scatter throughout the column and contribute to the emergent Comptonized spectrum. The upper surface of the thermal mound is located at radius $r=r_{\rm th}$, which is defined as the radius at which the Rosseland mean of the free-free optical depth, $\tau_\parallel^{\rm ff}$, measured from the top of the column, is equal to unity.

In general, the vertical variation of $\tau_\parallel^{\rm ff}$ is computed using the integral
\begin{equation}
\tau_\parallel^{\rm ff}(r) =\int_r^{r_{\rm top}}
\alpha_{\rm R}(r') \, dr' \ ,
\label{taurosseland}
\end{equation}
where $\alpha_{\rm R}$ is the Rosseland mean free-free absorption coefficient for fully-ionized hydrogen. Equation~(\ref{taurosseland}) implies that the Rosseland mean free-free optical depth at the top of the column is zero, so that
\begin{equation}
\tau_\parallel^{\rm ff}(r_{\rm top}) = 0 \ .
\label{eqn:alpha1}
\end{equation}
At the upper surface of the thermal mound, we have
\begin{equation}
\tau_\parallel^{\rm ff}(r_{\rm th}) =\int_{r_{\rm th}}^{r_{\rm top}} \alpha_{\rm R}(r') dr' = 1 \ .
\label{eqn:absorption optical depth equals unity}
\end{equation}
Inside the thermal mound, $\tau_\parallel^{\rm ff} >1 $, leading to an approximate balance between thermal emission and absorption, although the balance is not perfect due to the escape of photons through the sides of the accretion column. The various thermal transfer rates and corresponding timescales are further discussed in Section \ref{subsection:flow regions}.

\section{SOLVING THE COUPLED SYSTEM}
\label{section:solving the coupled ODE system}

The set of five fundamental hydrodynamical differential equations that must be solved simultaneously using {\it Mathematica} comprises Equations~(\ref{eqn:dardr}), (\ref{eqn:daidr}), (\ref{eqn:daedr}),  (\ref{eqn:dudr}), and (\ref{eqn:dEdr2}). It is convenient to work in terms of non-dimensional radius, flow velocity, sound speed, and total energy transport rate variables by introducing the quantities
\begin{equation}
\tilde r = \frac{r}{R_g} \ , \ \
\tilde v = \frac{v}{c} \ , \ \
\tilde a_i = \frac{a_i}{c} \ , \ \
\tilde a_e = \frac{a_e}{c} \ , \ \
\tilde a_r = \frac{a_r}{c} \ , \ \
\tilde{\mathscr{E}} = \frac{\dot E}{\dot M c^2} \ ,
\label{eqn:convertingr}
\end{equation}
where $R_g$ is the gravitational radius, defined by
\begin{equation}
R_g \equiv \frac{\textit{G} M_*}{c^2} \ .
\label{eqn:RG}
\end{equation}
The computational domain extends from the top of the accretion column, at radius $\tilde r = \tilde r_{\rm top}$, down to the stellar surface, at dimensionless radius $\tilde r=4.836$, assuming a canonical stellar mass $M_*=1.4\,M_\odot$ and radius $R_*=10\,$km. In terms of these non-dimensional quantities, Equations~(\ref{eqn:dardr}), (\ref{eqn:daidr}), (\ref{eqn:daedr}),  (\ref{eqn:dudr}), and (\ref{eqn:dEdr2}) take the form
\begin{align}
\frac{d \tilde a_r}{d \tilde r} &= \frac{\tilde a_r}{2} \left ( \frac{3}{\tilde r} + \frac{1}{\tilde v}\frac{d \tilde v}{d \tilde r} \right ) - \frac{\sigma_\parallel R_g}{2 m_{\rm tot} c} \frac{\dot M}{A} \frac{\gamma_r}{\tilde a_r} \left ( \tilde{\mathscr{E}} + \frac{\tilde v^2}{2}  + \frac{\tilde a_e^2}{\gamma_e - 1} + \frac{\tilde a_i^2}{\gamma_i - 1} + \frac{\tilde a_r^2}{\gamma_r - 1} - \frac{1}{\tilde r} \right ),  \label{eqn:RadiationODEtilde} \\
\frac{d \tilde a_i}{d \tilde r} &= \frac{(1 - \gamma_i) \tilde a_i}{2} \left ( \frac{3}{\tilde r} + \frac{1}{\tilde v}\frac{d \tilde v}{d \tilde r} + \frac{R_g}{c^2} \frac{A}{\dot M} \frac{\gamma_i \dot U_i}{\tilde a_i^2} \right ), \label{eqn:IonODEtilde} \\
\frac{d \tilde a_e}{d \tilde r} &= \frac{(1 - \gamma_e) \tilde a_e}{2} \left ( \frac{3}{\tilde r} + \frac{1}{\tilde v}\frac{d \tilde v}{d \tilde r} + \frac{R_g}{c^2} \frac{A}{\dot M} \frac{\gamma_e \dot U_e}{\tilde a_e^2} \right ), \label{eqn:ElectronODEtilde} \\
\frac{d \tilde{\mathscr{E}}}{d \tilde r} &= \frac{\tilde a_r^2}{c (\ell_2 - \ell_1) \gamma_r (\gamma_r - 1) \tilde v} \left ( \frac{R_*^3}{R_g \, \tilde r^3} \right)^{1/2} \min \left ( c, \frac{c}{\tau_\perp} \right) ,  \label{eqn:EnergyODEtilde} \\
\frac{d \tilde v}{d \tilde r} &= \frac{\tilde v}{\tilde v^2 -\tilde a_i^2 -\tilde a_e^2}
\left \{ \frac{3 \left ( \tilde a_i^2 +\tilde a_e^2 \right )}{\tilde r} - \frac{1}{\tilde r^2}
+\frac{\sigma_\parallel R_g}{m_{\rm tot} c} \frac{\dot M}{A}
\left ( \tilde{\mathscr{E}} + \frac{\tilde v^2}{2} + \frac{\tilde a_i^2}{\gamma_i-1}
+ \frac{\tilde a_e^2}{\gamma_e-1}  \right . \right . \nonumber \\
& \qquad \qquad \qquad \qquad \left . \left .  \ \ + \frac{\tilde a_r^2}{\gamma_r-1} - \frac{1}{\tilde r} \right )
+ \frac{R_g}{c^2} \frac{A}{\dot M}\left [ (\gamma_i - 1) \dot U_i + (\gamma_e - 1) \dot U_e \right ] \right \}  \ , \label{eqn:VelocityODEtilde}
\end{align}
where the column cross-sectional area $A$ is given by (see Equations (\ref{eqn:solid angle}) and (\ref{eqn:area}))
\begin{equation}
A(\tilde r) = \Omega r^2 = \frac{\Omega_* R_g^3}{R_*} \, \tilde r^3 \ .
\end{equation}
These relations are supplemented by Equations~(\ref{eqn:diffusion velocity}) and (\ref{eqn:Udottotal}), which are used to compute the perpendicular scattering optical thickness, $\tau_\perp$, and the energy exchange rates, $\dot U_i$ and $\dot U_e$, respectively.

Our task is to solve the five coupled hydrodynamic conservation equations (Equations ({\ref{eqn:RadiationODEtilde})-(\ref{eqn:VelocityODEtilde})) to determine the radial profiles of the dynamic variables $\tilde a_r, \tilde a_i, \tilde a_e, \tilde{\mathscr{E}}$, and $\tilde v $, subject to the boundary conditions discussed in Section~4. Once these profiles are available, the electron temperature $T_e(\tilde r)$ can be computed from the electron sound speed $\tilde a_e$ using the relation (see Equation (\ref{eqn:temperatureconversion}))
\begin{equation}
T_e(\tilde r) = \frac{m_{\rm tot} c^2}{\gamma_e k} \,  \tilde a_e^2(\tilde r) \ .
\label{eqn:Teequation}
\end{equation}
The solutions for $\tilde v(\tilde r)$ and $T_e(\tilde r)$ are used as input to the {\it COMSOL} finite element environment in order to compute the photon distribution function, $f(\tilde r,\epsilon)$, inside the column, which is the focus of Paper II.

Solving the set of five hydrodynamic ODEs and the associated photon transport equation requires the specification of six free parameters, with values that are determined by qualitatively comparing the computed theoretical spectrum with the observed phase-averaged photon spectrum for a given source, while at the same time satisfying all of the relevant boundary conditions. In addition to the six free parameters, the model also utilizes an additional thirteen auxiliary parameters, that are either computed using internal relations, or constrained by observations. We organize the various theoretical parameters into three groups, as discussed below, which we refer to as ``free,'' ``constrained,'' and ``derived.''

The six fundamental ``free'' model parameters, as listed in Table \ref{tab:free model parameters}, are the angle-averaged electron scattering cross-section, $\overline{\sigma}$, the scattering cross-section in the direction parallel to the magnetic field, $\sigma_\parallel$, the magnetic field strength at the magnetic pole, $B_*$, the inner and outer polar cap arc-radii, $\ell_1$ and $\ell_2$, respectively, and the incident radiation Mach number, $\Mach_{r0}$, which is used to set the radiation sound speed at the top of the column, $\tilde a_{r_{\rm top}}$, via the relation
\begin{equation}
\Mach_{r0} = \frac{|\tilde v_{\rm top}|}{\tilde a_{r,{\rm top}}} \ .
\label{eqn:starting Mr0}
\end{equation}

\begin{deluxetable}{cclc}
\tablewidth{0pt}
\tablecaption{Free Parameters \label{tab:free model parameters}}
\tablehead{\colhead{Number} & \colhead{Parameter} & \colhead{Description}}
\startdata
1 & $\overline{\sigma}$ & Angle-averaged scattering cross-section \\
2 & $\sigma_\parallel$ & Parallel scattering cross-section \\
3 & $\ell_1$ & Polar cap inner arc-radius \\
4 & $\ell_2$ & Polar cap outer arc-radius \\
5 & $\Mach_{r0}$ & Incident radiation Mach number \\
6 & $B_*$ & Stellar surface magnetic field strength \\
\enddata
\end{deluxetable}

The six ``constrained'' parameters used in our simulations, listed in Table \ref{tab:constrained input parameters}, comprise the stellar mass $M_*$, the stellar radius $R_*$, the source distance $D$, the X-ray luminosity $L_{\rm X}$, the accretion rate $\dot M$, and the scattering cross-section for photons propagating perpendicular to the magnetic field, $\sigma_\perp$. Rather than being free parameters, these quantities are specified using canonical values from observation and theory. We use the canonical values $M_* = 1.4 M_\odot$ and $R_* = 10$ km for our model calculations, and we set the scattering cross-section for photons propagating perpendicular to the magnetic field equal to the Thomson cross-section, $\sigma_{\perp}=\sigma_{\rm T}$ (e.g., Arons et al. 1987). The accretion rate $\dot M$ is derived from the observed X-ray flux, $F_{\rm X}=L_{\rm X}/(4 \pi D^2)$ by using Equations~(\ref{eqn:accretion luminosity definition}) and (\ref{eqn:observed luminosity definition}) to write
\begin{equation}
\dot M = \frac{4 \pi D^2 F_{\rm X} R_*}{G M_*} \ .
\label{eqn:LXfromD}
\end{equation}
The distance $D$ can be estimated using known associations with globular clusters (Frail \& Weisberg 1990), or, in some cases, via direct measurement using very long baseline interferometry (Frail \& Weisberg 1990).

\begin{deluxetable}{cclc}
\tablewidth{0pt}
\tablecaption{Constrained Parameters \label{tab:constrained input parameters}}
\tablehead{\colhead{Number} & \colhead{Parameter} & \colhead{Description}}
\startdata
7 & $R_*$ & Stellar radius\\
8 & $M_*$ & Pulsar mass\\
9 & $D$ & Distance to source\\
10 & $L_{\rm X}$ & X-ray luminosity \\
11 & $\dot M$ & Accretion rate\\
12 & $\sigma_\perp$ & Perpendicular scattering cross-section\\
\enddata
\end{deluxetable}

\begin{deluxetable}{cclc}
\tablewidth{0pt}
\tablecaption{Derived Parameters \label{tab:derived model parameters}}
\tablehead{\colhead{Number} & \colhead{Parameter} & \colhead{Description}}
\startdata
13 & $\tilde r_{\rm top}$ & Top of accretion column \\
14 & $\tilde v_{\rm top}$ & Incident free-fall velocity \\
15 & $\tilde a_{r,{\rm top}}$ & Incident radiation sound speed \\
16 & $\tilde a_{i,{\rm top}}$ & Incident ion sound speed \\
17 & $\tilde a_{e,{\rm top}}$ & Incident electron sound speed \\
18 & $\tilde{\mathscr{E}}_{\rm top}$ & Incident total energy flux \\
19 & $\tilde r_{\rm th}$ & Thermal mound radius \\
\enddata
\end{deluxetable}

The remaining seven ``derived'' parameters listed in Table \ref{tab:derived model parameters} are computed from the six fundamental free parameters $\overline{\sigma}, \sigma_\parallel, \ell_1, \ell_2, \Mach_{r0},$ and $B_*$ by utilizing the boundary conditions discussed in Section~\ref{sec:BOUNDARY CONDITIONS}. The coupled system of five ODEs is first-order, and therefore we need only specify boundary values for each of the five unknowns. We use the radius at the top of the accretion column, $\tilde r_{\rm top}$, computed using Equations (\ref{eqn:rtop}) and (\ref{eqn:convertingr}), to derive incident values for the five unknown variables $\tilde v_{\rm top}$, $\tilde a_{i,{\rm top}}$, $\tilde a_{e,{\rm top}}$, $\tilde a_{r,{\rm top}}$, and $\tilde{\mathscr{E}}_{\rm top}$ in the coupled conservation equations. The velocity at the top of the column is derived from the free-fall velocity, $v_{\rm top}$, given previously in Equation~(\ref{eqn:free-fall top}), which can be rewritten in the non-dimensional form
\begin{equation}
\tilde v_{\rm top}=-\sqrt{\frac{2}{\tilde r_{\rm top}}} \ .
\end{equation}
The incident radiation sound speed, $\tilde a_{r,{\rm top}}$, is computed from the value of the incident radiation Mach number, $\Mach_{r0}$, using Equation~(\ref{eqn:starting Mr0}).

We compute the value of the incident ion Mach number at the top of the accretion column, $\Mach_{i0}$, by solving the momentum equation (Equation (\ref{eqn:conservation of momentum equation})), using the method described in Appendix~A. The ion sound speed at the top of the column follows from the relation
\begin{equation}
\tilde a_{i,{\rm top}} =  \frac{|\tilde v_{\rm top}|}{\Mach_{i0}} \ .
\label{eqn:starting Mi0}
\end{equation}
Likewise, the incident electron sound speed, $\tilde{a}_{e,{\rm top}}$, is computed by converting Equation (\ref{eqn:electron sound speed to ion sound speed top}) to non-dimensional variables using Equations (\ref{eqn:convertingr}), which yields
\begin{equation}
\tilde a_{e,{\rm top}} = \left(\frac{\gamma_e}{\gamma_i} \right)^{1/2}\tilde a_{i,{\rm top}} \ .
\label{eqn:aetildetop}
\end{equation}
Similarly, the value for $\tilde{\mathscr{E}}_{\rm top}$ is determined by converting Equation (\ref{eqn:startingenergyflux}) to non-dimensional variables using Equations (\ref{eqn:convertingr}), yielding
\begin{equation}
\tilde{\mathscr{E}}_{\rm top} =  \left( \frac{\gamma_i}{1-\gamma_i} +
\frac{\gamma_e}{1-\gamma_e} \right) \frac{\tilde a_{i,{\rm top}}^{2}}{\gamma_i} +\left(\frac{1}{\tilde v_{\rm top}}+ \frac{4}{3} \right)
\frac{\tilde a_{r,{\rm top}}^{2}}{\gamma_r (1-\gamma_r)} \ .
\label{eqn:Etildetop}
\end{equation}
The thermal mound radius, $\tilde{r}_{\rm th}$ is computed using Equation (\ref{eqn:absorption optical depth equals unity}), in which the parallel absorption optical depth is set equal to unity.

\subsection{Computing the Photon Spectrum}

The computational domain for the calculation extends from the stellar surface, at dimensionless radius $\tilde r=4.836$, up to the top of the accretion column, at radius $\tilde r = \tilde r_{\rm top}$, where we have assumed a canonical stellar mass $M_*=1.4\,M_\odot$ and radius $R_*=10\,$km. The attainment of a completely self-consistent description of the hydrodynamic structure of the accretion column, along with the radiation spectrum, is achieved using an iterative procedure. The coupling between the hydrodynamical simulation performed in {\it Mathematica} and the spectrum calculation performed in {\it COMSOL} is made via three vectors of information which are passed between the two computational environments. In order to compute the dynamical structure in {\it Mathematica}, we require knowledge of the inverse-Compton temperature function, $g(\tilde r)$ (see Equation~(\ref{eqn:gfunction})). Conversely, in order to carry out the spectrum calculation in {\it COMSOL}, we require knowledge of the velocity and electron temperature profiles, $\tilde v(\tilde r)$ and $T_e(\tilde r)$, respectively.

The iteration procedure begins with a calculation of the ``$0^{\rm th}$'' hydrodynamical structure in {\it Mathematica}, which is generated by arbitrarily setting $g(\tilde r) = 1$, meaning that we are initially assuming that the inverse Compton temperature $T_{\rm IC}(\tilde{r})$ is exactly equal to the electron temperature $T_e(\tilde{r})$ for all $\tilde{r}$ along the column. Once the six free model parameters listed in Table~\ref{tab:free model parameters} are assigned provisional values, the system of five coupled ODEs is integrated in {\it Mathematica} to determine the first approximation of the dynamical structure of the column. The resulting accretion velocity profile, $\tilde v(\tilde r)$, and electron temperature profile, $T_e(\tilde r)$, are then exported from {\it Mathematica} and passed into the {\it COMSOL} multiphysics module in preparation for the computation of the phase-averaged radiation distribution inside the column.

The {\it COMSOL} multiphysics module is a computer environment that employs the finite element method (FEM) and is well-suited for solving the radiation transport equation, which is a second order, elliptical, nonlinear partial differential equation. {\it COMSOL} inputs the electron temperature and accretion velocity profiles from {\it Mathematica} and then solves the photon transport equation on a meshed grid using the boundary conditions discussed in Section~\ref{sec:BOUNDARY CONDITIONS}. The resulting photon distribution function $f(\tilde r,\epsilon)$ (photons cm$^{-3}$ erg$^{-3}$) and phase-averaged photon count rate spectrum $F_{\epsilon}(\tilde{\epsilon})$ (photons s$^{-1}$ cm$^{-2}$ keV$^{-1}$) are obtained and discussed in Paper II, where $\epsilon$ is the photon energy. All transport phenomena are calculated using $f(\tilde r,\epsilon)$, including the radiation flux $F_r$, the radiation energy density $U_r$, and the photon number density $n_{\rm ph}$. By exploiting the combined strengths of {\it Mathematica} and {\it COMSOL}, we are able to solve, for the first time to our knowledge, the complete self-consistent problem of spectral formation and radiation hydrodynamics in an X-ray pulsar accretion column. We briefly discuss some aspects of the dual-platform iteration and the related convergence criteria below, but we defer complete details on the {\it COMSOL} calculation to Paper~II.

\subsection{Cyclotron Absorption}
\label{subsection:cyclotron absorption}

Although we do not present detailed spectral results in this paper, it is important to highlight our method for treating cyclotron absorption here, since this process plays a significant role in determining the shape of the simulated spectrum, which is compared with the observational data in order to tie down the model parameters. In lieu of a detailed model for the formation of cyclotron absorption features in the envelopes of pulsar accretion columns, which has not been developed yet, we will treat the formation of the observed absorption features by supposing that the features are imprinted at a particular altitude, denoted by $r_{\rm{cyc}}$. Hence the centroid energy of the absorption feature is interpreted as the cyclotron energy corresponding to the dipole magnetic field strength at radius $r_{\rm{cyc}}$ in the column. We argue that this approach is reasonable, provided the cyclotron imprint radius $r_{\rm{cyc}}$ is close to the radius at which the X-ray luminosity per unit length along the column, $\mathscr{L}_r$, is maximized, where $\mathscr{L}_r dr$ is the energy emitted per unit time through the walls of the dipole-shaped volume of the accretion column between positions $r$ and $r+dr$.

We can derive an expression for $\mathscr{L}_{r}$ by noting that in our escape-probability formalism, the energy escaping through the walls of the accretion column between radii $r$ and $r+dr$ per unit time is given by
\begin{equation}
\mathscr{L}_{r} dr = \dot{U}_{\rm{esc}} A(r) dr \ ,
\end{equation}
where $\dot{U}_{\rm{esc}}$ is given by Equation~(\ref{eqn:energyescape}) and the cross-sectional area of the column is $A(r)=\Omega(r)r^2$ (see Equation~(\ref{eqn:area})). Solving for $\mathscr{L}_{r}$ yields
\begin{equation}
\label{eqn:maximum luminosity height}
\mathscr{L}_r = \frac{U_r(r)\Omega(r)r^2}{t_{\rm{esc}}(r)} \ .
\end{equation}
We denote the radius of maximum X-ray emission using $r_{\rm X}$. In our approach, we attempt to minimize the distance between $r_{\rm X}$ and the cyclotron imprint radius, $r_{\rm{cyc}}$. Out of the three sources treated here, Her X-1 is the only one in which the cyclotron absorption radius $r_{\rm{cyc}}$ is exactly equal to $r_{\rm X}$. In the other two sources, Cen X-3 and LMC X-4, the two radii deviate by about 10\%.

\subsection{Model Convergence}
\label{subsection:convergence}

Our method for determining the convergence of the solutions for the flow velocity $\tilde v(\tilde r)$, the sound speeds $\tilde a_i(\tilde r)$, $\tilde a_e(\tilde r)$, and $\tilde a_r(\tilde r)$, and the energy transport rate $\tilde{\mathscr{E}}(\tilde{r})$, is based on the comparison of successive iterates of the electron temperature, $T_e$, and the inverse-Compton temperature, $T_{\rm IC}$. We define the convergence ratios, $\mathscr{R}_e$ and $\mathscr{R}_{\rm IC}$, respectively, for the electron and inverse-Compton temperatures using
\begin{equation}
\mathscr{R}_e^{n+1} \equiv \frac{T_e^{n+1}}{T_e^n} \ , \qquad
\mathscr{R}_{\rm IC}^{n+1} \equiv \frac{T_{\rm IC}^{n+1}}{T_{\rm IC}^n} \ ,
\label{eqn:convergence}
\end{equation}
where the superscripts represent the iteration number for the corresponding solution vectors. The solutions are deemed to have converged when the vector of convergence ratios for both the electron and the inverse-Compton temperature profiles are within 1\% of unity across the entire computational grid.

As explained in Section~5.1, we obtain the solution for the ``$0^{\rm th}$" iteration for the dynamical structure by setting $g(\tilde r)=1$ across the grid in the {\it Mathematica} calculation, and we then pass the resulting velocity profile $\tilde v(\tilde r)$ and electron temperature profile $T_e(\tilde r)$ into the {\it COMSOL} platform in order to obtain the corresponding ``$0^{\rm th}$" iteration of the photon distribution function, $f(\tilde r,\epsilon)$. Once the solution for $f(\tilde r,\epsilon)$ has been obtained using {\it COMSOL}, the associated profile of the inverse-Compton temperature, $T_{\rm IC}(\tilde r)$, is computed using Equation~(\ref{eqn:Compton7}), which is then combined with the electron temperature profile $T_e(\tilde r)$ to obtain the new iteration of the temperature ratio function, $g(\tilde r)$, using Equation~(\ref{eqn:gfunction}). Subsequently, the new iterate for $g(\tilde r)$ is used as input into the {\it Mathematica} implementation to compute new results for the dynamical structure variables, and so on.

This iterative cycle is continued, and the convergence ratios between successive iterates are computed using Equation~(\ref{eqn:convergence}), until convergence is achieved, which operationally means that the convergence ratios for the electron and inverse-Compton temperature profiles differ from unity by less than 1\% at all radii in the column. In the end, once convergence is achieved, we have obtained a self-consistent set of results for the radiation distribution $f(\tilde r,\epsilon)$ and the five dynamical variables $\tilde v(\tilde r)$, $\tilde a_r(\tilde r)$, $\tilde a_i(\tilde r)$, $\tilde a_e(\tilde r)$, and $\tilde{\mathscr{E}}(\tilde r)$. In the following section, we discuss the application of the method to compute the structure of the accretion column and the photon spectrum for three specific accretion-powered X-ray pulsars.

\section{ASTROPHYSICAL APPLICATIONS}
\label{section:ASTROPHYSICAL APPLICATIONS}

We are now in a position to compute the spectrum of an X-ray pulsar based on our new physical model, incorporating realistic boundary conditions, along with the effects of radiation, ion, and electron pressures, strong gravity, bremsstrahlung emission and absorption, cyclotron emission and absorption, electron-ion thermal energy transfer, and a dipole magnetic field. In particular, the inclusion of Compton scattering allows us to perform a self-consistent study of the inverse-Compton temperature variation along the column. The bulk fluid surface stagnation boundary condition ensures that we capture the first-order Fermi energization of the radiation due to the strong compression of the gas as it comes to rest at the stellar surface. These features are included here for the first time, to our knowledge, in an X-ray pulsar simulation.

We will apply the model to three specific high-luminosity accretion-powered X-ray pulsars that span the range of luminosities $L_{\rm X} \sim 10^{37-38}\,\rm erg \ s^{-1}$, namely Her X-1, Cen X-3, and LMC X-4. The output includes detailed studies of the vertical profiles of all of the dynamical variables, as well as the escaping column-integrated X-ray spectrum produced by bulk and thermal Comptonization of bremsstrahlung, cyclotron, and blackbody seed photon sources. The theoretical X-ray spectra are compared qualitatively with the observed phase-averaged spectra for Her X-1, Cen X-3, and LMC X-4. Here we focus solely on the dynamical results, and we defer a discussion of the photon sources and spectral results to Paper~II.

The sequence of steps required to obtain a self-consistent solution for the dynamical structure and the radiation distribution was described in Section~\ref{section:solving the coupled ODE system}. The values obtained for the six fundamental model free parameters $(\overline{\sigma}, \sigma_\parallel, B_*, \ell_1, \ell_2, \Mach_{r0}$) are listed in Table~\ref{tab:free parameters for model sources} for each of the three sources treated here. The corresponding results obtained for the six constrained parameters are listed in Table~\ref{tab:constrained parameters for model sources}, and the values of the seven derived parameters are listed in Table~\ref{tab:derived parameters for model sources}. In Table \ref{tab:Diagnostic Model Parameters} we summarize a number of additional diagnostic (output) parameters that provide further insight into the nature of the model results obtained for each of the three sources.

\begin{deluxetable}{llll}
\tablewidth{0pt}
\tablecaption{Free Parameters for Her X-1, Cen X-3, and LMC X-4
\label{tab:free parameters for model sources}}
\tablehead{\colhead{Parameter} & \colhead{Her X-1} & \colhead{Cen X-3} &
\colhead{LMC X-4}}
\startdata
Angle-averaged cross-section $\bar{\sigma}/\sigma_{\rm T}$ & $2.60\times10^{-3}$ & $3.00 \times 10^{-3}$ & $2.50\times10^{-3}$ \\
Parallel scattering cross-section $\sigma_\parallel/\sigma_{\rm T}$ & $1.02\times10^{-3}$ & $7.51 \times 10^{-4}$ & $4.18\times10^{-4}$ \\
Inner polar cap radius $\ell_1$ (m)  & 0   & 657   & 547  \\
Outer polar cap radius $\ell_2$ (m)  & 125  & 750   & 650    \\
Incident radiation Mach $\Mach_{r0}$ & 4.07  & 6.15  & 2.76 \\
Surface magnetic field $B_{*}$ ($10^{12}\,$G) & 6.25 & 3.60 & 8.00 \\
\enddata
\end{deluxetable}

\begin{deluxetable}{lllll}
\tablewidth{0pt}
\tablecaption{Constrained Parameters for Her X-1, Cen X-3, and LMC X-4
\label{tab:constrained parameters for model sources}}
\tablehead{\colhead{Parameter} & \colhead{Her X-1} & \colhead{Cen X-3} & \colhead{LMC X-4} & \colhead{units}}
\startdata
$R_*$ & $10$ & $10$ & $10$ & km\\
$M_*$ & $1.4 \, M_\odot$ & $1.4 \, M_\odot$ & $1.4 \, M_\odot$ & g\\
$D$ & 5.0 & 8.0 & 55.0 & kpc\\
$L_{\rm X}$ & $2.00\times10^{37}$ & $2.82\times10^{38}$ & $3.89\times10^{38}$ & erg s$^{-1}$\\
$\dot M$ & $1.08 \times 10^{17}$ & $1.52 \times 10^{18}$ & $2.09 \times 10^{18}$ & g s$^{-1}$\\
$\sigma_\perp$ & $\sigma_{\rm T}$ & $\sigma_{\rm T}$ & $\sigma_{\rm T}$ & cm$^{2}$\\
\enddata
\end{deluxetable}

\begin{deluxetable}{llll}
\tablewidth{0pt}
\tablecaption{Derived Parameters for Her X-1, Cen X-3, and LMC X-4 \label{tab:derived parameters for model sources}}
\tablehead{\colhead{Parameter} & \colhead{Her X-1} & \colhead{Cen X-3} &
\colhead{LMC X-4}}
\startdata
Top of accretion column $r_{\rm top}$ (km) & $21.19$ & $24.40$ & $21.30$  \\
Incident free-fall velocity $v_{\rm top}/c$ & -0.442 & -0.412   & -0.441  \\
Incident radiation sound speed $a_{r,{\rm top}}/c$ & 0.109   & 0.067   & 0.160  \\
Incident ion sound speed $a_{i,{\rm top}}/c$ & $1.61\times10^{-3}$ & $9.45\times10^{-4}$ & $1.86\times10^{-3}$ \\
Incident electron sound speed $a_{e,{\rm top}}/c$ & $2.16\times10^{-3}$ & $1.27\times10^{-3}$ & $2.50\times10^{-3}$ \\
Incident total energy transport $\dot{E}_{\rm top}$ (erg s$^{-1}$)& $2.39 \times 10^{36}$   & $1.51 \times 10^{37}$   & $1.01 \times 10^{38}$  \\
Thermal mound radius $r_{\rm th}$ (km) & 10.00   & 10.59   &  10.53  \\
\enddata
\end{deluxetable}

\begin{deluxetable}{llll}
\tablewidth{0pt}
\tablecaption{Diagnostic Parameters for Her X-1, Cen X-3, and LMC X-4
\label{tab:Diagnostic Model Parameters}}
\tablehead{\colhead{Parameter} & \colhead{Her X-1} & \colhead{Cen X-3} &
\colhead{LMC X-4}}
\startdata
Maximum cap radius (m) & 223 & 761 & 633 \\
Radiation sonic radius $r_{\rm sonic}$ (km)  & 11.95  & 12.21  & 13.21\\
Cyclotron absorption radius $r_{\rm{cyc}}$ (km) & 11.74 & 10.94 & 14.26\\
Maximum emission radius $r_{\rm{X}}$ (km) & 11.74 & 12.02 & 12.94\\
Dipole turnover height $z_c$ (km)  & $2.46\times10^{4}$ & 688 & 914\\
Column length (km) & 11.19  & 14.40 & 11.30\\
Absorption column density $\rm{N_{H}}$ (cm$^{-2}$) & 19.72 & 22.20 &  21.97\\
Thermal mound $T_{\rm{th}}$ (K) & $6.91\times10^{7}$ & $5.73\times10^{7}$ & $7.59\times10^{7}$  \\
Surface $T_e$ (K) & $6.91\times10^{7}$ & $6.97\times10^{7}$ & $8.75\times10^{7}$  \\
Surface impact velocity $v_*/c$ & $8.44\times10^{-3}$ & $8.05\times10^{-3}$ & $9.80\times10^{-3}$\\
\enddata
\end{deluxetable}

\subsection{Her X-1}
\label{subsection:astrophysical application Her X-1}

Figure~\ref{fig:Her X-1 dynamical profiles} depicts the results obtained for the accretion column structure upon applying our model to Her X-1. The dynamical variables plotted include the bulk fluid velocity and the radiation sound speed, the gas and radiation pressures, the pressure gradients, the Mach numbers, the temperatures, the energy transport per unit mass, the bulk fluid density, and the parallel scattering and parallel absorption optical depths. We adopt for the source luminosity $L_{\rm X} = 2\times10^{37}$ erg s$^{-1}$ (Reynolds et al. 1997; Dal Fiume et al. 1998). The values of the six model free parameters in the Her X-1 simulation are $\bar{\sigma}/ \sigma_{\rm T}=2.600 \times10^{-3}$, $\sigma_{\parallel}/ \sigma_{\rm T}=1.024 \times10^{-3}$, $\ell_1=0\,$m, $\ell_2=125\,$m, $\Mach_{r0}=4.07$, and $B_*=6.25\times10^{12}\,$G. Note that the top of the accretion column in the graphs of Figure~\ref{fig:Her X-1 dynamical profiles} is located on the right side, and the stellar surface is located on the left side.

The accretion column for Her X-1 is completely filled with inflowing plasma, which makes this is the only completely filled column among the three sources we investigated here. This may be reasonable, since Her X-1 is a ``fast rotator,'' as discussed in Section~\ref{subsection:Accretion Dynamics and Hot Spot Size}. The upper limit for the outer radius is $r_0 \lesssim 223$\,m, given by Equation (\ref{eqn:maximum outer radius constraint}), which is almost double the 125\,m outer polar cap radius used in our model. In the case of Her X-1, the accretion column spans a length of 11.20\,km, and the bulk free-fall velocity at the top of the column (Equation (\ref{eqn:free fall velocity})) is equal to $0.442\,c$.

The radiation sound speed at the top of the column is derived using Equation (\ref{eqn:starting Mr0}), and the radiation sonic surface (where $\Mach_r=1$) is located at radius $r_{\rm sonic}=11.95$\,km, which is where the bulk fluid slows to less than the radiation sound speed. The onset of stagnation is most noticeable when the bulk fluid enters the extended sinking regime (Basko \& Sunyaev 1976), which begins approximately 700\,m above the surface, and is characterized by a gradually decelerating flow, accompanied by a corresponding increase in temperature, pressure, and density. Approximate stagnation occurs at the stellar surface, with a residual bulk velocity of $0.0084\,c$.

It is apparent from the Mach number profiles plotted in Figure 4 that the flow remains supersonic with respect to the gas at the lower boundary of our computational domain, which is located just above the stellar surface. Hence we would expect a final discontinuous shock transition to occur as the flow merges into the stellar crust. However, the amount of residual kinetic energy converted into radiation at the discontinuous shock is negligible compared to the energy loss associated with the radiation emitted farther up in the column. Similar behavior is also observed in the cases of LMC X-4 and Cen X-3. Hence our neglect of the discontinuous shock is reasonable for the luminous sources treated here. However, the effect of the discontinuous shock is likely to be more important in lower-luminosity sources such as X Persei (e.g., Langer \& Rappaport 1982).

The model for Her X-1 required 16 iterations before $T_e$ and $T_{\rm{IC}}$ stabilize to less than a 1\% change from the previous iteration. Electron and ion temperatures are in near thermal equilibrium throughout the column, and at the top of the column, we have $T_e = T_i = 1.41\times10^7$\,K. The inverse-Compton temperature at the top of the column, $T_{\rm{IC}}= 6.82\times10^7$\,K, is almost five times larger than the electron temperature, which is the largest temperature gap between the photons and gas at any radius in the column. The electron temperature at the stellar surface is found to be $6.91\times10^7$\,K. Further discussion of the temperature distributions and the related thermal and dynamical timescales is presented in Section~\ref{section:CONCLUSION}.

Energy transport per unit mass transport is plotted in Figure~\ref{fig:Her X-1 dynamical profiles} in terms of the dimensionless quantity $\dot E/(\dot M c^2)$. According to our sign convention, a negative value corresponds to energy flow downwards, towards the stellar surface (see Equation (\ref{eqn:free fall velocity})), and therefore the profile of the gravitational potential energy component, $\dot{E}_g$, is depicted as a positive value, given by
\begin{equation}
\frac{\dot E_g}{\dot M c^2} = \frac{G M_* \dot M}{r} \frac{1}{\dot M c^2} = \frac{1}{\tilde r} \ .
\label{eqn:gravitational potential energy transport per unit mass}
\end{equation}
At the stellar surface, the value of the dimensionless radius is $\tilde r = \tilde r_* = 4.836$, assuming canonical values for the stellar mass and radius, with $M_* = 1.4\,M_\odot$ and $R_*=10$\,km. The kinetic energy transport component dominates over the radiation component at the top of the column, while the two are nearly equal at the radiation sonic surface. However, in the sinking regime, the kinetic energy is negligible, and we see that the energy transport is dominated by radiation advection and diffusion.

According to Equation (\ref{eqn:total surface energy flux}), at the surface of the star, we expect the total energy transport rate to reduce to the gravitational component only, so that $\dot E/(\dot M c^2) = 1/\tilde r_* = 0.2068$. Hence the radiation energy flux should vanish at the stellar surface (the ``mirror'' condition), and that is the boundary condition we attempt to enforce. Her X-1 is the only source in which the mirror condition at the stellar surface is slightly relaxed. The Her X-1 results depicted in Figure~\ref{fig:Her X-1 dynamical profiles} indicate that the advective and diffusive components in Equation (\ref{eqn:radiation energy flux}) do not exactly cancel at the surface, and we obtain for the residual total energy transport rate $\dot E/(\dot M c^2) = 0.228$. This represents an error of $\sim 10\%$ from the purely gravitational component. Her X-1 is also the only source in which the radius at which the cyclotron absorption feature is imprinted, $r_{\rm cyc}=11.74\,$km, is exactly equal to the radius of maximum emission, $r_{\rm X}$. See Section~\ref{subsection:cyclotron absorption} for further details.

The scattering and absorption optical depths, given by Equations (\ref{eqn:absorption optical depth at thermal mound is equal to unity}) and (\ref{eqn:absorption optical depth equals unity}), respectively, are plotted in Figure~\ref{fig:Her X-1 dynamical profiles} for photons propagating parallel to the magnetic field. The top of the column is the last scattering surface before photons freely escape in the vertical direction, and therefore the scattering optical depths is low in that region, and gradually increases towards the bottom of the column. The scattering optical depth diverges exponentially in the sinking regime, reflecting the pileup of material at the stellar surface, where the plasma density becomes extremely large. On the other hand, the parallel absorption optical depth never reaches unity in the case of Her X-1. Therefore, the thermal mound must exist at the stellar surface and blackbody photons are produced at the electron surface temperature, which may help explain the positive radiation diffusion flux at the stellar surface for this source.

The results we obtain for the dimensions of the hot spot at the magnetic pole in Her X-1 are significantly different than those obtained by BW07. In particular, we find that the outer polar cap radius is 125\,m, whereas BW07 found that their cylindrical polar cap radius was $r_0=44\,$m, which is about three times smaller than our result. Furthermore, BW07 assumed a constant electron temperature of $6.25\times10^7$\,K, which is about $9.5$\,\% lower than our stellar surface temperature of $6.91\times10^7$\,K. Another significant difference is that our stellar surface $B$-field strength of $B_*=6.25\times10^{12}\,$G is nearly double the BW07 value of $B_*=3.80\times10^{12}\,$G. We believe that the differences between our results and those of BW07 probably reflect the fact that BW07 assumed constant values for $T_e$, $r_0$, and $B$, which means that they should be interpreted as average values in an actual accretion column. On the other hand, our model implements a realistic dipole magnetic field geometry, with varying electron and ion temperatures, and therefore our surface results would be expected to exceed the mean values taken from the BW07 model.

\begin{figure}[htbp]
\centering
\includegraphics[width=\textwidth]{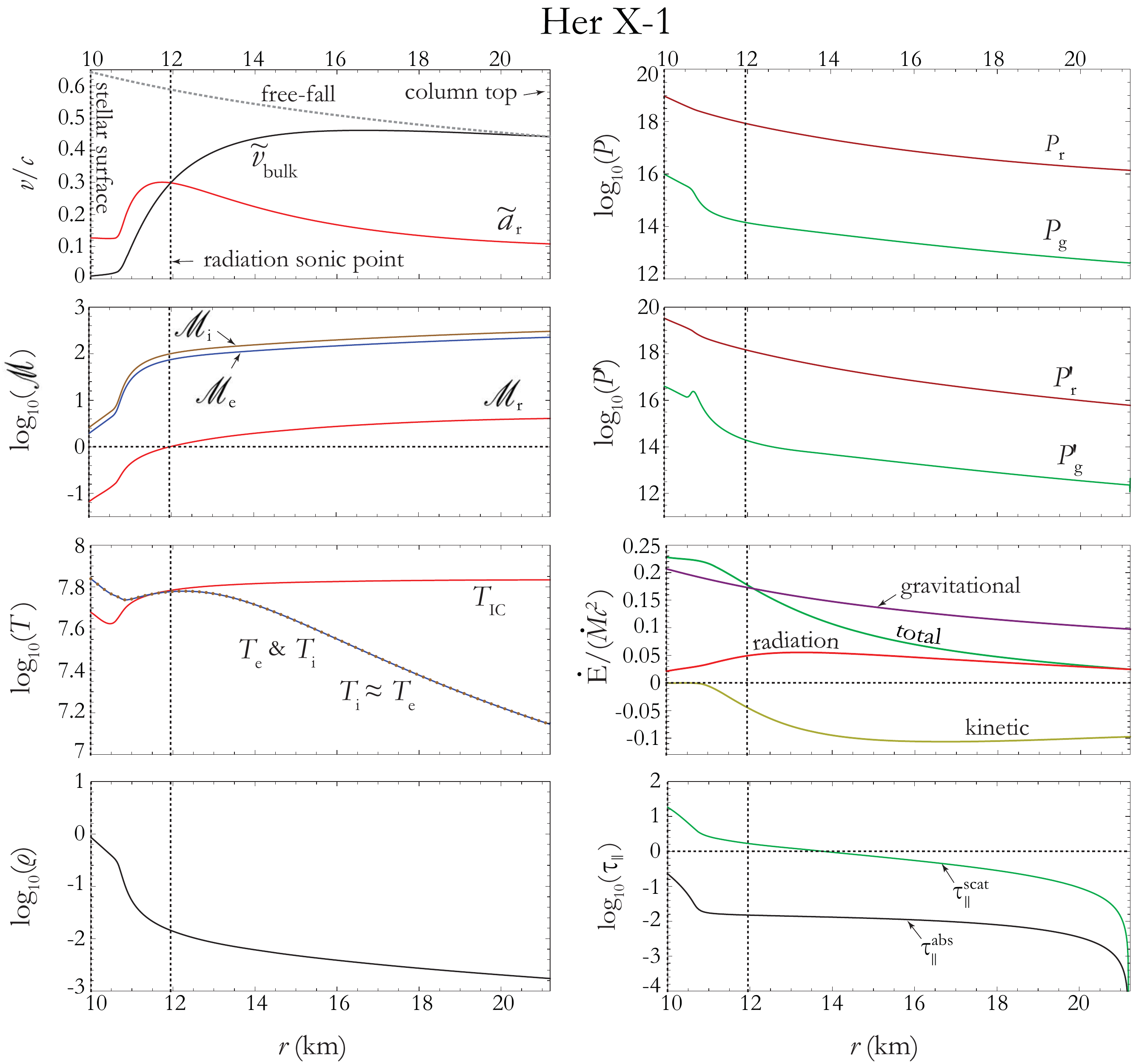}
\caption[Her X-1 Dynamical Profiles]{Model results for the dynamical profiles in the Her X-1 accretion column, based on the six free parameter values listed in Table \ref{tab:free parameters for model sources}. All quantities are plotted in cgs units except as indicated.}
\label{fig:Her X-1 dynamical profiles}
\end{figure}

\subsection{Cen X-3}

The profiles of the dynamical variables obtained in our application to Cen X-3 are depicted in Figure~\ref{fig:Cen X-3 dynamical profiles}. The six free parameter values used in the Cen X-3 simulation are $\bar{\sigma}/ \sigma_{\rm T}=3.000 \times10^{-3}, \sigma_{\parallel}/ \sigma_{\rm T}=7.510 \times10^{-3}, \ell_1=657 \, \textrm{m},\ell_2=750\,\textrm{m}, \Mach_{r0}=6.151, \textrm{and } B_*=3.6\times10^{12}\,$G. The source luminosity $L_{\rm X}=2.8\times10^{38}$ erg s$^{-1}$ is the same value used by BW07, which provides an opportunity to directly compare our model results with theirs using the same accretion rate. We find that the accretion column in Cen X-3 is a hollow cavity, with a thickness of 93\,m at the stellar surface. The upper limit for the outer radius is $r_0 \lesssim 761$\,m, according to Equation (\ref{eqn:maximum outer radius constraint}). The accretion column spans a length of 14.40\,km, and the bulk velocity at the top of the column is equal to the free-fall velocity of $0.412\,c$. The radiation sonic surface is located at radius $r_{\rm sonic}=12.21\,$km, and the bulk fluid enters the sinking regime at an altitude of 1.1\,km above the surface. A thermal mound exists for Cen X-3 at an altitude of 590\,m above the stellar surface, where the parallel optical depth exceeds unity, and the electron temperature is $5.73\times10^7$\,K. Approximate bulk stagnation occurs at the stellar surface, with a residual velocity of $0.0081\,c$ and a surface electron temperature equal to $6.97\times10^7$\,K, in contrast to the constant electron temperature $T_{e} = 3.40\times10^7$\,K used in the corresponding BW07 model. Eight iterations were required before $T_e$ and $T_{\rm{IC}}$ converged in our model. The stellar surface mirror condition is satisfied, so that the radiation energy flux essentially vanishes, and the total energy flux reduces to the gravitational component only. In the case of Cen X-3, the radius at which the cyclotron absorption feature is imprinted on the spectrum is $r_{\rm cyc}=10.94\,$km, whereas the radius of maximum emission is $r_{\rm X}=12.02\,$km.

\begin{figure}[htbp]
\centering
\includegraphics[width=\textwidth]{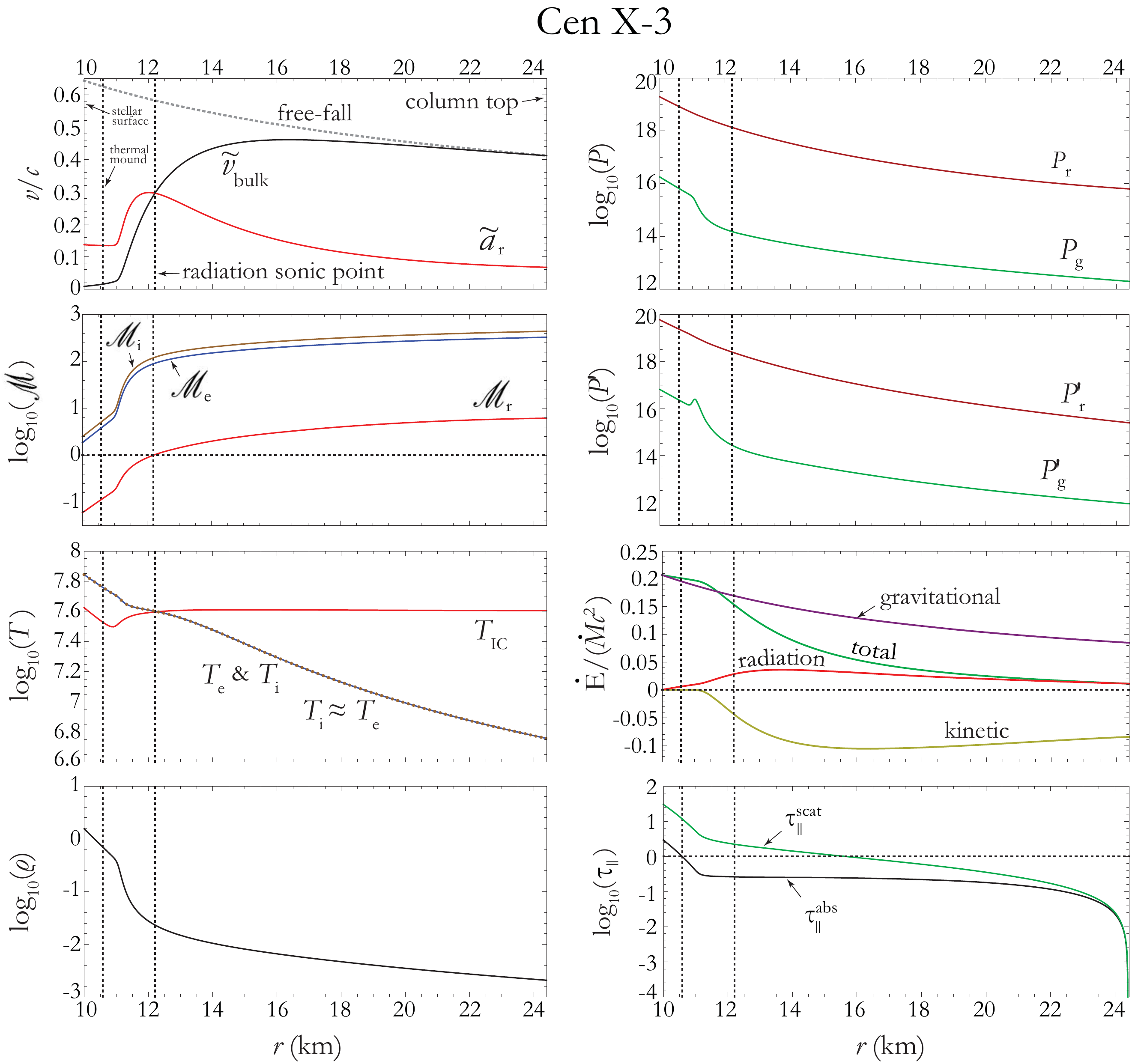}
\caption[Cen X-3 Dynamical Profiles]{Same as Fig.~\ref{fig:Her X-1 dynamical profiles}, except results are plotted for the Cen X-3 accretion column.}
\label{fig:Cen X-3 dynamical profiles}
\end{figure}

\subsection{LMC X-4}

Figure \ref{fig:LMC X-4 dynamical profiles} depicts the results we obtain for the dynamical profiles upon applying our model to LMC X-4, using for the source luminosity $L_{\rm X}=3.9\times10^{38}$ erg s$^{-1}$ (La Barbera et al. 2001). The six model free parameter values are $\bar{\sigma}/ \sigma_{\rm T}=2.500 \times10^{-3}, \sigma_{\parallel}/ \sigma_{\rm T}=4.176 \times10^{-3}, \ell_1=547\, \textrm{m},\ell_2=650\,\textrm{m}, \Mach_{r0}=2.761, \textrm{and } B_*=8.00\times10^{12}$\,G. The accretion column for LMC X-4 is hollow, exhibiting a geometry similar to that found in Cen X-3. The upper limit for the outer radius is $r_0 \lesssim 633$\,m, according to Equation (\ref{eqn:maximum outer radius constraint}). The accretion column spans a length of 11.30\,km, and the bulk velocity at the top of the column has a local free-fall velocity equal to $0.44\,c$. The radiation sonic point is located at radius $r_{\rm sonic}=13.21$\,km, and the bulk fluid enters the sinking regime at an altitude of 1.4\,km above the stellar surface. The thermal mound is located 530\,m above the surface, where the electron temperature is equal to $7.59\times10^7$\,K. Approximate stagnation occurs at the surface with a residual velocity of $0.0098\,c$ and a surface temperature equal to $8.75\times10^7$\,K, and the stellar surface mirror condition is satisfied. Our value for the electron surface temperature is somewhat higher than the BW07 model, which used the constant value $T_e = 5.90\times10^7$\,K. The model required eight iterations to converge $T_e$ and $T_{\rm{IC}}$, and the radius of maximum emission occurs at $r_{\rm X}=12.94\,$km, whereas the cyclotron absorption feature is imprinted at radius $r_{\rm cyc}=14.26\,$km.

\begin{figure}[htbp]
\centering
\includegraphics[width=\textwidth]{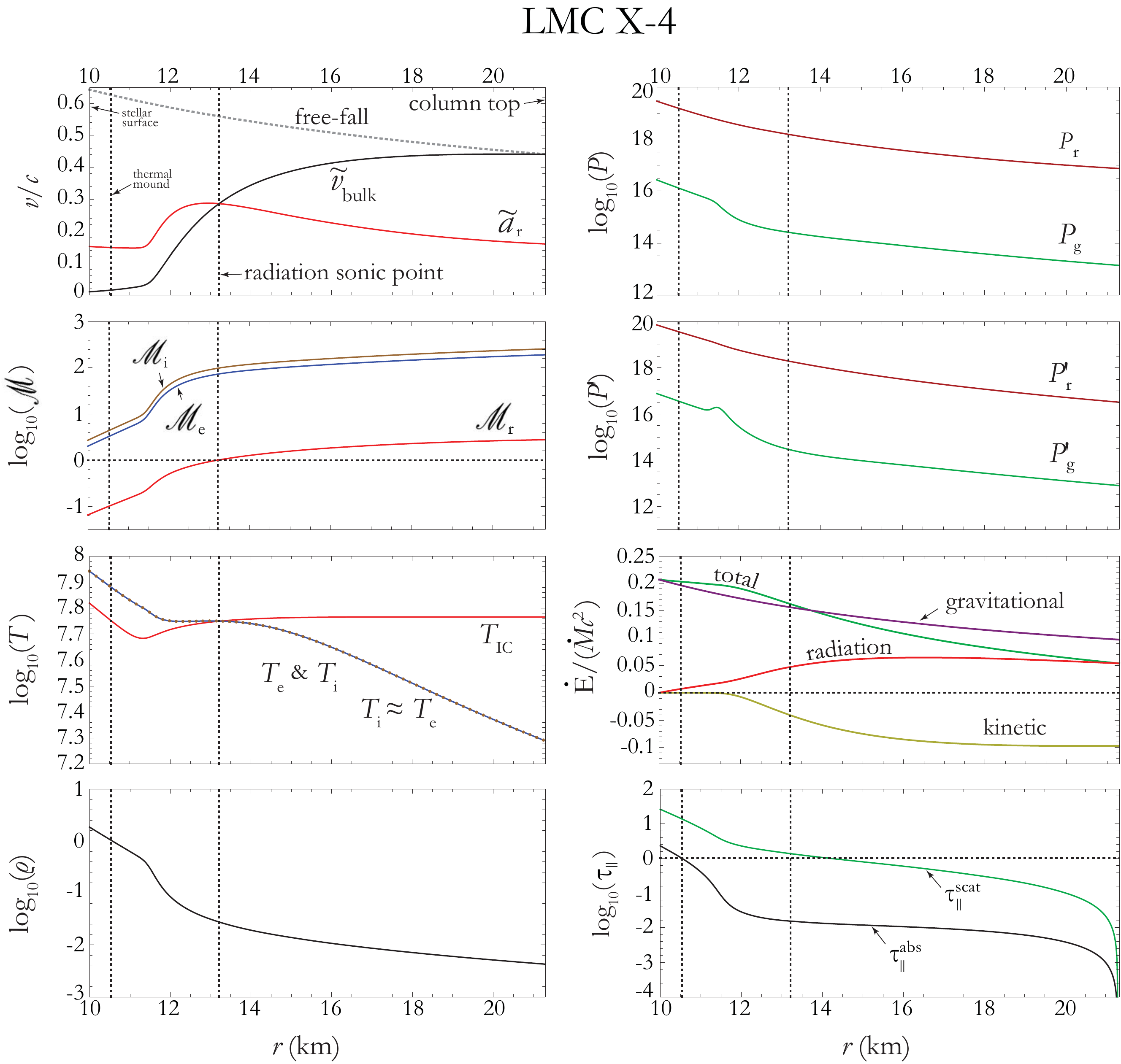}
\caption[LMC X-4 Dynamical Profiles]{Same as Fig.~\ref{fig:Her X-1 dynamical profiles}, except results are plotted for the LMC X-4 accretion column.}
\label{fig:LMC X-4 dynamical profiles}
\end{figure}

\section{DISCUSSION AND CONCLUSION}
\label{section:CONCLUSION}

Today, the general picture of pulsars as rapidly rotating neutron stars is widely accepted, but according to Werner Becker of the Max Planck Institute for Extraterrestrial Physics, ``The theory of how pulsars emit their radiation is still in its infancy, even after nearly forty years of work" (Becker 2006). The model developed here provides a realistic, self-consistent description of the radiation-hydrodynamical processes occurring within a dipole-shaped X-ray pulsar accretion column, including the effects of radiation, ion, and electron pressures, Newtonian gravity, bremsstrahlung emission and absorption, and cyclotron emission and absorption. The model also includes the dipole variation of the magnetic field, a rigorous calculation of the electron and ion temperatures, and a comprehensive set of rigorous physical boundary conditions. The model also includes a detailed treatment of thermal and bulk Comptonization, in terms of both the formation of the emergent X-ray spectrum, and the effect on the thermodynamic structure of the column. We find that by varying the six model free parameters $\overline{\sigma}, \sigma_\parallel, \ell_1, \ell_2, \Mach_{r0},$ and $B_*$, we can qualitatively fit the observed phase-averaged spectra for Her X-1, Cen X-3, and LMC X-4. Our focus in this paper is on the dynamical structure of the accretion column, and the detailed spectral results are presented in Paper~II. We review our main results in this section.

\subsection{Model Parameters}

Our model uses the canonically accepted values of $R_*=10^6$\,cm for stellar radius, $M_* = 1.4\,M_\odot$ for stellar mass, and $\sigma_\perp = \sigma_{\rm T} = 6.652\times10^{-25}\,\rm{cm}^2$ for the perpendicular electron scattering cross-section. Six free parameters uniquely determine the dynamical properties of the column, namely (1) the angle-averaged scattering cross-section $\bar{\sigma}$, (2) the inner polar cap arc-radius $\ell_1$, (3) the outer polar cap arc-radius $\ell_2$, (4) the radius at the top of the accretion column $\tilde r_{\rm top}$, (5) the incident radiation Mach number $\Mach_{r0}$, and (6) the stellar surface magnetic field strength, $B_*$, which is established at the magnetic pole. We carry out a self-consistent and iterative calculation of a set of five coupled conservation equations in order to self-consistently compute the hydrodynamic structure of the accretion column and the emergent radiation spectrum.

The dynamical structure is determined by using {\it Mathematica} to solve the dynamical equations, combined with appropriate physical boundary conditions. In this paper, we have presented results describing the detailed structure of the accretion columns in the well-known and high-luminosity X-ray pulsars Her X-1, Cen X-3, and LMC X-4. The solutions shown in Figures (\ref{fig:Her X-1 dynamical profiles}), (\ref{fig:Cen X-3 dynamical profiles}), and (\ref{fig:LMC X-4 dynamical profiles}) provide the radial profiles of the bulk flow velocity $\tilde{v}$, the radiation sound speed $\tilde{a}_r$, the ion sound speed $\tilde{a}_i$, the electron sound speed $\tilde{a}_e$, and the total energy transport rate per unit mass $\tilde{\mathscr{E}}$.

We find that the dynamical effects of gas pressure are negligible and that radiation pressure decelerates the gas to rest at the stellar surface in all three sources considered here, which all have relatively high accretion rates ($\dot{M}\gtrsim 10^{16}\,{\rm g\,s}^{-1}$). However, the inclusion of the gas energy equation is essential in our calculation of the ion and electron temperature profiles, which show significant deviations from the profile of the inverse-Compton temperature, especially near the top of the accretion column, where conditions are somewhat farther from equilibrium than in the deeper portions of the column.

A noticeable effect of increasing the surface magnetic field strength, $B_*$, in the model is an increase of the electron temperature at the stellar surface. We shall see in Paper II that this tends to harden the phase-averaged spectrum, and raises the energy of the exponential cutoff due to thermal Comptonization. The value of $B_{*}$ was determined by minimizing the distance between the location of the maximum emission from the walls of the column and the location at which the cyclotron absorption feature is imprinted on the spectrum. This is, admittedly, a rather crude criterion, but it is adequate for our purposes here, pending the availability of a generalized model that includes a rigorous treatment of the formation of the cyclotron absorption feature in the outer sheath of the column (Sch\"{o}nherr et al. 2008; Sch\"{o}nherr et al. 2014). Furthermore, our magnetic field follows the correct dipole variation with radius, whereas BW07 assumed a constant magnetic field. Since BW07 essentially utilized an average value of the magnetic field within the column, it is reasonable that our surface field value would exceed their (constant) value.

It should also be emphasized that the distance between the radius of maximum emission, $r_{\rm X}$, and the radius at which the cyclotron absorption feature is imprinted, $r_{\rm cyc}$, only agree closely in the case of Her X-1. In the other two sources, the disagreement between these two radii can be as large as $\sim 10$\%. A complete treatment of this issue will require the development of a more generalized code that simultaneously treats the hydrodynamics and the effect of cyclotron absorption occurring along the entire length of the column, which is beyond the scope of the present paper.

We find that the vertical extent of the accretion column is between 10\,km and 12\,km for all three sources. This is obviously comparable to the stellar radius, and therefore it is essential to implement the dipole variation of the magnetic field, both in order to compute the correct magnetic field variation, and also to properly compute the cross-sectional area of the column. Equation (\ref{eqn:rtop}) establishes the boundary condition at the upper surface of the column relating $r_{\rm{top}}$ and $\sigma_{\parallel}$, which is given by
\begin{equation}
\label{eqn:tau parallel at starting height in discussion section}
r_{\rm top} = \left[ r_c^{-3/2} + \frac{3 m_{\rm tot} (2 G M_*)^{1/2}} {2 \sigma_\parallel
\dot M} \frac{\Omega_*}{R_*} \right]^{-2/3} \ .
\end{equation}
The top of the cylindrical column in the BW07 model is given by combining their Equations (26) and (80) to yield
\begin{equation}
\label{eqn:rtop equation from BW07 model}
r_{\rm max}^{\rm BW} = \frac{R_*}{2} \left[\left( 1 + \frac{4 m_p G M_* \pi r_{0}^2}{\alpha c \sigma_{\parallel} \dot{M} R_{*}^2} \right)^{1/2}-1 \right] \ ,
\end{equation}
where $r_0$ is the radius of the cylinder and $\alpha\sim0.3-0.4$ is a constant. We compare Equation (\ref{eqn:tau parallel at starting height in discussion section}) to Equation (\ref{eqn:rtop equation from BW07 model}) and note the similarities with the purely cylindrical model from BW07 with respect to dependence on $r_0 = \ell_2$, $\dot{M}$, and $\sigma_{\parallel}$. Our model differs because we allow for a hollow column geometry that alters the value of $\Omega_*$ according to Equation (\ref{eqn:solid angle at stellar surface}). Additionally, Equation (\ref{eqn:rtop equation from BW07 model}) relies on the value of $\alpha$, which was required to be introduced in the BW07 velocity profile in order to solve the photon transport equation using the separation of variables method (Lyubarskii \& Sunyaev 1982).

The column-top radii for the three BW07 models using Equation (\ref{eqn:rtop equation from BW07 model}) are $r_{\rm max}^{\rm BW}$=13.72\,km (Her X-1), 29.97\,km (Cen X-3), and 32.44\,km (LMC X-4), respectively. The larger differences in accretion column length for Cen X-3 and LMC X-4 can be attributed to $B$-field geometry. Whereas the BW07 model is based strictly on a cylindrical geometry, our new model is based on a realistic dipole geometry which matches not only the free-fall velocity, but also the derivative of the free-fall velocity thereby satisfying the momentum conservation equation. The dynamical profiles for all three of our sources show that the radiation sonic surface and the length of the sinking regime increase proportionally to the source luminosity (Basko \& Sunyaev 1976).

\subsection{Free-Streaming Boundary Condition}

The free-streaming boundary condition at the top of the column can be verified by comparing the forces of gravity and radiation acting on the inflowing gas. The assumption at the top of the column is that the last photon-electron scattering events occur prior to photons freely escaping. Therefore, we expect the upward radiation force on the incoming electrons to be much smaller than the gravitational pull on the bulk fluid. The Newtonian gravitational force per electron-ion couple at the top of the column is
\begin{equation}
\label{eqn:Newton's Law of Gravitation}
F_g = - \frac{G M_* m_{\rm{tot}}}{r_{\rm{top}}^2} \ ,
\end{equation}
where $m_{\rm{tot}}=m_i+m_e$ is the combined masses of the electron and ion. The radiation force on incoming bulk fluid is equal to the product of the electron parallel scattering cross-section and the radiation pressure,
\begin{equation}
\label{eqn:radiation force on inflowing bulk fluid}
F_r = \sigma_{\parallel} P_r \ .
\end{equation}
We calculate the two forces from Equation (\ref{eqn:Newton's Law of Gravitation}) and Equation (\ref{eqn:radiation force on inflowing bulk fluid}) using the parallel scattering cross-section from Table \ref{tab:free parameters for model sources}, the column-top radii $r_{\rm top}$ from Table \ref{tab:derived parameters for model sources}, and the radiation pressure for each source, respectively, at the top of the columns shown in Figures \ref{fig:Her X-1 dynamical profiles}, \ref{fig:Cen X-3 dynamical profiles}, and \ref{fig:LMC X-4 dynamical profiles}. The force relationships calculated are shown in Table \ref{tab:Gravitational and Radiation Forces at Column Top}. The upward radiation force is dominated by the downward gravitational force. This result strengthens the free-streaming argument in which the top of the column is the last scattering surface.

A second method to verify the free-streaming surface is to compare the bulk fluid ram pressure and the radiation pressure. The ram pressure at the top of the column is
\begin{equation}
\label{eqn:ram pressure equation}
P_{\rm{ram}} = \rho v_{\rm ff}^2 \ .
\end{equation}
The pressure relationships calculated are shown in Table \ref{tab:Ram Pressure and Radiation Pressure at Column Top}. The ram pressure dominates at the top of the column for all three sources. Here, too, we conclude that the free-streaming condition is a valid assumption for all three sources.

\begin{deluxetable}{lcccc}
\tablewidth{0pt}
\tablecaption{Gravitational and Radiation Forces at Column Top
\label{tab:Gravitational and Radiation Forces at Column Top}}
\tablehead{\colhead{Source} & \colhead{$F_r$} & \colhead{$F_g$} & \colhead{units} & \colhead{$F_r/F_g$}}
\startdata
Her X-1 & 9.4 & 69.2 & $\times 10^{-12}$\,dyn & 0.14\\
Cen X-3 & 3.1 & 52.2 & $\times 10^{-12}$\,dyn & 0.06\\
LMC X-4 & 2.0 & 68.6 & $\times 10^{-12}$\,dyn & 0.03\\
\enddata
\end{deluxetable}

\begin{deluxetable}{lcccc}
\tablewidth{0pt}
\tablecaption{Ram Pressure and Radiation Pressure at Column Top
\label{tab:Ram Pressure and Radiation Pressure at Column Top}}
\tablehead{\colhead{Source} & \colhead{$P_r$} & \colhead{$P_{\textrm{ram}}$} & \colhead{units} & \colhead{$P_r/P_{\textrm{ram}}$}}
\startdata
Her X-1 & 0.14 & 3.05 & $\times 10^{17}$\,Ba & 0.05\\
Cen X-3 & 0.06 & 3.16 & $\times 10^{17}$\,Ba & 0.02\\
LMC X-4 & 0.73 & 7.43 & $\times 10^{17}$\,Ba & 0.10\\
\enddata
\end{deluxetable}

\subsection{Pick-Up Radius Variation}
\label{subsection:Pick-Up Radius}

As discussed in Section~\ref{subsection:entrainment}, the field lines connected to the inner and outer walls of the accretion column cross the accretion disk at the radii $R_{2,\,{\rm disk}}$ and $R_{1,\,{\rm disk}}$, respectively. Matter is picked up from the disk and entrained onto the magnetosphere at the Alfv\'en radius, $R_{\rm A}$. As the star rotates, the magnetic latitude in the disk plane, $\alpha$, oscillates between $\pm \varphi$, where $\varphi$ is the inclination angle between the magnetic and rotation axes of the star. As a result of the variation in $\alpha$, the Alfv\'en radius in the disk, $R_{\rm A}$, oscillates due to the oscillation of the magnetic field strength in the disk plane, $B_{\rm disk}$ (see Equations~(\ref{eqn:Alfven radius formula}) and (\ref{eqn:Beqmag})). In addition, the geometry of the inclined pulsar magnetosphere causes the inner and outer disk-crossing radii, $R_{2,\,{\rm disk}}$ and $R_{1,\,{\rm disk}}$, respectively, to oscillate as well. The combination of these oscillations causes matter to be fed into different parts of the accretion column as the star rotates. Hence, matter is fed to the inner wall of the accretion column when $R_{1,\,{\rm disk}}=R_{\rm A}$, and to the outer wall of the column when $R_{2,\,{\rm disk}}=R_{\rm A}$.

By using Equations~(\ref{eqn:Alfven radius formula}), (\ref{eqn:Beqmag}), and (\ref{eqn:R_equator2}) to evaluate $R_{\rm A}$, $R_{1,\,{\rm disk}}$, and $R_{2,\,{\rm disk}}$ as functions of the disk-plane latitude $\alpha$, we can attempt to determine the inclination angle of the system, $\varphi$, such that $R_{2,\,{\rm disk}} \lesssim R_{\rm A} \lesssim R_{1,\,{\rm disk}}$ during one spin of the star. We carry out this procedure for Cen X-3 and LMC X-4 in Figure~\ref{fig:capture}. The process also requires selecting a value for the normalization constant $\xi$ appearing in Equation~(\ref{eqn:Alfven radius formula}), such that the minimum value of the Alfv\'en radius equals the maximum value of $R_{2,\,{\rm disk}}$, corresponding to $\alpha=0$. We find that $\xi=1.21$ and $\xi=1.12$ for Cen X-3 and LMC X-4, respectively, as depicted by the red-dashed curves in Figure~\ref{fig:capture}. For comparison, we also plot the results obtained when $\xi=1.00$ for both sources, which are indicated by the orange-dashed curves. Once the value of $\xi$ is determined, the inclination angle $\varphi$ is obtained by requiring that $R_{\rm A}=R_{1,\,{\rm disk}}$ when $\alpha=\pm \varphi$. We find that $\varphi=22.6^\circ$ and $\varphi=26.0^\circ$ for Cen X-3 and LMC X-4, respectively. Our estimate of $\varphi=22.6^\circ$ for Cen X-3 is close to the estimate of $18^\circ$ provided by Kraus et al. (1996). Gas is transferred from the disk to the pulsar magnetosphere in the range of magnetic latitude labeled as the ``capture region'' in Figure~\ref{fig:capture}. We denote the mean values of the oscillating radii $R_{\rm A}$, $R_{1,\,{\rm disk}}$, and $R_{2,\,{\rm disk}}$ using $\left<R_{\rm A}\right>$, $\left<R_{1,\,{\rm disk}}\right>$, and $\left<R_{2,\,{\rm disk}}\right>$, respectively, and we present values for these quantities in Table \ref{tab:magnetic inclination angle for each source}. The Her X-1 values in Table \ref{tab:magnetic inclination angle for each source} were computed using $\xi=1.00$.

\subsection{Accretion Dynamics and Column Geometry}
\label{subsection:Accretion Dynamics and Hot Spot Size}

The size of the hot spot on the stellar surface must be understood in terms of the dynamics between accreted plasma gas and the neutron star magnetosphere. There is vast literature on the topic, which includes, but is not limited to, Lamb et al. (1973), Arons \& Lea (1976, 1980), Elsner \& Lamb (1976, 1977, 1984), Michel (1977a,b,c), Ghosh et al. (1977), Petterson (1977a,b,c), Ghosh \& Lamb (1978, 1979a,b), Lai (1999), Romanova et al. (2003), Pfeiffer \& Lai (2004), Ikhsanov et al. (2012), and Kulkarni \& Romanova (2013).

The general picture depends on the accretion scenario, which is often proposed in models pertaining to spherical (radial) accretion, or to inflow via a Keplerian disk. Mass transfer occurs across, or through, the magnetosphere, and is eventually aligned with the polar cap region via entrainment in the magnetic field lines, or, in the fast rotators, such as Her X-1, the plasma may be directly deposited far above the top of the accretion column from the plasma in a dense atmosphere. This picture is consistent with our results for Her X-1, which show that the accretion column is completely filled, whereas the columns for Cen X-3 and LMC X-4 are partially hollow. There are a number of additional factors to consider in these two differing topologies, which we discuss in further detail below.

Our model assumes the inflowing electrons and ions at the top of the accretion column are equilibrated, so that $T_{i0} \approx T_{e0}$ (Arons \& Lea 1976). This situation is treated by Equation (5) from Arons \& Lea (1976), which gives the infalling gas Mach number at the top of the column, for $r \ll R_{\rm{A}}$, as
\begin{equation}
\label{eqn:infalling gas Mach number}
\mathscr{M}_{\rm{infall}} =1.36 \times 10^9 \left( \frac{M_*}{M_{\odot}}\right)^{1/2} (r\,T_e)^{-1/2} \ ,
\end{equation}
where $M_*$ is the stellar mass, $M_{\odot}$ is one solar mass, $r$ is the radial distance (cm) from the X-ray source, and $T_e$ is the electron temperature (K) at the top of the column. Table \ref{tab:Infalling Gas Mach Number from Arons and Lea 1976} compares the infalling Mach numbers (Equation (\ref{eqn:infalling gas Mach number})) with the actual value of the incident ion Mach number using Equation (\ref{eqn:Mi0appendix}), where we set $T_{i0} = T_{e0}$. The error is less than 10\% for all three sources. We conclude that the large incident flow velocities in our models are expected, and momentum conservation and incident free-fall velocity are still rigorously implemented in our model.

\begin{deluxetable}{lccccc}
\tablecolumns{6}
\tablewidth{0pt}
\tablecaption{Infalling Gas Mach Number
\label{tab:Infalling Gas Mach Number from Arons and Lea 1976}}
\tablehead{
\colhead{Source}
& \multicolumn{1}{p{2cm}}{\centering $r_{\rm top}$ \\ km}
& \multicolumn{1}{p{2cm}}{\centering $T_{i0}$ \\ $10^7$\,K}
& \multicolumn{1}{p{2cm}}{\centering $\mathscr{M}_{\rm{infall}}$ \\ Eqn (\ref{eqn:infalling gas Mach number})}
& \multicolumn{1}{p{2cm}}{\centering $\mathscr{M}_{i0}$ \\ Eqn (\ref{eqn:Mi0appendix})}
& \multicolumn{1}{p{2cm}}{\centering error \\ \%}
}
\startdata
Her X-1 & 21.19 & 1.95 & 250 & 274 & 8.8 \\
Cen X-3 & 24.40 & 5.71 & 430 & 435 & 1.1 \\
LMC X-4 & 21.30 & 1.95 & 249 & 238 & 4.6 \\
\enddata
\end{deluxetable}

We shall review the fundamental orbital parameters for each source. The corotation radius, $R_{\rm{co}}$, is defined as the radius at which the Keplerian orbital period is equal to the star's spin period, $P_{\rm rot}$, so that (Lamb et al. 1973; Elsner \& Lamb 1977)
\begin{equation}
R_{\rm{co}} \equiv \left( \frac{G M_*}{\Omega_{\rm rot}^2} \right)^{1/3} = 1.5 \times 10^8 P_{\rm{rot}}^{2/3} \left( \frac{M_{*}}{M_{\odot}} \right)^{1/3} \, \rm{cm}\ ,
\label{eqn:corotation radius}
\end{equation}
where $\Omega_{\rm rot}=2 \pi/P_{\rm rot}$. The Keplerian angular velocity at the (mean) Alfv\'en radius is given by
\begin{equation}
\label{eqn:angular velocity at surfaces}
\Omega_{\rm{K}}(\left<R_{\rm{A}}\right>) = \left(\frac{G M_*}{\left<R_{\rm{A}}\right>^3}\right)^{1/2} \ .
\end{equation}
The influence of the stellar rotation on the accreting gas in the magnetosphere depends on the relationship between the angular velocities $\Omega_{\rm rot}$ and $\Omega_{\rm{K}}(\left<R_{\rm{A}}\right>)$. Slow rotators have $\Omega_{\rm rot} \lesssim \Omega_{\rm{K}}(\left<R_{\rm{A}}\right>)$, and therefore $R_{\rm{co}} \gtrsim \left<R_{\rm{A}}\right>$, which indicates that the force of gravity at $\left<R_{\rm{A}}\right>$ is larger than the centripetal force. The opposite condition is satisfied by fast rotators, in which $\Omega_{\rm rot} \gtrsim \Omega_{\rm{K}}(\left<R_{\rm{A}}\right>)$ and $R_{\rm{co}} \lesssim \left<R_{\rm{A}}\right>$. The rotation properties of each source are shown in Table \ref{tab:Stellar Rotation Properties}. The spin period, $P_{\rm{rot}}$, is 1.24\,s for Her X-1 (Vasco et al. 2013), 4.82\,s for Cen X-3 (Raichur \& Paul 2008b), and 13.5\,s for LMC X-4 (Levine et al. 2000). We see that Her X-1 is the only ``fast'' rotator, and therefore we expect that rotation will affect its accretion flow differently than in the slow rotators, such as Cen X-3 and LMC X-4 (Elsner \& Lamb 1976).

\begin{deluxetable}{lcccccc}
\tablecolumns{7}
\tablewidth{0pt}
\tablecaption{Stellar Rotation Properties
\label{tab:Stellar Rotation Properties}}
\tablehead{
\colhead{Source}
& \multicolumn{1}{p{2cm}}{\centering $P_{\rm{rot}}$ \\ s}
& \multicolumn{1}{p{2cm}}{\centering $\left<R_{\rm{A}}\right>$ \\ km}
& \multicolumn{1}{p{2cm}}{\centering $R_{\rm{co}}$ \\ km}
& \multicolumn{1}{p{2cm}}{\centering $\Omega_{\rm{K}}(\left<R_{\rm{A}}\right>)$ \\ rad s$^{-1}$}
& \multicolumn{1}{p{2cm}}{\centering $\Omega_{\rm rot}$ \\ rad s$^{-1}$}
& \colhead{Rotation}
}
\startdata
Her X-1 & 1.24 & 5334 & 1940 & 1.11 & 5.07 & fast \\
Cen X-3 & 4.82 & 1879 & 4780 & 5.29 & 1.30 & slow \\
LMC X-4 & 13.50 & 2535 & 9510 & 3.38 & 0.46 & slow \\
\enddata
\end{deluxetable}

\begin{figure}[htbp]
\centering
\includegraphics[width=\textwidth]{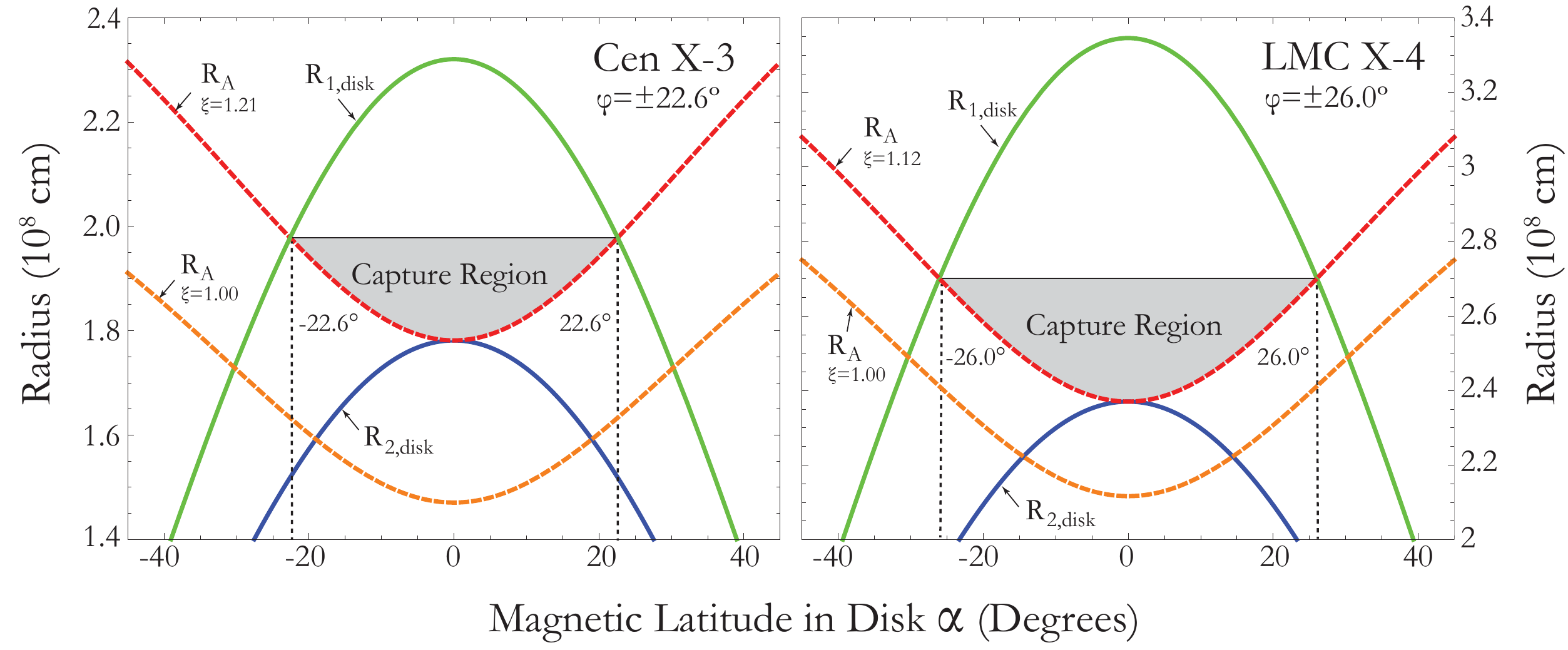}
\caption[capture]{Cen X-3 and LMC X-4 disk-crossing radii for the magnetic field lines connected to the inner and outer walls of the accretion column, denoted by $R_{1,\,{\rm disk}}$ and $R_{2,\,{\rm disk}}$, respectively, plotted as functions of the magnetic latitude, $\alpha$, in the plane of the accretion disk (Equations~(\ref{eqn:R_equator2})). Also plotted is the variation of the Alfv\'en radius, $R_{\rm A}$, obtained by combining Equations~(\ref{eqn:Alfven radius formula}) and (\ref{eqn:Beqmag}). See the discussion in the text.}
\label{fig:capture}
\end{figure}

The simulated rotation of Cen X-3, with a magnetic inclination of 22.6$^\circ$, over half of a spin period, is depicted in Figure \ref{fig:rotation plots of Alfven radius}. We define the phase $\beta=0^\circ$ as the direction to the companion star, so that the magnetic field axis and the axis of rotation are exactly aligned when viewed from the companion star's point of view. Therefore, when $\beta=0^\circ$, the magnetic pole is pointed towards the companion star, as nearly as possible. Figure \ref{fig:rotation plots of Alfven radius} displays eight instantaneous ``snapshots,'' with the rotation phase angle increasing by 22.5$^\circ$ each time. The plots illustrate how the Alfv\'{e}n radius, as well as the inner and outer disk-crossing radii, oscillate as the star rotates.

\begin{figure}[htbp]
\centering
\includegraphics[width=6.3in]{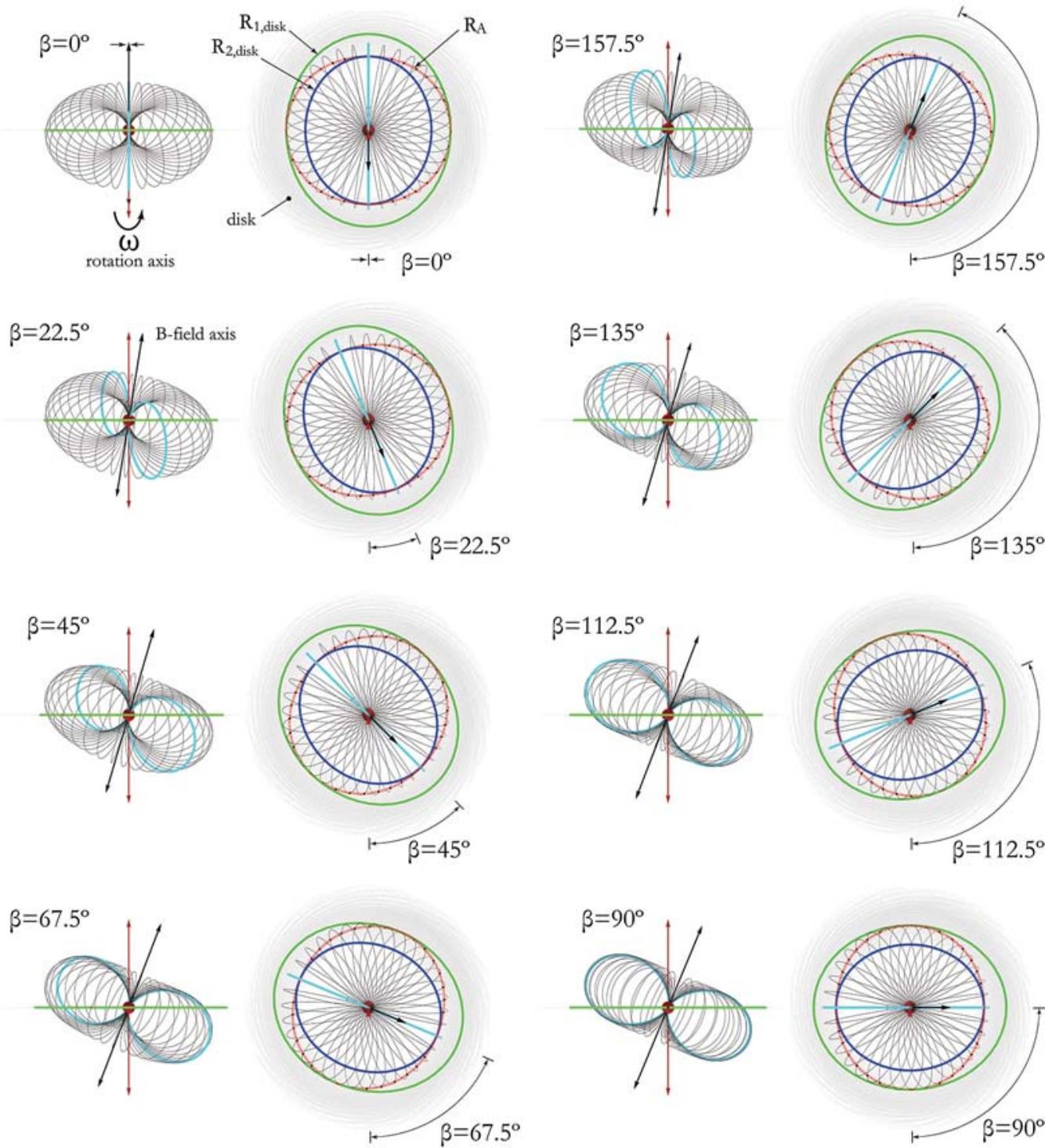}
\caption[rotation]{Rotation of Cen X-3 during half of the spin period. The angle $\beta=0^\circ$ corresponds to direction towards the companion star, so that the magnetic and rotation axes appear to be aligned. Eight instantaneous moments are depicted, in which the phase angle $\beta$ increases incrementally by 22.5$^\circ$, demonstrating how the stellar magnetic field sweeps through the accretion disk. Matter is captured by the magnetic field along the $R_{\rm A}$ curve.}
\label{fig:rotation plots of Alfven radius}
\end{figure}

Her X-1 has a completely filled column, and therefore the magnetic field associated with the inner polar angle ($\theta_1=0$) extends to the accretion capture radius, $r_{\rm{acc}}$, given by (Boroson et al. 2001)
\begin{equation}
\label{eqn:accretion capture radius}
1\times10^6 \lesssim r_{\rm{acc}} \equiv \frac{2 G M_*}{v_{\rm{rel}}^2} \lesssim 4\times10^6 \quad \rm{km},
\end{equation}
where $v_{\rm{rel}} = (v_{\rm{wind}}^2 + v_{\rm{ns}}^2)^{1/2}$, the orbital velocity $v_{\rm{ns}} \sim 169\,$km s$^{-1}$, and the wind velocity $v_{\rm{wind}} \sim 300 - 600$\,km s$^{-1}$. The results in Table \ref{tab:magnetic inclination angle for each source} suggest a high magnetic field inclination angle for Her X-1, in which the plane of the accretion disk is more likely aligned over the polar caps, rather than close to the stellar equatorial plane.

\begin{deluxetable}{lcccccc}
\tablecolumns{7}
\tablewidth{\textwidth}
\tablecaption{Magnetospheric Geometry and Disk Inclination Angle
\label{tab:magnetic inclination angle for each source}}
\tablehead{
\colhead{Source}
& \multicolumn{1}{p{2cm}}{\centering $\left<{R_{1,\,{\rm disk}}}\right>$ \\ km}
& \multicolumn{1}{p{2cm}}{\centering $\left<R_{2,\,{\rm disk}}\right>$ \\ km}
& \multicolumn{1}{p{2cm}}{\centering $\left<R_{\rm{A}}\right>$ \\ km}
& \multicolumn{1}{p{2cm}}{\centering $\theta_1$ \\ degrees}
& \multicolumn{1}{p{2cm}}{\centering $\theta_2$ \\ degrees}
& \multicolumn{1}{p{2cm}}{\centering $\varphi$ \\ degrees}
}
\startdata
Her X-1 & N/A & 32000 & 5331 & 0 & 0.72 & N/A\\
Cen X-3 & 2149 & 1650 & 1879 & 3.76 & 4.30 & 22.6\\
LMC X-4 & 3023 & 2142 & 2535 & 3.13 & 3.72 & 26.0\\
\enddata
\end{deluxetable}

In the hollow-column, slow rotators (Cen X-3 and LMC X-4), the supply of accreting plasma is confined to the magnetic field equatorial region (i.e., the angular momentum vectors of the $B$-field and accretion disk are nearly aligned), and the plasma may be forced to squeeze between the magnetic field lines in the equatorial plane via the Rayleigh-Taylor instability. In this scenario, the gas is eventually entrained onto the magnetic field as it flows towards the polar cap (Elsner \& Lamb 1976). On the other hand, in the fast rotators, such as Her X-1, the accretion column is completely filled, which may be the result of descent of the polar cusps (Arons \& Lea 1976; Michel 1977c; Elsner \& Lamb 1984), in which plasma enters the cusp region due to density buildup and the related pressure gradient. Michel (1977c) showed that the external gas pressure at the top of the polar cusp is expected to be at least five times greater than that at the equator, and therefore the magnetosphere shape for Her X-1, if it does have a larger plasma atmosphere above the polar caps, may be very different than those for Cen X-3 and LMC X-4.

We also note that the luminosity of Her X-1 is sufficiently high ($>10^{36}$\,erg s$^{-1}$) that reconnection of entrained magnetic fields may distort the large-scale structure of the magnetosphere to produce narrow open clefts (Arons \& Lea 1976), thereby allowing plasma to flood the full column width above the polar cap. Further, Pfeiffer \& Lai (2004) showed that accretion disks can be highly warped with saturated tilt angles at high latitudes where the inner radius of the disk may naturally rotate directly above the polar region. Finally, Ikhsanov \textit{et al.} (2012) conducted a study of spherical accretion scenarios in which the magnetic field of a companion star distorts the $B$-field flowing with the accreting material, thereby controlling the process via turbulent diffusion, which could potentially allow for accretion flow above the polar cap in fast rotators such as Her X-1.

\subsection{Energy Transport Timescales}

A thorough analysis of the timescale and energy rate dynamics would not be possible without the fundamental inclusion of gas pressure in the conservation equations. Recall from Section~\ref{subsection:equation of state} that the electrons are essentially confined along the magnetic field with a 1D Maxwell-Boltzmann energy distribution and an energy density per degree of freedom given by $(1/2) n_e k T_e$. The column dynamics are best understood by investigating the energy transfer rates and associated timescales in this formalism. The itemized list below provides definitions for eight of the key physical processes governing the thermodynamics of the energy transfer between the ions, electrons, and radiation. The timescales include the following:
\begin{enumerate}
\tablecaption{Timescales of Physical Processes inside the Accretion Column}
  \item Escape timescale:
        \begin{equation}\label{eqn:timescale list - escape time}
        t_{\rm{esc}} = \ell_{\rm esc}/w_\perp \ .
        \end{equation}
        See Equation~(\ref{eqn:energyescape}) for more discussion on the escape timescale. A photon will travel escape distance $\ell_{\rm esc}$, with diffusion velocity $w_{\perp}$, from the centerline of  the accretion channel to the exterior wall.
  \item Thermal bremsstrahlng absorption timescale:
        \begin{equation}\label{eqn:timescale list - absorption time}
        t_{\rm{abs}} = 1/(\alpha_{\rm R} \, c) \ .
        \end{equation}
        See Equation~(\ref{eqn:absorptioncoefficient}) for the Rosseland mean absorption coefficient $\alpha_{\rm R}$. The absorption timescale defines the average time before a photon is absorbed via bremsstrahlung thermal free-free absorption.
  \item Bulk Comptonization timescale:
        \begin{equation}\label{eqn:timescale list - bulk Comptonization}
        t_{\rm{bulk\;Comp}} = 1/(\vec\nabla \cdot \vec v) \ .
        \end{equation}
        Timescale over which the bulk fluid with velocity $\textbf{\textit{v}}$ is compressed and adds heat to the accreting gas. The divergence operator captures the compression dynamics.
  \item Electron scattering timescale:
        \begin{equation}\label{eqn:timescale list - scattering}
        t_{\rm{scat}} = 1/(n_e \sigmaT c) \ .
        \end{equation}
        Photons scatter isotropically off electrons with Thomson cross-section $\sigmaT$ inside the accretion column.
  \item Comptonization timescale:
        \begin{equation}\label{eqn:timescale list - Comptonization}
        t_{\rm{Comp}} = \frac{(1/2) \, n_e k T_e}{\dot U_{\rm Comp}} \ .
        \end{equation}
        Energy is added or removed from electrons due to interactions with photons within a Comptonization time interval $t_{\rm Comp}$. The rate of energy transfer, $\dot U_{\rm{Comp}}$, is given by Equation~(\ref{eqn:Udotcompton}).
  \item Bremsstrahlung emission timescale:
        \begin{equation}\label{eqn:timescale list - bremsstrahlung}
        t_{\rm{brem}} = \frac{(1/2) \, n_e k T_e}{\dot U_{\rm ff}} \ .
        \end{equation}
        Energy is removed from the electrons due to the bremsstrahlung cooling rate $\dot U_{\rm ff}$ given by Equation~(\ref{eqn:Udotbrem}).
  \item Cyclotron emission timescale:
        \begin{equation}\label{eqn:timescale list - Cyclotron}
        t_{\rm{cyc}} = \frac{(1/2) \, n_e k T_e}{\dot U_{\rm cyc}} \ .
        \end{equation}
        Energy is removed from the electrons due to the cyclotron cooling rate $\dot U_{\rm cyc}$ given by Equation~(\ref{eqn:Udotcyc}).
  \item The electron-ion equilibration timescale (based on Equations (32) and (103) from Arons et al. 1987):
        \begin{equation}\label{eqn:timescale list - ei equilibration}
        t_{\rm{ei}} =  3.2 \times 10^{-10} \left(\frac{n_e}{10^{22}\,\textrm{cm}^{-3}}\right)^{-1} \left(\frac{k T_{\rm{eff}}}{10\,\textrm{keV}}\right)^{3/2}\left(\frac{9}{\rm{ln}\,\Lambda}\right) \,\textrm{s} \ ,
        \end{equation}
        where $T_{\rm{eff}}$ is given by Equation (\ref{eqn:Teff}), and
        \begin{equation}
        \Lambda = 9.1-\frac{1}{2} \mathrm{ln} \left(\frac{n_e}{10^{22}\,\textrm{cm}^{-3}}\right) + \mathrm{ln}\left(\frac{k T_{\rm{eff}}}{10\,\textrm{keV}}\right) \ .
        \end{equation}
\end{enumerate}
An analysis of Equation (\ref{eqn:timescale list - ei equilibration}) shows that the electron-ion equilibration timescale is at least an order of magnitude smaller than all other timescales, which explains why the electron and ion temperatures are near thermal equilibrium throughout the column. The ionized gas is too hot and too dense for a significant deviation to occur as the electrons and ions thermally equilibrate more quickly than any other process.

\subsection{Flow Regions}
\label{subsection:flow regions}

The profiles for the timescales and energy transfer rates are shown in Figure~\ref{fig:Udot and timescales}. The three sources exhibit similar behavior in three basic regions, which we generalize here. The electron-ion equilibration timescale (see Equation (\ref{eqn:timescale list - ei equilibration})) is fast enough to ensure they are in near thermal equilibrium at all times. The corresponding energy transfer rate, $\dot{U}_{\rm{ei}}$, is orders of magnitude smaller than the four $\dot{U}$ terms shown in Figure \ref{fig:Udot and timescales} and is therefore not included. The three regions are described as follows:
\begin{enumerate}
\tablecaption{Timescales of Physical Processes inside the Accretion Column}
  \item Region 1 begins at the top of the column and extends over a majority of the column length. The cyclotron cooling timescale initially dominates the bremsstrahlung absorption timescale and is slightly faster than photon Comptonization. The inverse-Compton temperature is initially much larger than the gas temperature, and energy added to the gas by hotter photons is nearly offset by cyclotron emission. These two processes compete against each other as thermal bremsstrahlung absorption continually adds heat. Near the end of Region~1 the surplus heat reservoir in the photons is emptied while thermal bremsstrahlung absorption accelerates, thereby resulting in a role reversal between the photons and electrons at the entrance to Region~2.
  \item The gas begins to heat the photons via inverse-Compton scattering at the Region~2 entrance. Bulk fluid pressure and density rises rapidly as the accreting material builds-up near the surface. Electron scattering becomes the dominant timescale and bremsstrahlung absorption becomes the dominant heat transfer process. The energy transfer rates and the temperatures exhibit rising exponential logarithmic behavior inside the extended sinking regime.
  \item Region 3 begins at the top of the thermal mound for Cen X-3 and LMC X-4 where the parallel absorption optical depth exceeds unity. Thermal bremsstrahlung absorption and Comptonization dominate all other processes. Most photons are either absorbed or scatter within the mound and cannot escape. Mass density and pressure increase as the fluid stagnates near the stellar surface.
\end{enumerate}

We conclude that the Comptonization timescale and energy transfer rate determine two markedly different regions in the column. The upper region (Region 1) starts at the free-streaming surface and reaches deep into the column to between 1.5\,km and 2.5\,km above the stellar surface. It is characterized by slowly changing energy transfer from photons to the gas, where cyclotron and Comptonization processes dominate, and the relative positions among the various timescales is mostly unchanged. Beyond the column half-way point the bulk fluid density begins to play a very important role as the accreting material decelerates, fluid kinetic energy dissipates, and density rises to the point where scattering becomes the dominant timescale. The additional collisions result in accelerated energy addition to the gas due to bremsstrahlung thermal free-free absorption.

The lowest $\sim 500$\,m of Region 1 is characterized by slowing energy transfer from the photons to the electrons as they approach thermal equilibrium. The radiation sonic surface is located within a few hundred meters of the thermal mound surface. The dynamics change dramatically at the entrance to Region 2, where the gas temperature exceeds the inverse-Compton temperature. Here, the energy transfer rates are orders of magnitude larger than those in the upper region and the dominant heat mechanisms are bremsstrahlung thermal absorption and inverse-Compton scattering.

\begin{figure}[htbp]
\centering
\includegraphics[width=\textwidth]{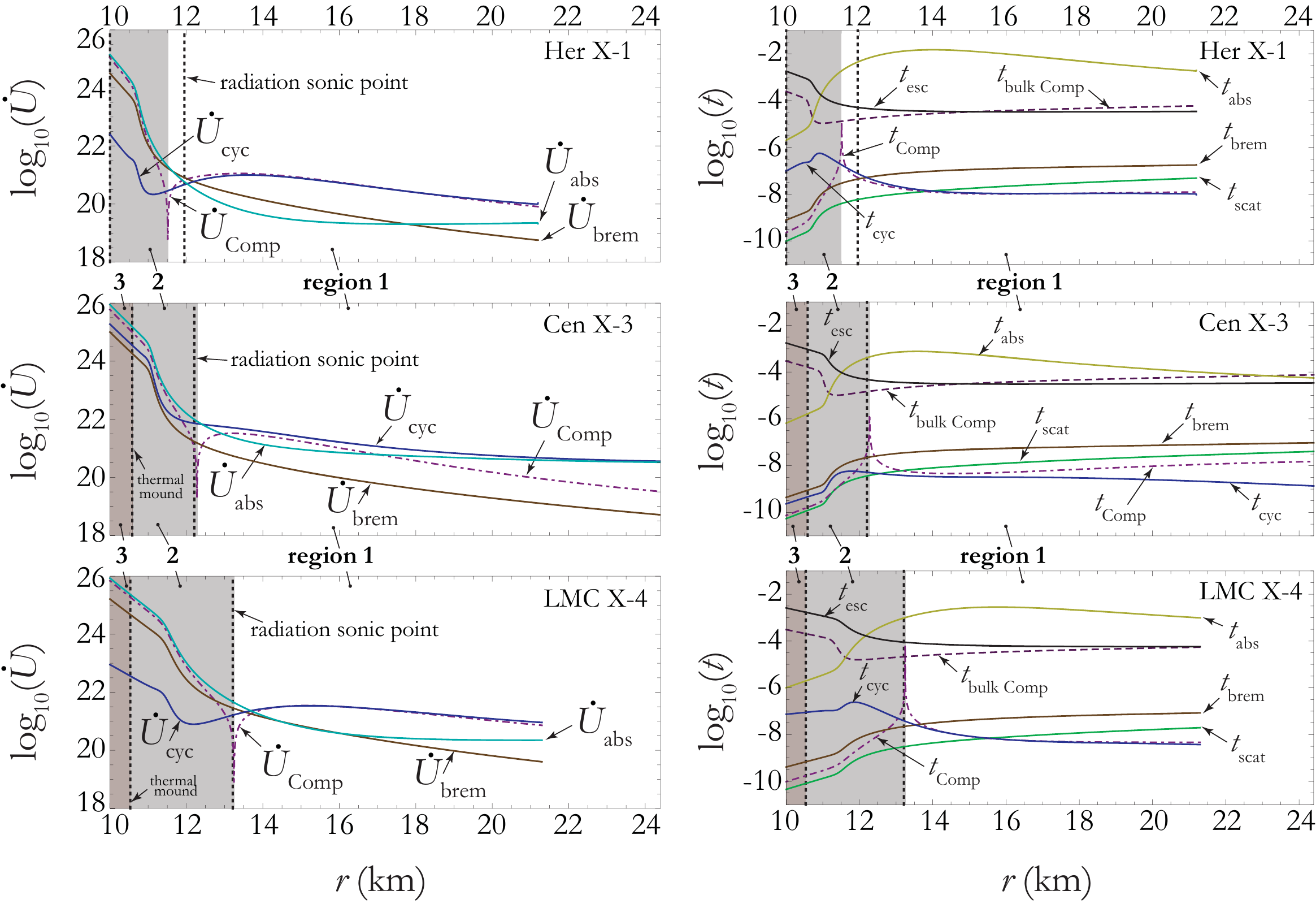}
\caption[Energy Transfer Rates]{Thermal coupling rates per unit volume, and associated timescales, for the accretion columns in Her X-1, Cen X-3, and LMC X-4, plotted in cgs units as functions of the radius from the center of the star. The accretion columns are divided into three regions according to the dominant timescales; see the discussion in the text.}
\label{fig:Udot and timescales}
\end{figure}

\subsection{Scattering Cross-Sections}

The values of the surface magnetic field, $B_*$, obtained in each of our three source models are larger than the corresponding values obtained in the BW07 models. The magnetic field strengths in the BW07 models were set as constants. The field strength in our model, however, quickly drops due to the dipole implementation, where $B \propto r^{-3}$, thereby resulting in a larger ratio $\epsilon/\epsilon_{\rm{cyc}}$ at the cyclotron energy $\epsilon_{\rm{cyc}}(r)$ for a photon at energy $\epsilon$. The difference in approaches becomes especially relevant when we compare the electron scattering cross-sections with the theoretical values from Equation (\ref{eqn:parallel}). We examine this more closely in Paper II where the average photon energy, $\bar{\epsilon}$, is computed along the column. Here, we compare our parallel and angle-averaged scattering cross-sections (labeled as WWB) with those used in the BW07 model, which are shown in Tables \ref{tab:Electron Scattering Cross-Section comparison (Parallel)} and \ref{tab:Electron Scattering Cross-Section comparison (Angle-Averaged)}. We notice our values are an order of magnitude larger. Canuto et al. (1971) showed that the magnitude of the electron scattering cross-section is reduced from the Thomson cross-section when $\epsilon < \epsilon_{\rm{cyc}}$, over all propagation angles $\theta$ with respect to the magnetic field direction, according to
\begin{equation}
\frac{\sigma(\theta)}{\sigma_{\rm T}} \propto \left[\frac{\epsilon}{B(r)}\right]^{2} \ .
\label{eqn:Canuto scattering relationship}
\end{equation}
We see from Equation (\ref{eqn:Canuto scattering relationship}) that the ratio $\sigma(\theta)/\sigma_{\rm T}$ in our model will be larger than those of BW07 because $B(r)$ is lower with increasing $r$, whereas the BW07 model maintained a constant $B$-field throughout.

\begin{deluxetable}{lccc}
\tablecolumns{4}
\tablewidth{0pt}
\tablecaption{Parallel Electron Scattering Cross-Section
\label{tab:Electron Scattering Cross-Section comparison (Parallel)}}
\tablehead{\colhead{Source} & \multicolumn{1}{p{2cm}}{\centering $\sigma_{\parallel}/\sigma_{\rm T}$ \\ WWB} & \multicolumn{1}{p{2cm}}{\centering $\sigma_{\parallel}/\sigma_{\rm T}$ \\ BW07} & \multicolumn{1}{p{2cm}}{\centering Ratio \\ \nicefrac{WWB}{BW07}}}
\startdata
Her X-1 & $2.60\times10^{-3}$ & $2.93\times10^{-4}$ & 8.9\\
Cen X-3 & $3.00\times10^{-3}$ & $4.51\times10^{-4}$ & 6.7\\
LMC X-4 & $2.50\times10^{-3}$ & $3.98\times10^{-4}$ & 6.3\\
\enddata
\end{deluxetable}

\begin{deluxetable}{lccc}
\tablecolumns{4}
\tablewidth{0pt}
\tablecaption{Angle-Averaged Electron Scattering Cross-Sections
\label{tab:Electron Scattering Cross-Section comparison (Angle-Averaged)}}
\tablehead{\colhead{Source} & \multicolumn{1}{p{2cm}}{\centering $\bar{\sigma}/\sigma_{\rm T}$ \\ WWB} & \multicolumn{1}{p{2cm}}{\centering $\bar{\sigma}/\sigma_{\rm T}$ \\ BW07} & \multicolumn{1}{p{2cm}}{\centering Ratio \\ \nicefrac{WWB}{BW07}}}
\startdata
Her X-1 & $1.02\times10^{-3}$ & $4.15\times10^{-5}$ & 24.6\\
Cen X-3 & $7.51\times10^{-4}$ & $8.30\times10^{-5}$ & 9.0\\
LMC X-4 & $4.18\times10^{-4}$ & $4.85\times10^{-5}$ & 8.6\\
\enddata
\end{deluxetable}

\subsection{Future Work}

The capabilities of our model, in its current state, permit a detailed physical study over the full length of an X-ray pulsar accretion column, including the radial dependences of many key phenomena, such as the photon emission, the hydrodynamic and thermodynamic structure, the imprint of the cyclotron absorption feature, and the stellar surface magnetic field strength. In this paper, we have modeled the dynamical behavior of three high-luminosity sources spanning a range of luminosities from $L_{\rm X} \sim 10^{37} - 10^{38}\,{\rm erg \ s}^{-1}$. We find that the gas pressure does not play a dynamically significant role in these sources. However, this question needs to be reexamined in the context of lower-luminosity sources, such as X Persei, in which the pressure of the gas could play a more significant role, as discussed by Langer \& Rappaport (1982). It is also likely that in these sources, the effect of the gas-mediated, discontinuous shock at the base of the column will be more significant than in the high-luminosity pulsars studied here (Becker et al. 2012). We plan to pursue these questions in future work.

Our model provides a new tool for estimating the surface magnetic field strength, and it may also facilitate studies of the variation of the cyclotron absorption energy $\epsilon_{\rm{cyc}}$ as the luminosity is varied (Mihara et al. 1995; Staubert et al. 2007; Becker et al. 2012). In particular, our model can be used investigate observed changes in the Her X-1 pulse phase averaged cyclotron line energy (Staubert et al. 2007) by changing the mass flow rate. More recent observations by Staubert et al. (2014) show a reduction in the line energy over a span of sixteen years. We can potentially perform a parameter study using our model to yield possible explanations for this behavior. This may also include experimental modification of the boundary conditions due to stellar surface magnetic field compression or other anomalies such as energy transport into or out of the star (Payne \& Melatos 2007; Mukherjee et al. 2013).

We also plan to eventually implement energy-dependent electron scattering cross-sections, relativistic corrections for gravity, implementing a cyclotron absorption term to observe the imprint of the cyclotron resonant scattering feature in the calculated spectrum, using the energy-dependent bremsstrahlung thermal free-free absorption coefficient instead of the Rosseland mean coefficient, and adding a second spatial dimension to observe 2D dynamical behavior. We also intend to couple the five dynamical conservation equations and the photon transport PDE into a single-platform simulation within the {\it COMSOL} finite element environment. This may also facilitate studies of the effect of an imbedded gas-mediated shock, thereby allowing us to observe how the electron and ion temperatures react to the presence of a density discontinuity near the base of the accretion column. This would permit a direct comparison with the work of Langer \& Rappaport (1982) and Canalle et al. (2005), where the discontinuous shock has a profound effect on the thermodynamics in the column.

\appendix
\section{ION SOUND SPEED AT COLUMN TOP}

At the top of the column, the incident value for the ion sound speed $\tilde a_i$ (see Table \ref{tab:free parameters for model sources}) is derived from the column-top ion Mach number $\Mach_{i0}$ and incident flow velocity $\tilde v_{\rm top}$, given by (where $\tilde{v}<0$ indicates flow is towards the stellar surface)
\begin{equation}
\tilde a_{i,{\rm top}} = - \frac{\tilde v_{\rm top}}{\Mach_{i0}} \ .
\label{eqn:incident sound speed}
\end{equation}
To arrive at $\Mach_{i0}$, we begin with ion and electron energy conservation in the comoving frame using Equation~(\ref{eqn:gasenergydensity1})
\begin{equation}
\frac{d U_{i,e}}{d r} = -\left( \frac{1}{v}\frac{d v}{d r} + \frac{3}{r} \right) \gamma_{i,e}U_{i,e} +
\frac{\dot U_{i,e}}{v} \ ,
\label{eqn:gasenergydensity2}
\end{equation}
where we have used the following substitution for the mass density gradient $d \rho / d r$.
\begin{equation}
\frac{1}{\rho} \frac{d \rho}{d r}
 = - \frac{1}{v} \frac{d v}{d r}
- \frac{3}{r} \ ,
\end{equation}
which can be derived using Equation (\ref{eqn:massdensity}).

The electrons and ions are assumed to have the same temperature at the top of the column ($T_e \approx T_i$), therefore the energy exchange rate between the two is zero ($\dot U_i^{\rm top} = - \dot U_{\rm ei}^{\rm top}=0$). (Radiative processes due to the ions are negligible compared to electron radiative processes.) We convert energy density to pressure using $P=(\gamma-1) U$ and combine the two species (ions and electrons) from Equation~(\ref{eqn:gasenergydensity2}) to obtain the total gas pressure gradient at the top of the accretion column. Converting to dimensionless distance $\tilde r$ and flow velocity $\tilde v$ we obtain
\begin{equation}
\frac{d P_g}{d \tilde r} \bigg |_{\tilde r_{\rm top}} = \left ( \frac{d
P_i}{d \tilde r} + \frac{d P_e}{d
\tilde r} \right ) \bigg |_{\tilde r_{\rm top}} = \left [ (\gamma_i P_i + \gamma_e P_e) \left(
-\frac{1}{\tilde v}\frac{d \tilde v}{d \tilde r}-\frac{3}{\tilde r}
\right) + \frac{R_g (\gamma_e-1)}{c \tilde v} \dot U_e \right ] \bigg |_{\tilde r_{\rm top}} \ ,
\label{eqn:gas pressure gradient}
\end{equation}
where the energy transfer terms within $\dot U_e$ are described with Equation~(\ref{eqn:Udottotal}) in Section~\ref{section:gasenergyequation}. This gives us the equation for the gas pressure gradient needed later.

We now investigate how the free-streaming boundary condition affects the derivation of the momentum equation. The free-streaming condition discussed in Section~\ref{sec:BOUNDARY CONDITIONS} is given by
\begin{equation}
- \frac{c}{3 n_e \sigma_\parallel} \frac{dU_r}{dr} = c \, U_r \qquad r=r_{\rm top} \ .
\label{eqn:freestreaming2}
\end{equation}

Converting radiation energy density $U_r$ to radiation pressure $P_r$ in Equation (\ref{eqn:freestreaming2}), we can write the radiation pressure gradient with respect to the dimensionless radius, $\tilde r$, at the top of the column as
\begin{equation}
\frac{d P_r}{d \tilde r} \bigg |_{\tilde r_{\rm top}} = -3 n_e \sigma_\parallel R_g P_r \bigg |_{\tilde r_{\rm top}} \ .
\label{eqn:RadiationPressure1}
\end{equation}
We use the momentum equation from Equation (\ref{eqn:conservation of momentum equation}) and convert to dimensionless units to obtain the steady-state equation, given by
\begin{equation}
\frac{d \tilde v}{d \tilde r} = \frac{A}{\dot{M} c} \left(
\frac{d P_g}{d \tilde r} + \frac{d
P_r}{d \tilde r} \right) - \frac{1}{\tilde r^2
\tilde v} \ ,
\label{eqn:momentum1}
\end{equation}
where total pressure is the sum of gas and radiation pressures ($P_{\rm tot} = P_g + P_r$). By substituting for the gas pressure gradient ($d P_g/d \tilde r$) using Equation (\ref{eqn:gas pressure gradient}), the radiation pressure gradient from Equation (\ref{eqn:RadiationPressure1}), and substituting for flow velocity $\tilde v$ and gradient $d \tilde v / d \tilde r$  using the free-fall values assumed at the top of the column
\begin{equation}
\tilde v_{\rm top} = -\left( \frac{2}{\tilde r_{\rm top}} \right)^{1/2} \ ,
\label{eqn:utilde1}
\end{equation}
and
\begin{equation}
\frac{1}{\tilde v} \frac{d \tilde v}{d \tilde r} \bigg |_{\tilde r = \tilde r_{\rm top}} = -
\frac{1}{2 \tilde r_{\rm top}} \ ,
\end{equation}
we further reduce Equation (\ref{eqn:momentum1}) to the following:
\begin{equation}
\left [ -\frac{5}{2} \frac{\gamma_i P_i + \gamma_ e P_e}{\tilde r} - \frac{R_g(\gamma_e-1)}{c} \left (
\frac{\tilde r}{2} \right )^{1/2} \dot U_e\ - 3 n_e \sigma_\parallel R_g P_r \right ] \bigg |_{\tilde r = \tilde r_{\rm top}} = 0 \ .
\label{eqn:starting Mi0B}
\end{equation}

We obtain the final form for $\Mach_{i0}$ by using Equations~(\ref{eqn:soundspeeds}) to convert ion and electron pressures to their respective sound speeds, Equations~(\ref{eqn:convertingr}) to change to non-dimensional quantities, Equations~(\ref{eqn:massdensity}) and (\ref{eqn:numberdensity}) for $\rho$ and $n_e$ substitutions, Equation~(\ref{eqn:aetildetop}) which gives the relationship at the top of the column between $\tilde a_e$ and $\tilde a_i$, Equations~(\ref{eqn:starting Mr0}) and (\ref{eqn:starting Mi0}) to convert to Mach numbers, Equation~(\ref{eqn:utilde1}) for the bulk velocity at the top of the column, and Equation~(\ref{eqn:solid angle}) to substitute for $\Omega$. Finally, substituting for the specific heat coefficients ($\gamma_e=3$,
$\gamma_i=5/3$, and $\gamma_r=4/3$) we obtain
\begin{equation}
\left ( \frac{9}{4} \frac{\sigma_\parallel \, \dot M}{m_{\rm tot} c \, \Omega
R_g} \frac{1}{\Mach_{r0}^2} \frac{\tilde v}{\tilde r^2} -
\frac{7}{\Mach_{i0}^2} \frac{\tilde v^2}{\tilde r} - \frac{2 \Omega
R_g^3}{\dot M c^2} \, \tilde r^2
\dot U_e \right ) \bigg |_{\tilde r = \tilde r_{\rm top}} = 0\ ,
\label{eqn:appendix Mi0 equation}
\end{equation}
which can be rearranged to yield
\begin{equation}
\Mach_{i0} =\left ( \frac{9}{28} \frac{\sigma_\parallel \, \dot M}{m_{\rm tot} c \, \Omega
R_g} \frac{1}{\Mach_{r0}^2} \frac{1}{\tilde r \tilde v}  - \frac{2}{7}\frac{\Omega
R_g^3}{\dot M c^2} \frac{\tilde r^3}{\tilde v^2}
\dot U_e \right)^{-1/2} \bigg |_{\tilde r = \tilde r_{\rm top}} \ ,
\end{equation}
or using the value at the top of column for $\tilde v_{\rm top}$ (Equation (\ref{eqn:utilde1})), $\sigma_\parallel$ (Equation (\ref{eqn:tau parallel at starting height})), and substituting for $\Omega$ (Equation (\ref{eqn:solid angle})), we obtain
\begin{equation}
\Mach_{i0} =\left( -\frac{27}{56} \frac{1}{\Mach_{r0}^2} \left [ \frac{1}{1 - \left (\tilde r_{\rm top} / \tilde r_c \right )^{3/2}} \right ] - \frac{1}{7} \frac{R_g^4 \, \Omega_*}{\dot M c^2 R_*} \tilde r_{\rm top}^5 \dot U_{e,{\rm top}} \right)^{-1/2} \ .
\label{eqn:Mi0appendix}
\end{equation}
We see that the initial ion Mach number $\Mach_{i0}$ depends only on the choice of our free parameters $\tilde r_{\rm top}$, $\Omega_*$, and $\Mach_{r0}$, and on $\dot U_{e,{\rm top}}$. However, since $\dot U_{e,{\rm top}}$ is, in general, a function of the electron temperature $T_{e,{\rm top}}$, it is implicitly a (complicated) function of $\Mach_{i0}$ and Equation (\ref{eqn:Mi0appendix}) must be solved using a root finder.

{}

\label{lastpage}

\end{document}